\documentclass[rmp,aps,twocolumn,floatfix]{revtex4}

\usepackage{epsfig,psfrag}
\usepackage{color}
\usepackage{graphicx}
\usepackage{dcolumn}
\usepackage{bm}
\usepackage{textcomp}

\begin{document}



\title{Josephson and Andreev transport through quantum dots}

\author{A. Mart\'{\i}n-Rodero and A. Levy Yeyati\\
Departamento de F\'{i}sica Te\'{o}rica de la Materia
Condensada C-05\\
Universidad Aut\'{o}noma de Madrid, E-28049;
Madrid, Spain}

\begin{abstract}
In this article we review the state of the art on the transport properties of 
quantum dot systems connected to superconducting and normal electrodes.
The review is mainly focused on the theoretical achievements although
a summary of the most relevant experimental results is also given.
A large part of the discussion is devoted to the single level Anderson type models
generalized to include superconductivity in the leads, which already contains
most of the interesting physical phenomena. Particular attention is paid to
the competition between pairing and Kondo correlations, the emergence of
$\pi$-junction behavior, the interplay of Andreev and resonant tunneling, 
and the important role of Andreev bound states which characterized the
spectral properties of most of these systems. 
We give technical details on the several different analytical and numerical
methods which have been developed for describing these properties.
We further discuss the recent theoretical efforts devoted to extend
this analysis to more complex situations like multidot, multilevel or
multiterminal configurations in which novel phenomena is expected to
emerge. These include control of the localized spin states by a Josephson current
and also the possibility of creating entangled electron pairs by
means of non-local Andreev processes. 
\end{abstract}

\maketitle

\tableofcontents

\section{Introduction}
\label{intro}

The field of electronic transport in nanoscale devices is experiencing a fast
evolution driven both by advances in fabrication techniques and by the interest in
potential applications like spintronics or quantum information processing.
Within this context quantum dot (QD) systems are playing a central role. These
devices have several different physical realizations including semiconducting 
heterostructures, small metallic particles, carbon nanotubes or other molecules
connected to metallic electrodes. In spite of this variety a very attractive
feature of these devices is that they can usually be described theoretically
by simple "universal-like" models characterized by a few parameters. In addition
to their potential applications, these systems provide a unique test-bed for
analyzing the interplay of electronic correlations and transport properties
in nonequilibrium conditions.

Electron transport in semiconducting QDs 
has been studied since the early 90's when phenomena like Coulomb blockade (CB) was first observed \cite{kastner93}.
It soon became clear that QDs could allow to study the effect in transport properties of
basic electronic correlations phenomena like the
Kondo effect as suggested in early predictions \cite{glazman88,lee88}. These predictions were first
tested in metallic nanoscale junctions containing magnetic impurities \cite{ralph94}. 
However, a definitive breakthrough in the field came with the
observation of this effect in semiconducting QDs by Goldhaber et al. \cite{goldhaber98}
and Cronenwett et al. \cite{cronenwett98}. 
A great advantage of these devices is to offer the possibility of controlling the
relevant parameters, thus allowing a more direct comparison with the theoretical predictions.
Since then the effect of Kondo correlations in electronic transport has been observed in
several physical realizations of QDs based on carbon nanotubes (CNTs) \cite{nygard00}
and big molecules like fullerenes \cite{fullerenes}.

In parallel to these advances the study of superconducting (SC) transport in nanoscale devices
has also experienced a great development. 
From a theoretical point of view, with the advent of mesoscopic physics, a more
detailed understanding of superconducting transport was developed around the central
concept of coherent Andreev reflection (AR) \cite{andreev,beenakker-RMP}. This concept has allowed to unify the 
description of superconducting transport in different types of structures like normal metal-superconductor (N-S),
S-N-S junctions and superconducting quantum point contacts (SQPC).
Due to the multiple AR (MAR) mechanism the spectral density of systems 
like S-N-S or SQPCs is characterized by the presence of the so-called Andreev bound states (ABS) inside the
superconducting gap. These states are sensitive to the superconducting phase difference and are
thus current-carrying states which usually give the dominating contribution to the Josephson effect.

In recent years it has become feasible to produce hybrid systems combining different physical 
realizations of QDs well contacted to superconducting electrodes (for a review see \cite{defran10}).
Superconducting transport through QDs provides the interesting possibility to explore
the interplay of the AR mechanism and typical QD phenomena like CB and Kondo effect. 
The central aim of this review article is to discuss the main advances which have
taken place on this issue during the last years.

\begin{figure}
\begin{center}
\includegraphics[scale=0.7]{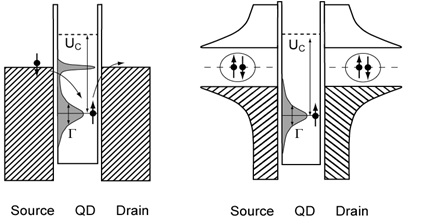}
\caption{Schematic representation of a single spin-degenerate level QD connected to normal (left panel) and
superconducting (right panel) leads with a large charging energy $U_c$. In the normal case the local density of states 
(LDOS) in the dot exhibits
the typical form corresponding to the Kondo regime with a narrow resonance at the Fermi energy and a
a broad resonance (of width $\Gamma$) below it. In the superconducting case the Kondo resonance 
(assumed to be narrower than the superconducting gap) disappears due to the competition with the
pairing correlations in the leads. Courtesy of C. Sch\"onenberger.} 
\label{kondo-pairing}
\end{center}
\end{figure}

A usual assumption in these studies is that a basic description of the main properties of 
these hybrid systems  can be provided by the Anderson model and its generalizations
to include SC leads, orbital degeneracy, etc.
The single level model applies when the dot
level spacing $\delta \epsilon$ is larger than all other relevant energy scales.
In the normal state the model allows to describe in a unified
way CB and the Kondo effect both in and out of equilibrium conditions \cite{revival-Kondo}. 
With two superconducting leads interesting new
physics already appear in the equilibrium case. Due to the Josephson effect, electron transport is
possible without an applied bias voltage and the model describes the
competition between Kondo effect and induced pairing within the dot.
Figure \ref{kondo-pairing} illustrates this competition: depending on the ratio between 
the Kondo temperature, $T_K$, and the superconducting order parameter, $\Delta$, there is a
phase transition between a Kondo dominated spin-singlet ground state to
a degenerate magnetic ground state. This transition is accompanied by a
reversal of the sign of the Josephson current. Thus in the magnetic
case the S-QD-S system constitutes a realization of the so-called
$\pi$-junction \cite{glazman89,spivak91}. A more detailed understanding of this
transition describing the appearance of intermediate phases with metastable states 
was achieved more recently \cite{rozhkov99,veci03}. 
The realization of a $\pi$-junction in QD systems should distinguished from the
similar phenomena in SFS junctions, where F denotes a ferromagnetic material
\cite{golubov04}.

Another basic situation which has been extensively explored (both
theoretically and experimentally) is the N-QD-S case. This situation
has been mainly analyzed in the linear transport regime in which 
it exhibits and interesting interplay between Kondo behavior and
resonant Andreev reflection. In contrast to the S-QD-S case this
system does not exhibit a quantum phase transition but there is
instead a crossover from a Kondo dominated regime for large $T_K/\Delta$
to a singlet superconducting regime in the opposite limit.

A third paradigmatic situation which has been studied is
the voltage biased S-QD-S system. This situation
constitutes a much more demanding task for the theory due to
the need of describing properly the strong out-of-equilibrium 
distribution which is generated by the infinite series of 
multiple Andreev reflection (MAR) processes together with the
effects of Coulomb interactions. The problem becomes 
simpler in two limiting cases: 1) when Coulomb interactions
are small and treated in a mean field approximation thus
allowing to analyze
the interplay of MAR and resonant tunneling and
2) when the Coulomb energy is the larger energy scale
in the problem and
the contribution of MAR processes are largely suppressed.

More recent developments include the study of
several QDs (connected either in series or in parallel)
coupled to one or more SC electrodes. 
In these situations there is a competition between 
not only the Kondo and the SC correlations but also
the possible magnetic coupling of the spins localized
within the dots. This could open the possibility to control the 
spin state of the dots system by means of the Josephson current. 

In addition there is a growing interest in
analyzing transport in these hybrid structures in a 
multiterminal configuration. One of the aims of 
these studies is the detection and 
control of non-local Andreev processes, which offer the
possibility of producing entangled electron pairs \cite{recher01}.

This review article is organized as follows:
in Section \ref{model-formalism} we introduce the basic theoretical models
used to describe the hybrid QD systems, which are largely based on the Anderson model
and its generalizations. In this section we also give a brief summary of the
application of nonequilibrium Green functions techniques for the calculation of
the electronic properties within these type of models. Section \ref{SQDS-eq} 
is devoted to review the main results for the S-QD-S systems in equilibrium,
i.e. in the dc Josephson regime. We first discuss the case of a non-interacting
resonant level which is useful to illustrate the emergence of ABSs and its
contribution to the Josephson current. In the subsequent subsections we 
give account of the different theoretical methods which have been used
to analyze the effect of interactions in the dc Josephson regime. 
A main issue which is discussed in this section are the phase diagrams
describing the transition to the $\pi$-state as a function of the model parameters.
We also give a brief account of the existing experimental results
for S-QD-S devices in this regime. The case of a QD coupled to both a normal and
a superconducting leads (N-QD-S) is addressed in Section \ref{NQDS}. Most of the
results obtained for this systems correspond to the linear regime with different
levels of approximation to include the Coulomb interactions. We also briefly
mention existing results for the non-linear regime and the few experiments 
which have been reported of this case up to date. Section \ref{SQDS-neq} 
is devoted to the voltage biased S-QD-S system. We first give some technical
details on the calculations for the non-interacting case in order to 
illustrate how to deal with the out of equilibrium MAR mechanism. We also 
comment in this section the few existing results including interactions
in this regime and give an account of the related experiments.  
Finally, in Section \ref{multi} we discuss several different situations which
go beyond the single-level two-terminal case discussed in the previous sections.
These include: multidot systems connected either in parallel or in series,
the multilevel situation and setups in a multiterminal configuration. 
We conclude this article with a brief discussion of related issues not included
in the present review and of topics which, in our view, deserve to be further
analyzed in the near future.

\section{Basic models and formalism}
\label{model-formalism}

The minimal model for a QD coupled to metallic electrodes in the regime
where $\delta \epsilon$ is sufficiently large to restrict the analysis to
a single spin-degenerate level is provided by the single level Anderson model \cite{anderson61},
with the Hamiltonian $H=H_L+H_R+H_T+H_{QD}$ where $H_{QD}$ corresponds 
to the uncoupled dot given by
\begin{equation}
H_{QD} = \sum_{\sigma} \epsilon_0 c^{\dagger}_{0\sigma} c_{0\sigma} +
U n_{0\uparrow} n_{0\downarrow} ,
\label{HQD}
\end{equation}
where $c^{\dagger}_{0\sigma}$ creates and electron with spin $\sigma$ on the dot level
located at $\epsilon_0$ and $U$ is the local Coulomb interaction for two electrons
with opposite spin within the dot ($n_{0\sigma}= c^{\dagger}_{0\sigma}c_{0\sigma}$).
On the other hand, $H_{L,R}$ describe the uncoupled left and right leads which can be either
normal or superconducting. In this last more general case, they are usually represented
by a BCS Hamiltonian of the type
\begin{equation}
H_{\nu} = \sum_{k\sigma} \xi_{k,\nu} c^{\dagger}_{k\sigma,\nu} c_{k\sigma,\nu} +
\sum_{k} \left(\Delta_{\nu} c^{\dagger}_{k\uparrow,\nu}c^{\dagger}_{-k\downarrow,\nu} + 
\mbox{h.c.} \right),
\label{Hnu}
\end{equation}
where $c^{\dagger}_{k\sigma,\nu}$ creates an electron with spin $\sigma$ at the
single-particle energy level $\xi_{k,\nu}$ of the lead $\nu=L,R$
(usually referred to the lead chemical potential, i.e. $\xi_{k,\nu} = \epsilon_{k,\nu} - \mu_{\nu}$) 
and $\Delta_{\nu}=|\Delta_{\nu}|\exp{(i\phi_{\nu})}$ is the (complex) superconducting order parameter on lead $\nu$.
Finally, $H_T$ describes the coupling between the QD level to the leads and has the form
\begin{equation}
H_{T} = \sum_{k\sigma,\nu} \left(V_{k,\nu} c^{\dagger}_{k\sigma,\nu} c_{0\sigma} + \mbox{h.c.} \right).
\label{HT}
\end{equation}

In order to reduce the number of parameters it is usually assumed that the normal density of states
of the leads $\rho_{\nu}(\omega)$ is a constant in the range of energies around the Fermi
level of the order of the superconducting gap and that the $k$ dependence of the
hopping elements $V_{k\nu}\simeq V_{\nu}$ can be neglected within this range. The coupling to the leads
is then characterized by a single parameter $\Gamma_{\nu} = \pi \rho_{\nu} |V_{\nu}|^2$, 
which can be interpreted as the normal {\it tunneling rate} from the dot to the leads.

Within the above model $\Delta_{\nu}=0$ would correspond to the normal state. 
For vanishing $\Gamma_{\nu}$ the model is in the so-called atomic limit
which is characterized by sharp peaks in the spectral density at $\epsilon_0$
and $\epsilon_0+U$. This limit corresponds to the Coulomb blockade regime
in an actual QD where the conductance is strongly suppressed except at the
charge degeneracy points.
When the couplings to the leads increase ($\Gamma_{\nu}$ become larger than temperature)
virtual processes allow the charge and spin in the dot to fluctuate and
a resonance at around the Fermi energy appears close to half-filling
due to the Kondo effect.
This simple model thus already captures the most relevant Physics of ultrasmall QDs with well
separated energy levels, like the crossover from the Coulomb blockade to the Kondo regime
as the temperature is lowered.

The simplicity of this model has allowed to obtain
exact results in the equilibrium case by means of the Bethe ansatz \cite{wiegmann}. 
The most basic of these
results is the expression for the Kondo temperature \cite{hewson93}
\begin{equation}
T_K = \sqrt{\frac{U\Gamma}{2}} \exp{\left(-\frac{\pi|\epsilon_0(\epsilon_0+U)|}{2U\Gamma}\right)},
\end{equation}
where $\Gamma = \Gamma_L + \Gamma_R$.

This temperature characterizes the crossover from the so-called local moment regime for
$T \gg T_K$ to the regime where Kondo correlations between the localized spin within the
QD and the spin of the electrons in the leads sets in. Although this physics is basically
well understood since the 70's for the case of magnetic impurities in metals, its 
consequences for transport in artificial nanostructures has started to be developed
much more recently specially driven by the advances in fabrication techniques. 
In this respect while the linear transport properties are well understood still open
questions remain regarding the non-equilibrium regime.

In the superconducting case another energy scale, associated with the superconducting
gap, appears bringing additional complexity to the problem, whose description is
in fact the scope of this review. Even in the equilibrium situation the Anderson 
model with superconducting leads contains the non-trivial Physics associated to
the Josephson effect. A relevant parameter is then provided by the 
superconducting phase difference $\phi=\phi_{L}-\phi_{R}$. 

In order to analyze the electronic and transport properties of a general
superconducting system in the presence of interactions and in a non-equilibrium 
situation it is convenient to use Green function techniques. 
The Keldysh formalism provides the basic tools for this purpose.

Due to the presence of superconducting correlations it is convenient to 
introduce the Nambu spinor field operators $\Psi_j, \Psi^{\dagger}_j$,
with $\Psi^{\dagger}_j = \left( c^{\dagger}_{j\uparrow}, c_{j\downarrow} \right)$ where
$j=k\nu,0$ denotes the $\nu=L,R$ electrodes and the dot level respectively.
The different terms in the model Hamiltonian of Eqs. (\ref{HQD}), (\ref{Hnu}) and (\ref{HT}) can
then be written as
\begin{eqnarray}
H_{QD} &=& \Psi_0^{\dagger} \hat{h}_0 \Psi_0 + U n_{0\uparrow}n_{0\downarrow} \nonumber \\
H_{\nu} &=& \sum_k \Psi_{k\nu}^{\dagger} \hat{h}_{k\nu} \Psi_{k\nu} \nonumber \\
H_{T} &=& \sum_{k,\nu} \left(\Psi_{k\nu}^{\dagger} \hat{V}_{k\nu} \Psi_{0} + \mbox{h.c.} \right) \;\; ,
\end{eqnarray}
where $\hat{h}_0 = \epsilon_0 \tau_3$, $n_{0\sigma} = \frac{1}{2} \Psi^{\dagger}_0 \left[\tau_0 + \mbox{sign}(\sigma) \tau_3\right] \Psi_0$,
$\hat{h}_{k\nu} = \xi_{k\nu}\tau_3 + \mbox{Re}{\Delta_{\nu}} \tau_1 + \mbox{Im} {\Delta_{\nu}}\tau_2$, $\tau_{i=0,1,2,3}$ being the
Pauli matrices defined in Nambu space.

Starting from these spinor field operators, 
generalized single-particle propagators can be defined
along the Keldysh closed time loop as

\begin{equation}
\hat{G}^{\alpha\beta}_{j,j'}(t,t') = -i \langle T_c \left[ \Psi_j(t_{\alpha}) 
\Psi^{\dagger}_{j'}(t'_{\beta}) \right] \rangle  \;\; ,
\end{equation}
where $\alpha,\beta \equiv \pm$ denote the two branches in the Keldysh contour. These propagators allow to calculate
in a straightforward way most of the relevant quantities like the mean charge and the superconducting order parameter
within the dot as well as the mean current through it, which are given by

\begin{eqnarray}
n_0(t) &=& i \mbox{Tr}\left(\tau_3 \hat{G}^{+-}_{00}(t,t)\right)- 1  \\
\mbox{Re}{\Delta_0}(t) &=& U \mbox{Tr}\left(\tau_1 \hat{G}^{+-}_{00}(t,t)\right) \nonumber \\
I_{\nu}(t) &=& \frac{e}{\hbar} \sum_k \mbox{Tr}\left(\tau_3 \left[
\hat{V}_{k\nu} \hat{G}^{+-}_{k\nu,0}(t,t) -\hat{V}_{\nu k} \hat{G}^{+-}_{\nu k,0}(t,t)\right]\right) . \nonumber
\label{mean-quantities}
\end{eqnarray}

These expressions are formally exact but of little use unless the single-particle
propagators are known. Fully analytical and exact results can only be obtained in the
non-interacting case. In the presence of interactions numerical methods allow to 
obtain exact results in the equilibrium case. In a more general case one is bound
to find reasonable approximations for these propagators valid for a restricted 
range of parameters. It is usually convenient to express these approximations
in terms of a self-energy $\Sigma$ which is related to the propagators by
the usual Dyson equation 

\begin{equation}
\breve{\hat{G}}_{00} = \breve{\hat{G}}^{(0)}_{00} + \breve{\hat{G}}^{(0)}_{00}
\breve{\hat{\Sigma}}_{00} \breve{\hat{G}}_{00} ,
\end{equation}
where $\breve{\hat{G}}^{(0)}$ denotes the unperturbed  propagators corresponding to 
an appropriately defined non-interacting Hamiltonian $H_0$
(the $\breve{}$ symbol indicates matrix structure in Keldysh space) and where integration
over internal times is implicitly assumed. 
Different approximations for the self-energy associated with electron-electron interactions
are discussed in the forthcoming sections. The analysis of the problem is greatly simplified
in the stationary case where Fourier methods can be applied both in the equilibrium 
and in the non-equilibrium situation. We shall start discussing in the next section the simplest
possible case of an equilibrium situation. In this case further simplification arises
from the possibility of expressing all Keldysh propagators in terms of the retarded-advanced propagators
and the Fermi equilibrium distribution function, $n_F(\omega) = 1/\left[1 + \exp{\beta (\omega-\mu)}\right]$, 
where $\mu$ is the 
chemical potential and $\beta = 1/k_BT$ as

\begin{equation}
\hat{G}^{+-}(\omega) = n_F(\omega) \left[G^a(\omega) - G^r(\omega)\right] \;\; .
\end{equation}

Thus, for instance, the mean current can be written as

\begin{equation}
I_{\nu} = \frac{e}{h} \sum_k \int d\omega n_F(\omega) \mbox{Tr}\left[V_{k\nu} \mbox{Re}\left(G^a_{k\nu,0}-G^r_{k\nu,0}\right)\right].
\label{current}
\end{equation}

\section{Equilibrium properties of quantum dots with superconducting leads}
\label{SQDS-eq}

For a nanoscale system coupled to superconducting electrodes a finite
current can flow even in the absence of an applied bias voltage due to the Josephson
effect. We shall illustrate this effect starting from the non-interacting situation
$(U = 0)$ within the single level Anderson model introduced above.
In this case the advanced-retarded Green function of the coupled dot can be
expressed as

\begin{widetext}
\begin{equation}
G^{a,r}_{00}(\omega) = \left( \begin{array}{cc} 
\omega - \epsilon_0 - \Gamma_L g^{a,r}_L - \Gamma_R g^{a,r}_R & 
\Gamma_L e^{i\phi_L} f^{a,r}_L + \Gamma_R e^{i\phi_R} f^{a,r}_R \\
\Gamma_L e^{-i\phi_L} f^{a,r}_L + \Gamma_R e^{-i\phi_R} f^{a,r}_R &
\omega + \epsilon_0 - \Gamma_L g^{a,r}_L - \Gamma_R g^{a,r}_R \end{array} 
\right)^{-1} ,
\label{retarded-advanced-GF}
\end{equation}
\end{widetext}
where $f^{a,r}_{L,R}=|\Delta_{L,R}|/\sqrt{|\Delta_{L,R}|^2 - (\omega \pm i0^+)^2}$
and $g^{a,r}_{L,R}=-(\omega \pm i0^+)f^{a,r}/|\Delta_{L,R}|$ are the dimensionless 
BCS green functions of the uncoupled leads. 
For simplicity we focus below on the case where both leads are of the same material for which
$|\Delta_L|=|\Delta_R|=\Delta$. 

\begin{figure}
\begin{center}
\includegraphics[scale=0.4]{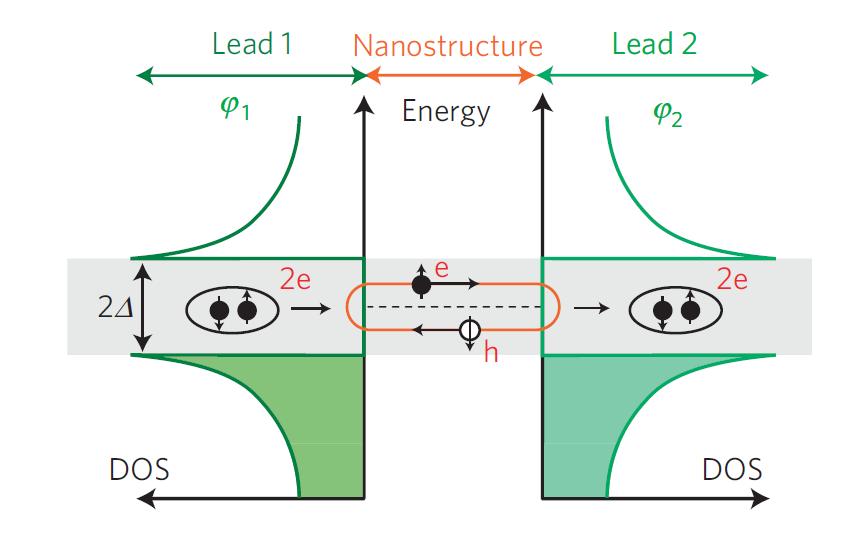}
\caption{Schematic representation of the physical mechanism responsible for the formation of
ABs in a generic nanostructure coupled to superconducting leads. Reprinted by permission
from Macmillan Publishers Ltd: Nature Physics \cite{pill10}, copyright (2010).}
\label{MAR-schematic}
\end{center}
\end{figure}

The spectral density associated with this model exhibits bound states within the superconducting
gap (i.e $|\omega| \leq \Delta$). Physically, the ABSs arise from
virtual multiple Andreev reflection processes at the interface between the dot and each of the leads. 
In such processes and for energies inside the gap,
the electrons (holes) incident towards the leads are reflected back as holes (electrons), as
illustrated in Fig. \ref{MAR-schematic}. The condition for the appearance of ABSs is that the 
accumulated phase in the closed trajectory be a multiple of $2\pi$, which is equivalent 
to satisfying the equation

\begin{eqnarray}
D(\omega) &\equiv& \left[\omega - \epsilon_0 - \Gamma g(\omega) \right]
\left[\omega + \epsilon_0 - \Gamma g(\omega) \right] \nonumber\\
&&- |\Gamma_L e^{i\phi_L} + \Gamma_R e^{i\phi_R}|^2f(\omega)^2 = 0 \;\; ,
\label{ABSequation}
\end{eqnarray}
where we have used that the  
dimensionless BCS Green functions $g,f$ become real for energies inside the superconducting gap.
It can be shown \cite{rozhkov00} that this equation has two real roots inside the gap
$\pm \omega_s$, i.e. symmetrically located with respect to the Fermi energy.  

An essential property of the ABSs is that they correspond to current-carrying states. 
In fact, due to its dependence on the superconducting phase difference they have associated
a Josephson current $i_s(\phi) =  2e/\hbar \left(\partial \omega_s/\partial \phi\right)$. The
total Josephson current is obtained by adding the contribution of all states with finite
occupation. Thus, at zero temperature only the lower ABS contributes and 
there is an additional contribution from the continuous spectrum $\omega < -\Delta$ 
which we discuss below.

\begin{figure}
\begin{center}
\includegraphics[scale=0.35]{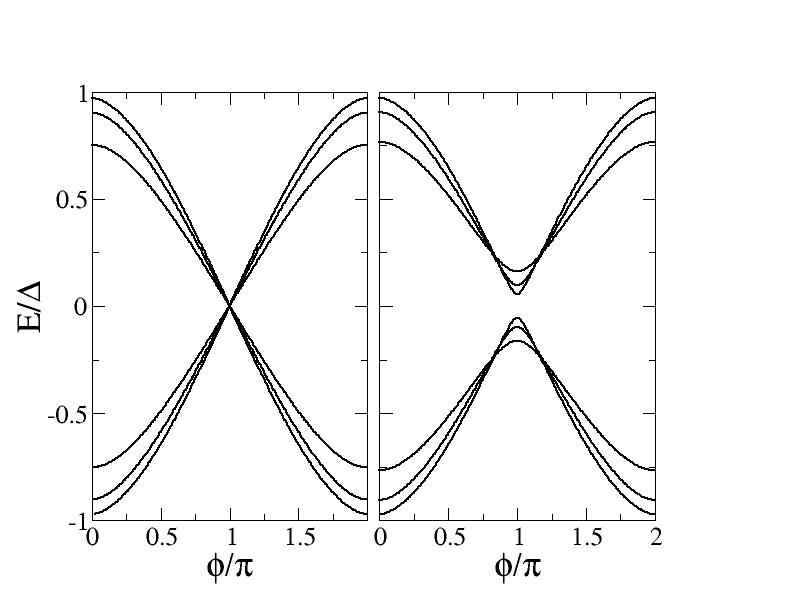}
\caption{Andreev bound states for a non-interacting S-QD-S with different $\Gamma/\Delta$ values:
1.0, 2.0 and 4.0. Left panel corresponds to the e-h symmetric case ($\epsilon_0=0$) and
the right panel to a case with $\epsilon_0 = 0.5 \Delta$.}
\label{figureABSs}
\end{center}
\end{figure}

The ABS equation (\ref{ABSequation}) becomes particularly simple for an electron-hole
and left-right symmetric case (i.e. $\epsilon_0 =0$ and $\Gamma_L = \Gamma_R$)
when it can be reduced to 

\begin{equation}
\omega \pm \Delta \cos{\phi/2} + \frac{\omega \sqrt{\Delta^2 - \omega^2}}{\Gamma} = 0,
\label{ABS-symmetric}
\end{equation}
which for $\Gamma \gg \Delta$ has the simple solutions $\omega_s \simeq \pm \tilde{\Delta}
\cos{\phi/2}$, where $\tilde{\Delta}$ is a reduced gap parameter which for $|\phi| \ll 1$ is given by
$\tilde{\Delta} = \Delta \left[1 - 2(\Delta/\Gamma)^2\right]$. The ABSs for this case 
tend to the ones of a perfectly transmitting one channel superconducting contact
$\pm \Delta \cos{\phi/2}$, the main qualitative difference being the reduced amplitude 
of their dispersion detaching them from the gap edges at $\phi = 2n\pi$. This is illustrated
in Fig. \ref{figureABSs}(a).

In the absence of electron-hole symmetry (i.e. $\epsilon_0 
\ne 0$) a finite internal gap between the upper and lower ABSs appears as in the
case of a non-perfect transmitting one channel contact. When
$\Delta/\Gamma \rightarrow 0$ the ABSs for this case are given by $\omega_s =
\pm \Delta \sqrt{1 - \tau \sin^2{(\phi/2)}}$, where $\tau = 1/(1 + (\epsilon_0/\Gamma)^2)$
is the normal transmission at the Fermi energy. Outside this limiting case the ABSs exhibit
both the internal gap and the detachment from the continuum states at $\phi= 2n\pi$,
as it is illustrated in Fig. \ref{figureABSs}(b).

\begin{figure}[htb!]
\begin{center}
\includegraphics[scale=0.3]{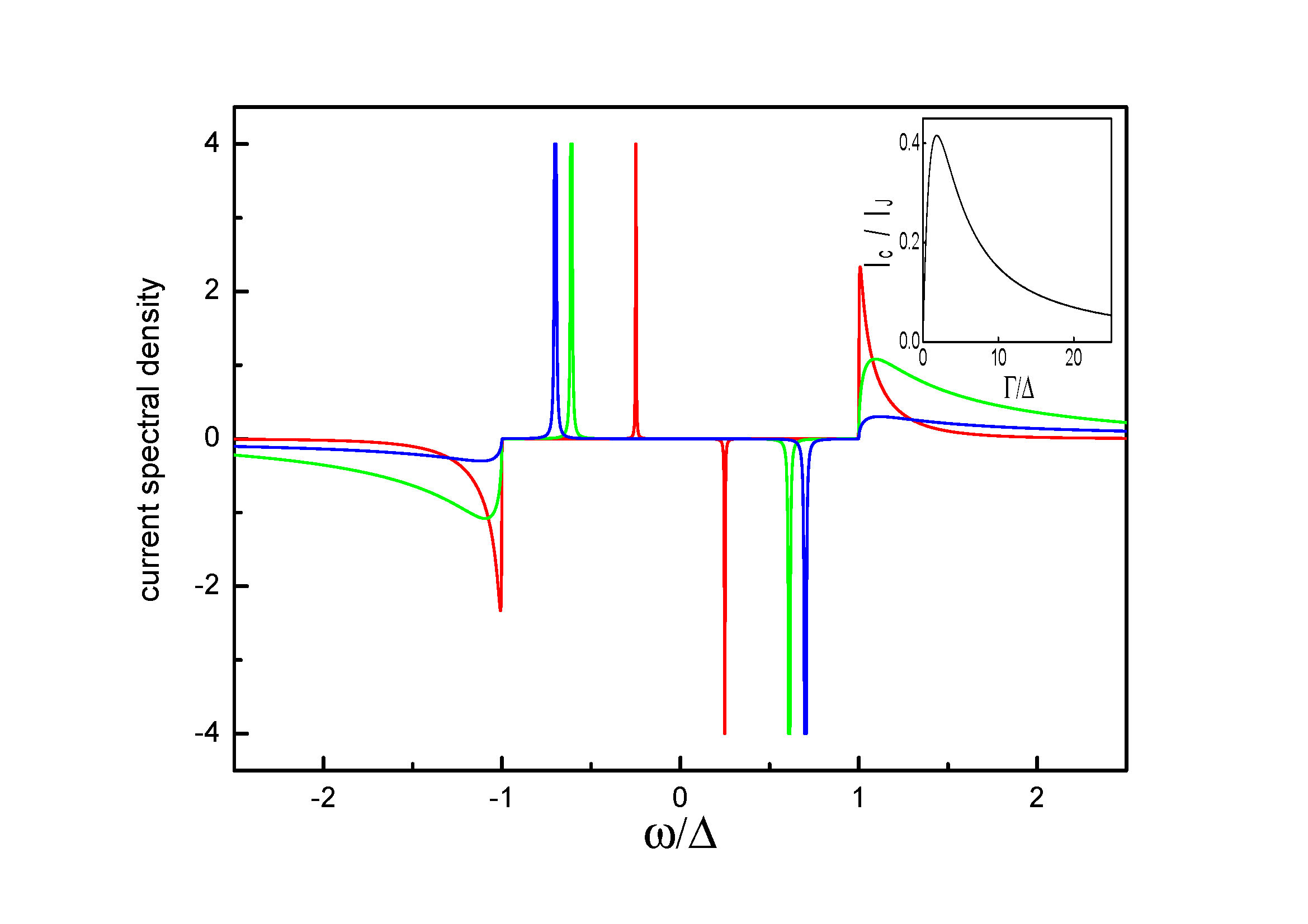}
\caption{Current spectral density for the non-interacting S-QD-S system with $\epsilon_0$ and
$\phi=1.5$ for increasing values of $\Gamma/\Delta=$ 0.5 (red), 4 (green)  and 16 (blue). 
A small broadening has been introduced to help visualizing the ABs contribution, which 
has been truncated for the sake of clarity. 
The inset shows the relative contribution
of the continuous spectrum $I_c$ compared to the total value $I_J$.}
\label{current-density}
\end{center}
\end{figure}

An interesting issue to comment is that the contribution from the states in the
continuous spectrum becomes negligible when $\Delta/\Gamma \rightarrow 0$. In this limit
the zero temperature current-phase relation (CPR) is simply given by
\begin{equation}
i_s(\phi) = \frac{e\Delta}{2\hbar}\frac{\tau \sin{\phi}}{\sqrt{1 - \tau \sin^2{(\phi/2)}}} \;\; ,
\end{equation}
which is the CPR of a one channel contact with transmission $\tau$. 
For finite $\Delta/\Gamma$ there is a contribution from the continuum states.
The expression for the total Josephson current in the non-interacting case can be
derived from Eq. (\ref{current}), which yields the following compact form

\begin{equation}
I = \frac{8e}{h} \Gamma_L \Gamma_R \sin{\phi} \int d\omega n_F(\omega) \mbox{Im} \left(\frac{f^r_L f^r_R}{D^r} \right) .
\label{current-non-int}
\end{equation}

The current density in Eq. (\ref{current-non-int}) contains both the contribution from the
ABSs (region $|\omega| < \Delta$) and from the continuous spectrum $|\omega|>\Delta$.
The behavior of the current density as a function of $\Delta/\Gamma$ is depicted in
Fig. \ref{current-density}. As can be observed the contribution from the continuous spectrum
has the opposite sign compared to the one arising from the ABS. The inset shows that 
this contribution 
becomes negligible in the limit $\Delta/\Gamma \rightarrow 0$ and
reaches a maximum for $\Delta/\Gamma \sim 2$. 

In the rest of this section we shall discuss the different theoretical approaches to include the 
effect of interactions in the dc Josephson effect through single level QD models. We also
include a subsection on experimental results.

\subsection{Cotunneling approach}
\label{cotunneling}

From a theoretical point of view the simplest approach to account for 
the effect of interactions in the Josephson current is to perform a perturbative
expansion to the lowest non-zero order in the tunnel Hamiltonian. This so-called
cotunneling approach was first used by Glazman and Matveev \cite{glazman89}, who
predicted the onset of the $\pi$-junction behavior by this method. More precisely
they obtained for the $U \rightarrow \infty$ limit

\begin{equation}
I(\phi) = \lambda \frac{e}{\hbar}\frac{\Gamma_L \Gamma_R}{\Delta} F\left(\frac{|\epsilon_0|}{\Delta}\right) \sin \phi ,
\end{equation}
where $\lambda$ changes its value from 2 ($\epsilon_0 >0$) to -1 ($\epsilon_0 <0$), thus
describing the transition to the $\pi$-phase, the function $F(x)$ having the form

\begin{equation}
F(x) = \frac{1}{\pi^2} \int \frac{dt_1 dt_2}{(\cosh t_1 + \cosh t_2)(x + \cosh t_1)(x + \cosh t_2)} . \nonumber\\
\end{equation}

This approximation is clearly not valid for describing the Kondo regime ($T_K \gg \Delta$) which
requires non-perturbative approaches like the ones discussed in following subsections. 

\subsection{Mean field and variational methods}
\label{mean-field}

Another simple approximate methods to deal with interactions are those
of a mean field type like the Hartree-Fock approximation (HFA) or the slave-boson
mean field (SBMF). In spite of their simplicity these approximations are able to
capture important qualitative features due to interactions in certain limits. 

We start by analyzing the HFA. In the context of magnetic impurities in
superconductors this method was first applied by Shiba \cite{shiba73}, 
while for the analysis of the Josephson effect it was first considered in Ref. \cite{rozhkov99}
and further analyzed in \cite{veci03}. Within this approximation 
electrons with a given spin ``feel" a static potential due to the average
occupation of the opposite spin, which corresponds to a simple constant
self-energy $(\Sigma_{00})_{11} = U <n_{0\downarrow}>$ and
$(\Sigma_{00})_{22} = -U <n_{0\uparrow}>$. In principle within the same level
of approximation there appears a non-diagonal self-energy taking into
account the induced pairing within the dot due to proximity effect,
which can be written as $(\Sigma_{00})_{12} = U <c^{\dagger}_{0\uparrow} c^{\dagger}_{0\downarrow}>$.
The effect of this  non-diagonal contribution, 
which was not included in Ref. \cite{rozhkov99}, was analyzed in Ref. \cite{veci03}.
The determination of the self-energy in the HFA requires a self-consistent 
calculation by using Eqs. (\ref{mean-quantities})
which cannot be performed analytically in general.  

The most significant result within the HFA is the appearance of a 
broken symmetry state in which the dot acquires a finite magnetic moment
(i.e. $<n_{0\uparrow}> \ne <n_{0\downarrow}>$) for certain ranges of parameters.
In this respect one should be cautious in principle as the HFA is known
to predict also broken symmetry states for the same model with normal leads
\cite{anderson61}, which are known to be spurious. However, for the superconducting case
ground states with a finite magnetization do exist for certain parameter range
as commented in the introduction. As it is shown below the HFA gives a rather good
estimate of the magnetic ground state energy in the regions where it is the most
stable phase.

\begin{figure}
\begin{center}
\includegraphics[scale=0.6]{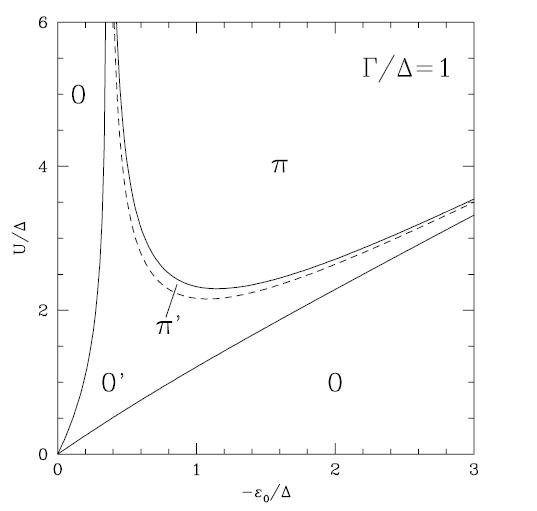}
\caption{Phase diagram in the $U,-\epsilon_0$ plane for $\Gamma/\Delta=1$ obtained
using the HFA \cite{rozhkov99}). The phases are classified
into $0$, $0'$, $\pi'$ and $\pi$ as explained in the text. Reprinted figure with permission
from A.V. Rozhkov and D. Arovas, Physical Review Letters {\bf 82}, 2788, 1999 \cite{rozhkov99}.
Copyright (1999) by the American Physical Society.}
\label{phase-diagram-HFA}
\end{center}
\end{figure}

The general properties of the ground state within the HFA are most conveniently
displayed by a phase-diagram like the one in Fig. \ref{phase-diagram-HFA}.
In this diagram the notation "0", "$0'$", "$\pi'$" and "$\pi$" corresponds to
the different ground state symmetries. Thus, "0" corresponds to the 
non-magnetic case for all values of $\phi$ (the absolute energy minimum being located
at $\phi=0$), while the "$\pi$" denotes that
the magnetic solution is the most stable for all $\phi$ values
(with the absolute minimum at $\phi=\pi$). On the other hand,
"$0'$" and "$\pi'$" refer to intermediate situations with mixed magnetic and
non-magnetic solutions as a function of $\phi$, the name indicating  
whether the absolute energy minimum corresponds to a non-magnetic or a magnetic
solution.
From Fig. \ref{phase-diagram-HFA} the broken symmetry ground states are predicted
to appear around the $\epsilon_0 = -U/2$ line, which corresponds to the half-filled
case for sufficiently large $U$, i.e. $U > \Gamma, \Delta$. It is worth noticing that
for normal leads this region corresponds to the deep Kondo regime, which anticipates
an interesting interplay between both effects in the superconducting case beyond the
HFA.	

\begin{figure}
\begin{center}
\includegraphics[scale=0.55]{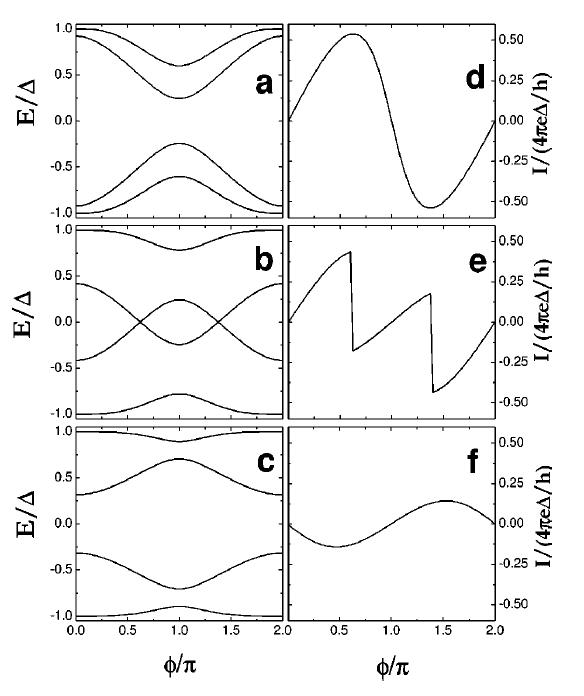}
\caption{Evolution of the ABSs and Josephson current in the ``toy" model of Ref. \cite{veci03} for
$\epsilon_0/\Gamma = -0.5$ and increasing $E_{ex}/\Gamma$ parameter: 0.25 (upper panels), 0.75 (middle panels) and
1.5 (lower panels). Reprinted figure with permission
from E. Vecino {\it et al.}, Physical Review B {\bf 68}, 035105, 2003 \cite{veci03}.
Copyright (2003) by the American Physical Society.}
\label{ABS-toy-model}
\end{center}
\end{figure}

Further insight into the HFA solution can be provided by a "toy" model
introduced in Ref. \cite{veci03} (A similar model was analyzed in Ref. \cite{benjamin07}). 
In this simplified model the finite 
magnetization which appears in the HFA is simulated by means of an exchange
field parameter, $E_{ex}$, corresponding to the splitting of the diagonal dot levels for each 
spin, i.e. $\epsilon_{0\sigma} = \epsilon_0 + \sigma E_{ex}$. 
The analysis of the Andreev states within this toy model is similar to the
one given at the beginning of this section for the non-interacting case, and becomes particularly simple in the limit 
$\Delta \ll \Gamma$ in which they adopt the analytical expression

\begin{eqnarray}
\left(\frac{\omega_{\pm}}{\Delta}\right)^2 =
\frac{\cos^2{\phi/2} + 2 E^2 + Z^2(Z^2 + \sin^2{\phi/2})
\pm 2 X S(\phi)}
{Z^4  + 2 (X^2 + E^2) + 1}, \nonumber\\
\label{estados}
\end{eqnarray}
where $E = \epsilon_0/2\Gamma$, $X = E_{ex}/2\Gamma$ and $Z^2 =
X^2 - E^2$ and $S(\phi)$ is given by

\[S(\phi) = \sqrt{Z^2 \cos^2{\phi/2} + E^2 + \sin^2{\phi}/4} . \]

This expression clearly shows that the effect of the exchange field is to 
break the spin degeneracy producing an splitting of the ABSs. Consequently
for $E_{ex} \ne 0$ one could in principle observe up to four ABSs in the
spectral density. The evolution of these states with increasing $E_{ex}$ is
shown in Fig. \ref{ABS-toy-model} together with the corresponding Josephson
current. While for $E_{ex} < \Gamma$ the splitting is small and states corresponding
to different spin orientation do not cross, for increasing $E_{ex}$ the upper
and lower states closer to the Fermi energy begin to cross yielding a current-phase
relation of $0'$ or $\pi'$ character. Eventually for sufficiently large $E_{ex}$ these
two states completely interchange position with a complete reversal of the sign of
the current ($\pi$-phase). It should be mentioned that although in this toy model
the spin degeneracy is artificially broken, it nevertheless qualitatively 
simulates the behavior of the exact solution in the magnetic phase. 

Another simple approach of a mean field type is provided by the slave boson mean field
approximation (SBMFA). This method was introduced by Coleman for the normal Anderson model 
\cite{coleman84,hewson93}. It is based in the introduction of auxiliary boson fields $b^{\dagger}_0,b_0$ which 
act as projectors onto the empty impurity state. At the same time fermion creation and annihilation
operators $f^{\dagger}_{0\sigma}, f_{0\sigma}$ are introduced for describing the singly occupied states. In order to
get rid of the doubly occupied states in the $U \rightarrow \infty$ these operators should satisfy
the completeness relation
\begin{equation}
b^{\dagger}_0b_0 + \sum_{\sigma} f^{\dagger}_{0\sigma}f_{0\sigma} = 1
\label{constraint}
\end{equation}

In terms of these operators the terms $H_{QD}$ and $H_T$ become 
\begin{eqnarray}
H_{QD} &=& \sum_{\sigma} \epsilon_0 f^{\dagger}_{0\sigma}f_{0\sigma} \nonumber \\
H_T &=& \sum_{k\sigma,\nu} \left(V_{k\nu} c^{\dagger}_{k\nu,\sigma} b^{\dagger}_0 f_{0\sigma} + \mbox{h.c.} \right).
\end{eqnarray}

So far this transformation is exact in the $U \rightarrow \infty$ limit. 
Specific diagrammatic methods to obtain the impurity self-energy in this slave boson
formulation have been developed \cite{bickers87}.
Within this formulation the simplest solution is provided by the mean field approximation
in which the boson operator is treated as a c-number. In fact the dot Hamiltonian reduces in this case to
\begin{eqnarray}
H^{MF}_{QD} &=& \sum_{\sigma} \epsilon_0 f^{\dagger}_{0\sigma}f_{0\sigma} + \lambda \left( |b_0|^2  + \sum_{\sigma}
f^{\dagger}_{0\sigma}f_{0\sigma} - 1 \right) , \nonumber\\
&&
\end{eqnarray}
where $\lambda$ is a Lagrange multiplier associated to the constraint (\ref{constraint}). The problem becomes
equivalent to a non-interacting impurity model with renormalized parameters $\tilde{\epsilon}_0 = \epsilon_0 + \lambda$
and $\tilde{\Gamma}_{\nu} = |b_0|^2 \Gamma_{\nu}$. Self-consistency is achieved by minimizing the system
free energy.

Strictly speaking the mean field approximation is only valid in the $N\rightarrow \infty$ limit
where $N$ is the level degeneracy of the Anderson model (for the single level case N=2). 
However, the mean field approximation yields a reasonably good description of quantities like
the Kondo temperature in the normal case \cite{hewson93}.
When applied to the Anderson model with superconducting electrodes the SBMFA is only valid in the
regime $T_K \gg \Delta$ as it is not able to describe the transition to the $\pi$-phase when
$T_K \sim \Delta$. In the regime $T_K \gg \Delta$ the ABSs as described by the SBMFA corresponds
to the non-interacting case with renormalized parameters $\tilde{\epsilon}_0$ and $\tilde{\Gamma}$.
In this way the ABSs within the SBMFA in this regime would be given by $\omega_s(\phi) 
= \pm \sqrt{1 - \tilde{\tau}\sin^2 \phi/2}$ with $\tilde{\tau} = 4T_K^2/(\tilde{\epsilon}_0^2 + 4T_K^2)$. 
In principle, the self-consistent effective parameters in the superconducting state can differ from those in the
normal state. However, in the limit $T_K \gg \Delta$ in which the approximation is supposed
to work this difference can be neglected. The SBMFA in the $U \rightarrow \infty$ limit
has only been applied for the case of superconducting leads in a few references: 
Avishai et al. \cite{avis03} for analyzing the dc current with an applied bias voltage,
and in Ref. \cite{levy03} for studying the dynamics of Andreev states in the Kondo regime.
Both works correspond to the non-equilibrium situation which will be discussed in Section \ref{SQDS-neq}.

For a proper description of the phase-diagram within a mean-field slave boson approach a
finite-$U$ version of the method, like the one
introduced by Kotliar and Ruckenstein \cite{kotliar86}, is necessary. Within this method
the number of auxiliary boson fields is extended up to four, denoted by $e,p_{\sigma}$ and
$d$, which project into the empty, singly occupied (with either spin orientation) and
doubly occupied dot states respectively. These operators must satisfy the constraints

\begin{eqnarray}
\sum_{\sigma} p^{\dagger}_{\sigma} p_{\sigma} + e^{\dagger}e + d^{\dagger}d =1 \nonumber \\
c^{\dagger}_{0\sigma}c_{0\sigma} = p^{\dagger}_{\sigma} p_{\sigma} + d^{\dagger}d .
\label{constraints-finiteU}
\end{eqnarray}

For recovering the non-interacting limit it is necessary to introduce also an 
auxiliary operator $z_{\sigma}=(1-d^2-p_{\sigma}^2)^{-1/2}
(e p_{\sigma}+ p_{\bar{\sigma}}d)
(1-e^2-p_{\bar{\sigma}}^2)^{-1/2}$, in terms of which the model Hamiltonian becomes

\begin{widetext}
\begin{eqnarray}
H &=& H_L+ H_R +
\sum_{\sigma}\epsilon_0 \hat{f}_{0\sigma}^{\dagger}\hat{f}_{0\sigma}+
U d^{\dagger} d +
\sum_{k \nu,\sigma} \left( V_{k\nu} z_{\sigma}^{\dagger}
\hat{f}_{0\sigma}^{\dagger} c_{k\nu \sigma}+h.c. \right) \nonumber\\
&&-\lambda \left(e^{\dagger}e +  d^{\dagger}d + \sum_{\sigma}p_{\sigma}^{\dagger}p_{\sigma}-1 \right) 
-\sum_{\sigma}\lambda_{\sigma} \left(f_{0\sigma}^{\dagger}f_{0\sigma}
-p_{\sigma}^{\dagger}p_{\sigma}-d^{\dagger}d \right) ,
\end{eqnarray}
\end{widetext}
where $\lambda$ and $\lambda_{\sigma}$ are the Lagrange multipliers associated with the constraints (\ref{constraints-finiteU}).
Again, in a mean field approximation, the auxiliary fields are treated as c-numbers to be determined self-consistently. 

The type of phase-diagram that is obtained within the finite-U SBMFA will be analyzed in subsection \ref{diagonalization}, 
for the zero band-width limit which allows a comparison with exact diagonalizations.
As it is shown in that subsection the finite-U SBMFA tends to underestimate the stability of the $\pi$-phase
in contrast with the HFA, which typically overestimates it. 

Another relatively simple approach is provided by the use a variational wave-function. 
This approach was used by Rozhkov and Arovas \cite{rozhkov00} extending previous works
\cite{varma76} in which variational wave-functions were proposed for analyzing the
normal Kondo problem. In their work Rozhkov and Arovas propose different many-body variational 
states in the $U\rightarrow \infty$ limit for the singlet and the doublet states, looking
for the their relative stability. They find a transition between both ground states for
$\Delta/T_K \sim 2$ and also predict the appearance of the intermediate phases
$0'$ and $\pi'$ in addition to the pure $0$ and $\pi$ ones.  

\subsection{Diagrammatic approaches}
\label{diagrammatic}

\subsubsection{Perturbation theory in the Coulomb interaction}

A natural extension over the HFA is provided by applying diagrammatic perturbation theory in 
the Coulomb parameter $U$. Already at the level of second order one can obtain an approximation
for the self-energy which is able to capture part of the interplay between Kondo effect and
pairing interactions. This approximation has been applied both for the
S-QD-S case in equilibrium \cite{mats01,veci03}, as well as for the N-QD-S case \cite{cuevas01,yama10}.

\begin{figure}
\begin{center}
\includegraphics[scale=0.7]{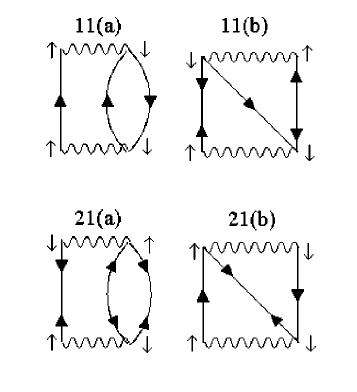}
\caption{Feynmann diagrams for the second-order self-energy in the Anderson model with
superconducting electrodes. Reprinted figure with permission
from E. Vecino {\it et al.}, Physical Review B {\bf 68}, 035105, 2003 \cite{veci03}.
Copyright (2003) by the American Physical Society.}
\label{diagrams-2nd-order}
\end{center}
\end{figure}

The diagrams contributing to the second-order self-energy in the superconducting Anderson model
are depicted in Fig. \ref{diagrams-2nd-order}. The first diagrams (denoted as 11(a)) is equivalent to 
the one appearing in the normal case, describing interaction of an electron with an 
electron-hole pair with opposite spin. The other diagrams include anomalous superconducting 
propagators and are therefore characteristic of the superconducting state. The presence of these
propagators gives several effects: the appearance of non-diagonal elements of the self-energy in
Nambu space, and the presence of diagrams like 11(b) and 21(b) in Fig. \ref{diagrams-2nd-order}
which corresponds to a double-exchange process. Finally, diagram 21(a) describes the interaction
of a Cooper pair with fluctuations in the pairing amplitude within the dot.

Formally, these diagrams can be computed from the full Green functions of the non-interacting
case by means of the expressions \cite{veci03}

\begin{widetext}
\begin{eqnarray}
\Sigma^{r(2)}_{11,a}(\omega) & = & \frac{U^2}{(2\pi i)^3}
\int d\epsilon_1 \int d\epsilon_2 \int
d\epsilon_3  \;
\frac{G^{(0)+-}_{11}(\epsilon_1) G^{(0)+-}_{22}(\epsilon_2)
G^{(0)-+}_{22}(\epsilon_3) \,\, + \,\, G^{(0)-+}_{11}(\epsilon_1)
G^{(0)-+}_{22}(\epsilon_2) G^{(0)+-}_{22}(\epsilon_3)}
{\omega - \epsilon_1 - \epsilon_2 + \epsilon_3 + i0^+} \nonumber,
\label{sigma11a}
\end{eqnarray}

\begin{eqnarray}
\Sigma^{r(2)}_{11,b}(\omega) &=& \frac{U^2}{(2\pi i)^3}
\int d\epsilon_1 \int d\epsilon_2 \int
d\epsilon_3 \;
\frac{G^{(0)+-}_{12}(\epsilon_1) G^{(0)+-}_{21}(\epsilon_2)
G^{(0)-+}_{22}(\epsilon_3) \,\, + \,\, G^{(0)-+}_{12}(\epsilon_1)
G^{(0)-+}_{21}(\epsilon_2) G^{(0)+-}_{22}(\epsilon_3)}
{\omega - \epsilon_1 - \epsilon_2 + \epsilon_3 + i0^+} \nonumber,
\label{sigma11b}
\end{eqnarray}

\begin{eqnarray}
\Sigma^{r(2)}_{21,a}(\omega) &=& - \frac{U^2}{(2\pi i)^3}
\int d\epsilon_1 \int d\epsilon_2 \int
d\epsilon_3 \; 
\frac{G^{(0)+-}_{21}(\epsilon_1) G^{(0)+-}_{12}(\epsilon_2)
G^{(0)-+}_{21}(\epsilon_3) \,\, + \,\, G^{(0)-+}_{21}(\epsilon_1)
G^{(0)-+}_{12}(\epsilon_2) G^{(0)+-}_{21}(\epsilon_3)}
{\omega - \epsilon_1 - \epsilon_2 + \epsilon_3 + i0^+} \nonumber,
\label{sigma21a}
\end{eqnarray}

\begin{eqnarray}
\Sigma^{r(2)}_{21,b}(\omega) &=& \frac{U^2}{(2\pi i)^3}
\int d\epsilon_1 \int d\epsilon_2 \int
d\epsilon_3  \;
\frac{G^{(0)+-}_{22}(\epsilon_1) G^{(0)+-}_{11}(\epsilon_2)
G^{(0)-+}_{21}(\epsilon_3) \,\, + \,\, G^{(0)-+}_{22}(\epsilon_1)
G^{(0)-+}_{11}(\epsilon_2) G^{(0)+-}_{21}(\epsilon_3)}
{\omega - \epsilon_1 - \epsilon_2 + \epsilon_3 + i0^+} \nonumber.
\label{sigma21b}
\end{eqnarray}
\end{widetext}

The evaluation of these expressions for a general range of parameters requires
a significant numerical effort. An efficient algorithm can be implemented to evaluate 
these expressions based on Fast Fourier transformations, as discussed in \cite{czycholl}.

\begin{figure}
\begin{center}
\includegraphics[scale=0.65]{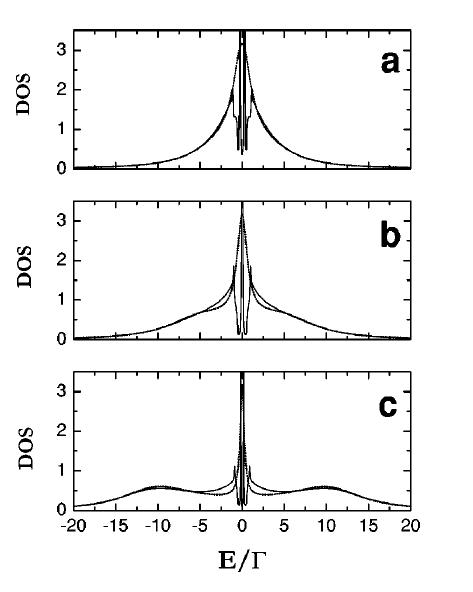}
\caption{Evolution of the DOS in the S-QD-S system in equilibrium within the second-order
self-energy approximation for and electron-hole symmetric case with $\Delta/\Gamma=0.1$.  
The $U/\Gamma$ parameter takes the values 2.5 (a), 5 (b) and 10 (c). 
Reprinted figure with permission
from E. Vecino {\it et al.}, Physical Review B {\bf 68}, 035105, 2003 \cite{veci03}.
Copyright (2003) by the American Physical Society.}
\label{spectral-density-2nd-order}
\end{center}
\end{figure}

In the limit $\Delta/\Gamma \ll 1$ Kondo correlations dominate over
pairing ones. The results of the second-order self-energy approach for the half-filled case
capture the main features of the onset of Kondo correlations in the spectral density when
$U > \Gamma$. This is illustrated in Fig. \ref{spectral-density-2nd-order} taken
from Ref. \cite{veci03}, which shows its evolution for increasing $U/\Gamma$. 
The spectral density is similar to the one found in the normal state except for the 
superimposed features inside the superconducting gap.
The overall shape
evolves from the single Lorentzian broad resonance for $U <\Gamma$ to the three peaked 
structure characteristic of the Kondo regime when $U > \Gamma$. In this regime the
width of the central Kondo peak is set by the scale $T_K$, which in the present 
approximation is given by $T_K \sim \Gamma/(1 - \alpha_0)$, where

\begin{equation}
\alpha_0 = \frac{\partial \Sigma_{11}}{\partial \omega}(0) \simeq -\left(\frac{U}{2\pi\Gamma}
\right)^2 \left(3 - \frac{\pi^2}{4}\right) \; ,
\end{equation}
thus coinciding with the perturbative result in the normal state (see Ref. \cite{yamada75}).
Although this perturbative approach fails to yield the exponential behavior of $T_K$ for
large $U/\Gamma$, it provides a reliable description of the spectral density for moderate
values of this parameter \cite{ferrer86}.

The second-order self-energy allows also to analyze the renormalization of the ABSs due to
the presence of Coulomb interactions. For values of $U/\Gamma < 10$ the renormalized ABSs
maintain approximately the $\sim \cos{\phi/2}$ behavior of the non-interacting case but
with a narrower dispersion set by $\omega_s(0) \simeq \Delta \left[ 1 - (U/U_0)^2 \right]$,
where $(U_0/\Gamma)^2 = (\Gamma/\Delta) \pi^2/(2 \pi + 2)$.

On the other hand, when $T_K \sim \Delta$ a transition to the $\pi$-phase
is expected. Within the second-order self-energy approach the transition can be identified
by allowing for a breaking of the spin-symmetry in the initial non-interacting problem and
searching for self-consistency. Rather than imposing the consistency condition of the HFA, i.e.
$\tilde{\epsilon}_{0\sigma} = \epsilon_0 + U <n_{0\bar{\sigma}}>$ in Ref. \cite{veci03} it was
imposed that the effective dot level for each spin-orientation be determined by 
the charge-consistency condition, i.e. $<n_{0\sigma}> = <n^0_{0\sigma}>$, where $<n^0_{0\sigma}>$
is the dot charge corresponding to the broken-symmetry non-interacting Hamiltonian. 
Such a procedure was shown to eliminate the unstable behavior of perturbation theory 
when developed from the HFA \cite{yeyati93}.

\subsubsection{NCA approximation}

\begin{figure}
\begin{center}
\includegraphics[scale=0.12]{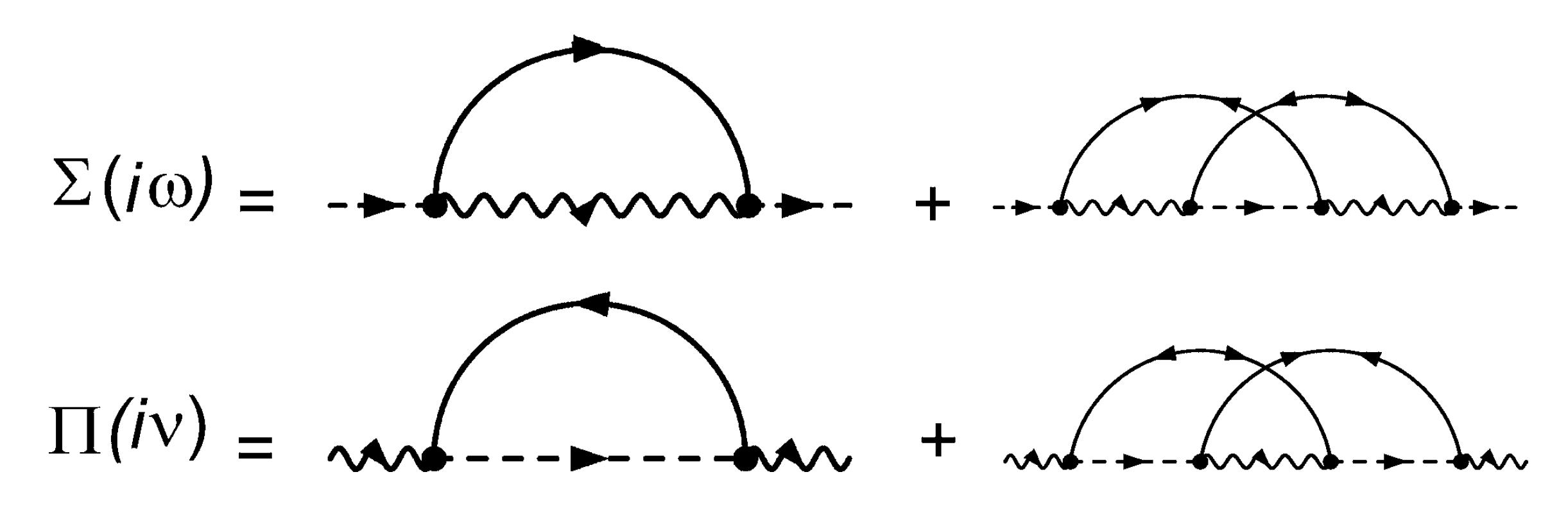}
\caption{Fermion (top) and boson (bottom) self-energy diagrams in the NCA approximation extended to the superconducting case.
Reprinted figure with permission
from G. Sellier {\it et al.}, Physical Review B {\bf 72}, 174502, 2005 \cite{sell05}.
Copyright (2005) by the American Physical Society.}
\label{diagram-NCA}
\end{center}
\end{figure}

Within the diagrammatic approximations one can include the so-called non-crossing approximation
(NCA). In this case an infinite order resumation of the perturbation theory is performed starting
from the $U \rightarrow \infty$ slave boson representation of the Anderson Hamiltonian 
\cite{bickers87}. To the lowest order in $1/N$ (where $N=2$ is the spin degeneracy) the
family of diagrams in this resumation is represented in Fig. \ref{diagram-NCA}. The
dashed lines correspond to the fermion propagators and the wavy lines to the slave bosons.
In the normal case the NCA include only the first two diagrams in the Dyson equation for 
the fermion and boson propagators. The extension to the superconducting case was proposed
in Ref. \cite{cler00} and corresponds to including the anomalous propagators for describing
multiple Andreev reflection processes (last two diagrams in the fermion and boson self-energies 
represented in Fig. \ref{diagram-NCA}).

In the normal case the NCA theory has been shown to yield reliable results for temperatures down below $T_K$ 
\cite{bickers87,hewson93} in spite of certain pathologies like its failure to fulfill the
Friedel sum-rule. The self-consistent extension for the superconducting case by Clerk et.
al. \cite{cler00-2} is also formally exact to order $1/N$ and it is thus expected to 
yield reasonable results even in the presence of MAR processes. 

\begin{figure}
\begin{center}
\includegraphics[scale=0.4]{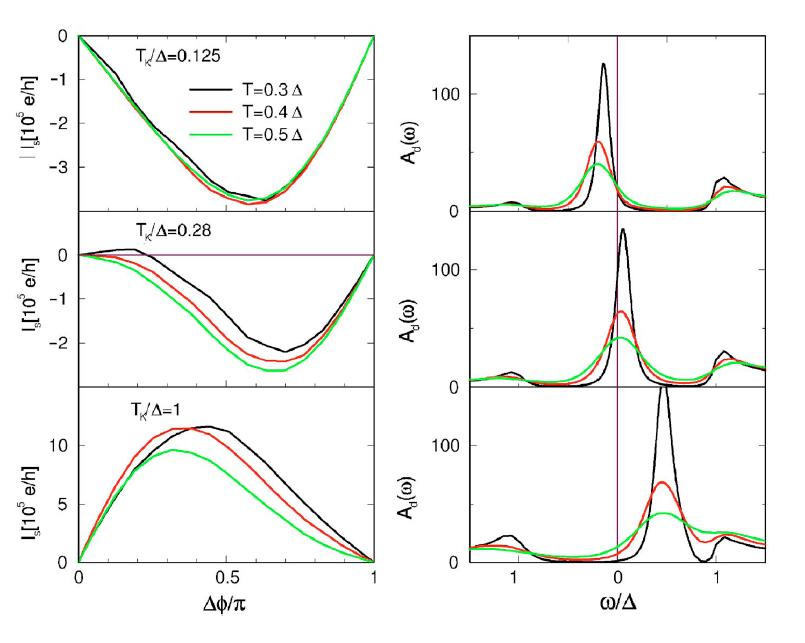}
\caption{Josephson current and subgap LDOS for the equilibrium S-QD-S model within the NCA for
three values of $T_K/\Delta$ and three different temperatures. Reprinted figure with permission
from G. Sellier {\it et al.}, Physical Review B {\bf 72}, 174502, 2005 \cite{sell05}.
Copyright (2005) by the American Physical Society.}
\label{curr-ldos-NCA}
\end{center}
\end{figure}

Further analysis of the NCA applied to the S-QD-S system was provided in Ref. \cite{sell05}.
Their results for the Josephson current and LDOS in the superconducting gap region are summarized in Fig. \ref{curr-ldos-NCA}.
The fact that the calculations are performed for temperatures which are a quite large
fraction of $\Delta$ yields very broad resonances for the subgap states. As can be observed in 
Fig. \ref{curr-ldos-NCA}, only one broad resonance can be clearly resolved within the gap. 
These results are in contrast to what is obtained using exact numerical methods as
will be discussed in Sect. \ref{NRG}.
In this approximation the transition to the $\pi$-phase appears as a 
smooth crossover which can be associated to the crossing of this resonance through the Fermi energy. 

\subsubsection{Real time diagrammatic approach}

Another technique which has been applied to the study of quantum dots coupled
to superconducting leads is the real time diagrammatic approach first introduced
by K\"onig et al. \cite{konig96}  for the normal Anderson model.
The main idea of this technique is to integrate out the fermionic degrees of freedom
of the electrodes leading to a reduced description of the density matrix 
projected on the Hilbert space of the isolated dot states. In the superconducting case
this reduced density matrix also depends on the 
number of Cooper pairs in the leads relative to some chosen reference. 
The aim of the technique is to determine the
time evolution of this reduced density matrix in the Keldysh contour thus allowing
to consider both equilibrium and non-equilibrium situations. 
In Ref. \cite{governale08} the method has been applied to the equilibrium S-QD-S
case obtaining results in agreement with those of Ref. \cite{glazman89} in the cotunneling
limit. The method has been mainly applied to analyze the properties of quantum dots
connected to both normal and superconducting leads in multiterminal configurations
out of equilibrium, an issue which will be commented in Sect. \ref{multi}.

\subsection{Diagonalization by numerical methods}
\label{diagonalization}

Within this subsection we will review methods which attempt a direct diagonalization of the
superconducting Anderson model, either by truncating the initial Hilbert space using
physical arguments valid for certain parameter region
or by using the Numerical Renormalization group (NRG) method. 

\subsubsection{Exact diagonalization for the large $\Delta$ limit}

An exact diagonalization of the model is possible in the limit $\Delta \rightarrow \infty$.
In this limiting case the Hilbert space of the problem is automatically reduced to
states spanned by the different electronic configuration of the dot levels.
The effect of the superconducting leads appears 
as a pairing term between the electrons within the dot. The effective Hamiltionian for the truncated
Hilbert space becomes \cite{veci03,tana07,meng09}

\begin{eqnarray}
H^{eff} &=& 2\Gamma \cos{\phi/2} \left(c_{0\uparrow}c_{0\downarrow} + 
c^{\dagger}_{0\downarrow} c^{\dagger}_{0\uparrow}\right) + \epsilon_0 \sum_{\sigma} n_{0\sigma} \nonumber\\
&&+ U n_{0\uparrow}n_{0\downarrow} .
\end{eqnarray}

The eigenvalues of this reduced Hamiltonian can be determined straightforwardly by noting 
the decoupling of subspaces with even and odd number of electrons. The ground state for the even case
(corresponding to total spin $S=0$) is a linear combination of the empty and doubly occupied
dot state with an energy

\begin{equation}
E_{0(S=0)}(\phi) = \epsilon_0 + U/2 - \sqrt{(\epsilon_0 +U/2)^2 + 4\Gamma^2 \cos^2{\phi/2}}.
\end{equation}

On the other hand, the odd number subspace simply corresponds to a single uncoupled spin 
with energy $E_{0(S=1/2)}=\epsilon_0$. The transition to the magnetic ground state thus
occurs for $E_{0(S=1/2)} = E_{0(S=0)}(\phi)$. In the simpler electron-hole symmetric case
$(\epsilon_0 = -U/2)$ this condition reduces to $2\Gamma\cos{\phi/2} = U/2$ and thus the
full $\pi$ state appears for $\Gamma<U/4$. This simple model already gives a rough qualitative
account of the $0-\pi$ quantum phase transition.

A further step in the idea truncating the Hilbert space is performed in the so called zero
band-width limit \cite{affleck00,veci03,berg07}. In this approximation the 
superconducting leads are represented by a single
localized level (which formally corresponds to the limit of vanishing width of the leads
spectral density). This approximation is justified when the superconducting gap is large compared
to the other energy scales in the problem, and thus can be considered as a refinement with respect
to the previous approach. The corresponding Hamiltonian can be written as $H= H_d + H_T + H_L + H_R$, 
with 

\begin{eqnarray}
H_{\nu} &=& \sum_{\sigma} \epsilon_{\nu} c^{\dagger}_{\nu\sigma} c_{\nu\sigma} +
\left( \Delta_{\nu} c_{\nu\uparrow} c_{\nu\downarrow} + \mbox{h.c.} \right) \nonumber\\
H_{T} &=& \sum_{\nu,\sigma} \left( V_{\nu} c^{\dagger}_{\nu\sigma} c_{0\sigma} + \mbox{h.c.}  \right) ,
\end{eqnarray}
where $\nu=L,R$ denotes the left-right sites describing the leads in this approximation.

Although the total number of particles is not a good quantum number,
their parity is conserved as in the previous case. This allows to reduce the initial 
64 states in the Hilbert space to a subspace of 20 states for even parity with a total spin
$z$-component $S_z=0$ and 15 states for odd parity with $S_z=\pm 1/2$. These values of $S_z$
are the ones corresponding to the ground state in each subspace with total spin
$S=0$ and $S=1/2$. 
In addition to providing a qualitative description of the phase diagram of the
full model, this simplified calculation can furthermore be useful as a test for comparing different
approximation methods. 

\begin{figure}
\begin{center}
\includegraphics[scale=0.35]{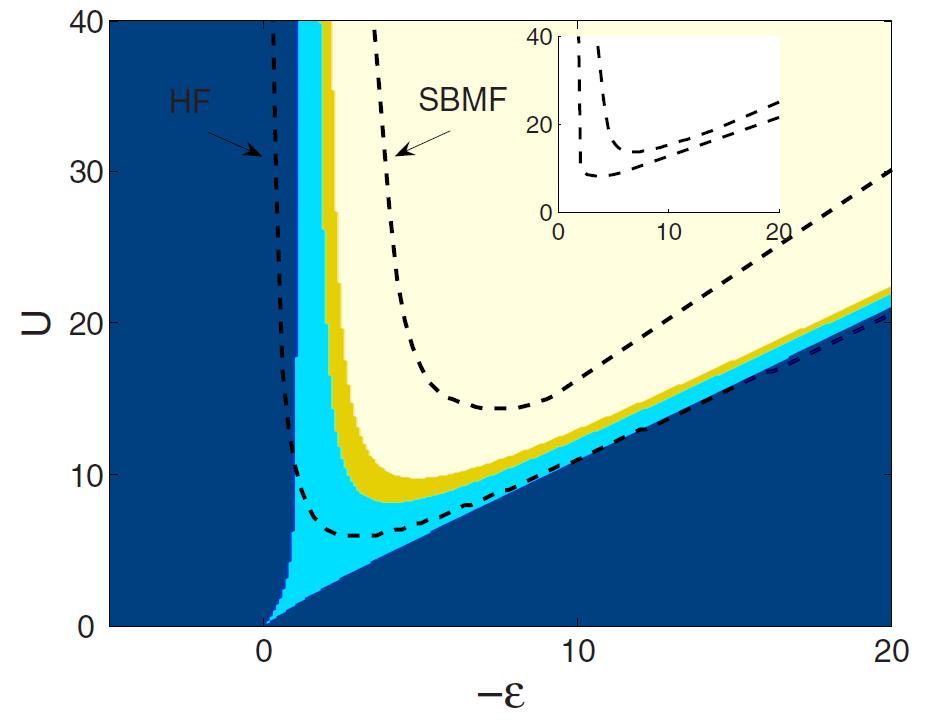}
\caption{Phase diagram of the S-QD-S system in the ZBW model for the leads obtained 
by exact diagonalization for $\Delta=V_L=V_R$, taken as unit of energy. The dashed lines correspond
to the boundary between the 0 and $\pi$ regions within the HFA (lower line) and the finite-U SBMFA
(upper line). The inset show the corresponding results for these boundaries within the full model. 
Reprinted figure with permission
from F.S. Bergeret {\it et al.}, Physical Review B {\bf 76}, 174510, 2007 \cite{berg07}.
Copyright (2007) by the American Physical Society.}
\label{phase-diagram-ZBW}
\end{center}
\end{figure}

The phase diagram obtained within this approximation was discussed in Refs. \cite{veci03,berg07}
and is shown in Fig. \ref{phase-diagram-ZBW} for $\Gamma \equiv V = \Delta$, with $V_L=V_R=V$. 
As can be observed, the overall diagram is very similar to the one shown before for the
HFA (Fig. \ref{phase-diagram-HFA}) exhibiting the four phases $0,0',\pi'$ and $\pi$ in the
same sequence. For a more direct comparison it is necessary to perform the HFA of the ZBW model,
which as an exactly solvable model also provides a stringent test of the approximation. The
lower broken line in Fig. \ref{phase-diagram-ZBW} indicates the boundary of the $\pi$-phase
within the HFA. It can be noticed that the HFA overestimates the stability of this magnetic phase.
On the other hand, it is also possible to test the finite-U SBMF approximation in this ZBW model. The
corresponding boundary for the $\pi$-phase is indicated by the upper broken line in Fig.
\ref{phase-diagram-ZBW}. In opposition to the HFA this approximation overestimates the stability
of the $0$ phase, the exact boundary therefore lying in between the two different 
mean-field approximations. It is interesting to point out that a similar difference between
both approximations is also found for the full model (see inset in Fig. \ref{phase-diagram-ZBW}).
One would then expect that the exact boundary for the full model lays in between these two.

\begin{figure}
\begin{center}
\includegraphics[scale=0.3]{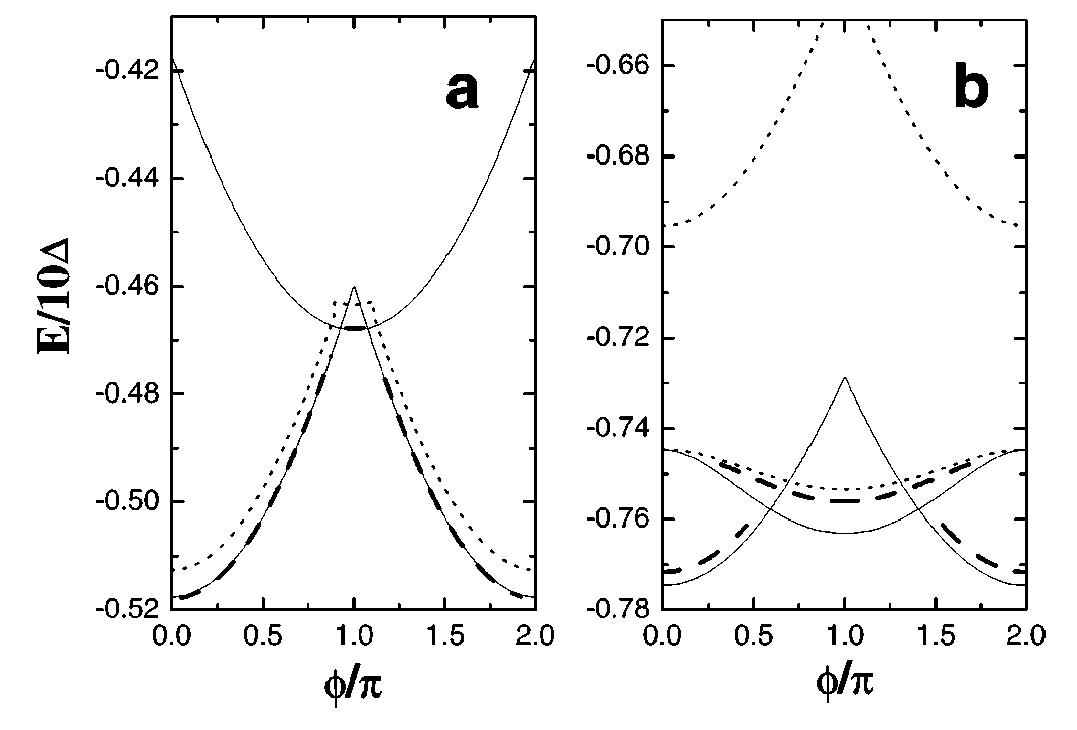}
\caption{Ground state energy for the $S=0$ and $S=1/2$ states of the S-QD-S system in the ZBW model for the leads.
Full lines correspond to the exact results, dotted lines to the HFA and dashed lines to the second-order self-energy
approximation. Reprinted figure with permission
from E. Vecino {\it et al.}, Physical Review B {\bf 68}, 035105, 2003 \cite{veci03}.
Copyright (2003) by the American Physical Society.}
\label{comparison-ZBW-2nd-order}
\end{center}
\end{figure}

The ZBW model can also be used to test approximate methods beyond the mean field ones. In Ref. \cite{veci03}
this was done for the second-order self-energy approximation. As it is illustrated in Fig.
\ref{comparison-ZBW-2nd-order} for the symmetric case $(\epsilon_0 = -U/2)$
the second-order self-energy approach  matches quite well the exact ground state energy 
for $U/\Delta$ values up to $\simeq 2.5$. It is interesting to note that for larger $U$ values
even when the HFA already yields a full $\pi$-state, the second-order approximation 
predicts a mixed ground state in agreement with the exact solution.

\subsubsection{Numerical Renormalization Group (NRG)}
\label{NRG}

The NRG method is based on the ideas of Wilson on logarithmic
discretization for magnetic impurity problems \cite{wilson75} and was first applied to the
Anderson model by Krishna-murthy et al. \cite{krishna80}. The idea behind the method is 
to discretize the energy levels in the leads on a logarithmic grid of energies 
$\Lambda^{-n}$ (with $\Lambda > 1$ and $1\le n \le N \rightarrow \infty$) with exponentially
high resolution on the low-energy excitations. This discretization allows then to map
the impurity model into a linear "tight-binding" chain with hopping matrix elements
decaying as $\Lambda^{-n/2}$ with increasing site index $n$. The sequence of Hamiltonians
which is constructed by adding a new site in the chain is then diagonalized iteratively.
As the number of states grows exponentially an adequate truncation scheme is required.

The NRG scheme has been first generalized to the case of an Anderson impurity in a superconducting
host by Yoshioka and Ohashi \cite{yosh00} and implemented by several authors to analyze
the S-QD-S model with a finite phase difference \cite{choi04,ogur04,tana07,lim08}. 
For the left-right symmetric case (i.e. $\Delta_L=\Delta_R$ and $\Gamma_L=\Gamma_R$)
the sequence of Hamiltonians can be written as \cite{choi04}

\begin{eqnarray}
\tilde{H}_{N+1} &=& \sqrt{\Lambda} \tilde{H}_N + \xi_N \sum_{\mu,\sigma} 
\left(f^{\dagger}_{\mu,N+1,\sigma}
f_{\mu,N+1,\sigma} + \mbox{h. c.} \right) \nonumber \\
&& - \Lambda^{N/2} \sum_{\mu} \tilde{\Delta}_{\mu}
\left(f^{\dagger}_{\mu,N+1,\uparrow} f^{\dagger}_{\mu,N+1,\downarrow} + \mbox{h. c.}
\right) ,
\end{eqnarray}
where the initial Hamiltonian is given by

\begin{eqnarray}
\tilde{H}_0 &=& \frac{1}{\sqrt{\Lambda}} \left[ \tilde{H}_{QD} + 
\sum_{\mu=e,o} \sum_{\sigma} \tilde{V}_{\mu} \left(c^{\dagger}_{0\sigma} f_{\mu,0,\sigma}
+ \mbox{h. c.} \right) \right. \nonumber\\
&& \left. - \sum_{\mu} \tilde{\Delta}_{\mu} \left(f^{\dagger}_{\mu,0,\sigma}
f_{\mu,0,\sigma} + \mbox{h. c.} \right) \right] .
\end{eqnarray}

The fermion operators $f_{\mu,N,\sigma}$ correspond to an effective tight-binding chain 
resulting from the logarithmic discretization and the canonical transformation 
into the even-odd linear combination of original left-right states in the leads and

\begin{eqnarray}
\tilde{H}_{QD} &\equiv& \chi \frac{H_{QD}}{D} \; \; \;, \;\;\; \tilde{\Delta}_{\mu} \equiv
\chi \frac{\Delta_{\mu}}{D}  \\
\tilde{V}_e &=& \chi \sqrt{\frac{2\Gamma}{\pi D}} \cos{\phi/4} \;\; , \; \;
\tilde{V}_o = -\chi \sqrt{\frac{2\Gamma}{\pi D}} \sin{\phi/4} , \nonumber
\end{eqnarray}
with $\chi = 2/(1 + 1/\Lambda)$ and $D$ being an energy cut-off in the
leads spectral density. The original Hamiltonian is recovered in the
limit $H/D = \lim_{\rightarrow \infty} \tilde{H}_N/(\chi \Lambda^{(N-1)/2})$.

\begin{figure}
\begin{center}
\includegraphics[scale=0.35]{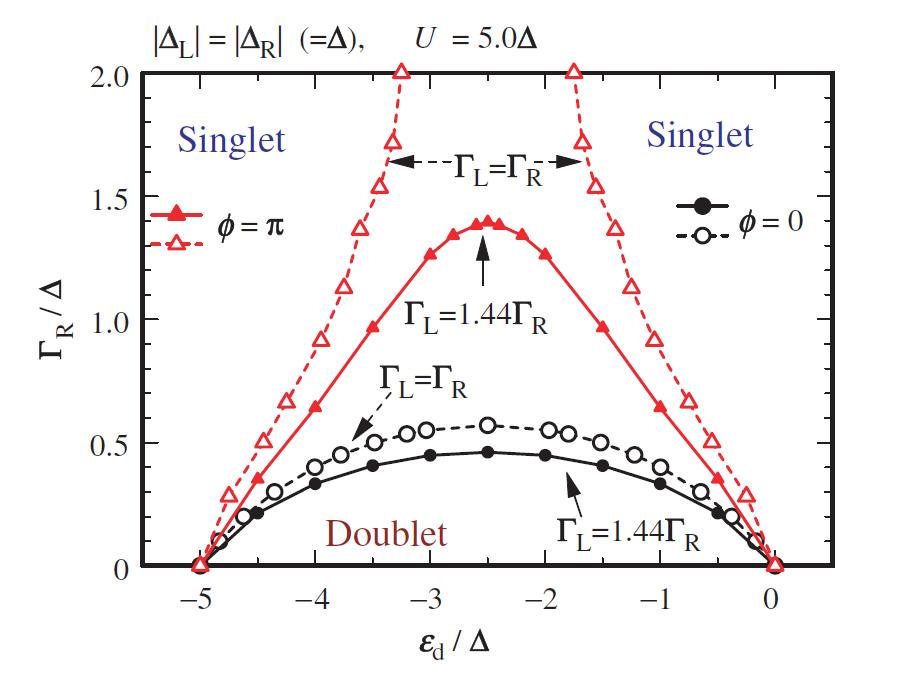}
\caption{Phase diagram of the S-QD-S system in the $\Gamma_R,\epsilon_0$ plane for fixed $U$ and 
different values of $\Gamma_L/\Gamma_R$ obtained using the NRG method. 
Reprinted figure with permission
from Y. Tanaka {\it et al.}, New Journal of Physics {\bf 9}, 115, 2007 \cite{tana07}.
Copyright (2007) by IOP Publishing Ltd.}
\label{phase-diagram-NRG-Tanaka}
\end{center}
\end{figure}

The NRG method was applied to analyzing the Josephson current
in a S-QD-S system in Ref. \cite{choi04}, this work confirming the predicted $0-\pi$ quantum phase
transition at $\Delta \sim T_K$ for the electron-hole symmetric case. 
It should be mentioned that more recent calculations \cite{karr08} using NRG obtain 
Josephson currents which are approximately a factor 2 larger than the ones of
\cite{choi04}. 
On the other hand, Oguri et al. \cite{ogur04} used NRG to analyze this model in the case of 
$|\Delta_L| \gg |\Delta_R|$. In this case the model can be exactly
mapped into a single channel model consisting on the right lead coupled to the Anderson
impurity with a local pairing $\Delta_d \equiv \Gamma_L e^{i\phi_L}$,
thus allowing a simpler implementation of the NRG algorithm. Further work for the $\Gamma_L 
\neq \Gamma_R$ case although with $\Delta_L=\Delta_R$ by Tanaka et al. \cite{tana07} 
confirmed the presence of intermediate $0'-\pi'$ phases even in the left-right asymmetric
case. A characteristic phase diagram obtained for this case is shown in Fig. 
\ref{phase-diagram-NRG-Tanaka}.

In addition to the ground state properties, NRG methods have been applied in an attempt to
clarify the structure of the subgap ABSs. In Ref. \cite{lim08} the spectral density inside
the gap obtained from the NRG algorithm was analyzed, showing that a pair of ABSs
located symmetrically respect to the Fermi energy is present in the $U \rightarrow \infty$ 
case. This is in contrast to the NCA results discussed previously (shown in Fig. 
\ref{curr-ldos-NCA}) where a single broad resonance appears. Similar conclusions are
obtained in Ref. \cite{baue07} although for the single lead case and for finite $U$.
A word of caution should be said regarding the analysis of the ABSs in this last
work in which the relation $\Sigma_{22}(\omega) = -\Sigma_{11}(-\omega)$ is assumed
in their Eq. (8) for the states inside the gap. This relation would not be strictly valid for
the doublet ground state when choosing a given spin orientation. In this case the quasi-particle
excitation energies would become spin dependent and the electron-hole symmetry would be broken.
This would allow in principle to have up to 4 ABSs inside the gap as predicted both by the Hartree-Fock
approximation and in the exact $\Delta \rightarrow \infty$ limit. Of course, 
in the $\pi$ phase the spin is not frozen but is fluctuating. In this sense the above relation 
between the self-energy components would be recovered when averaging over the $S_z=1/2$ and 
$S_z=-1/2$ states.
We believe in any case that a more detailed analysis of the ABSs using the NRG method is still lacking.

\subsection{Functional renormalization group}
\label{fRG}

\begin{figure}
\begin{center}
\includegraphics[scale=0.25]{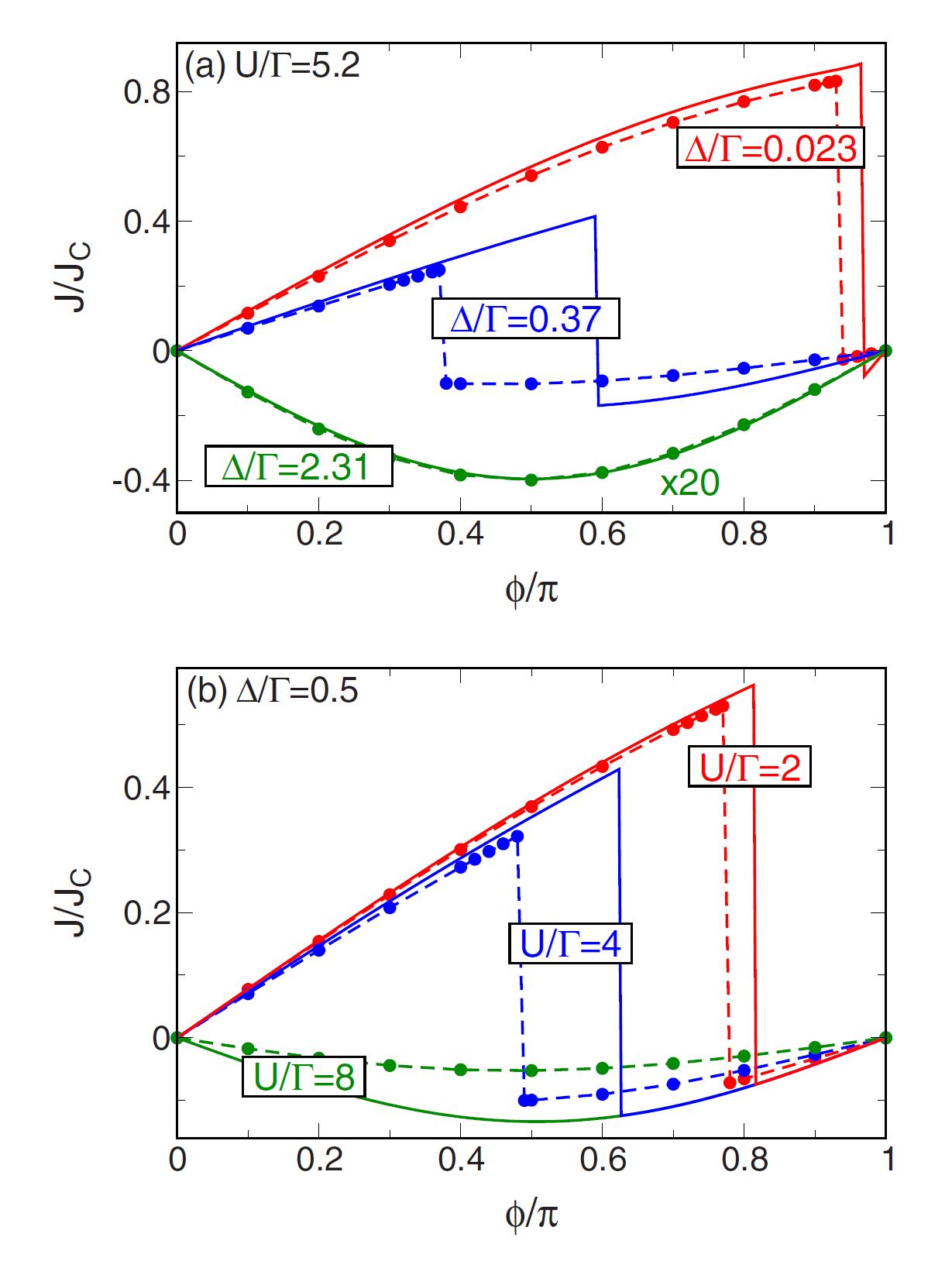}
\caption{Current phase-relations for the S-QD-S system obtained using the fRG approach truncated
at the HF level for different values of $U/\Gamma$ and $\Delta/\Gamma$. For comparison the results
obtained using the NRG method are also plotted (indicated by the filled dots).
Reprinted figure with permission
from C. Karrash {\it et al.}, Physical Review B {\bf 77}, 024517, 2008 \cite{karr08}.
Copyright (2008) by the American Physical Society.}
\label{current-fRG}
\end{center}
\end{figure}

The functional renormalization group (fRG) method is based on the application of an RG analysis 
to the diagrammatic expansions in terms of electronic Green functions. This is an 
approximate method whose accuracy depends on the initial diagrams used in the evaluation of 
the electron self-energies. The starting point is the introduction of an energy cut-off
$\Lambda$ into the Matsubara non-interacting Green-functions 

\[ G^{0,\Lambda}(i\omega) = \Theta(|\omega|-\Lambda) G^0(i\omega) \]
Using these propagators the $n-$particle vertex functions acquires a $\Lambda$ dependence.
The flow equations are determined differentiating these vertex functions with respect to $\Lambda$
which are then solved iteratively for increasing $\Lambda$. In Ref. \cite{karr08} the
method has been applied to the S-QD-S system employing a truncation scheme which keeps 
only diagrams corresponding the the static Hartree-Fock approximation. 
The $\Lambda$-dependent Green function used in Ref. \cite{karr08} was of the form,

\begin{equation}
G^{\Lambda}(i\omega) = \left( \begin{array}{cc}
i \tilde{\omega} - \epsilon_0 - \Sigma^{\Lambda} & \tilde{\Delta}(i\omega) - \Sigma^{\Lambda}_{\Delta} 
\\
\tilde{\Delta}(i\omega) - \Sigma^{\Lambda}_{\Delta} & i\tilde{\omega} + \epsilon_0 + \Sigma^{\Lambda}
\end{array} \right)^{-1} ,
\end{equation} 
where $i\tilde{\omega} = i\omega ( 1+ \sum_{\mu} \Gamma_{\mu} g(i\omega))$ and
$\tilde{\Delta}(i\omega) = \sum_{\mu} \Gamma_{\mu} f(i\omega) e^{i\phi_{\mu}}$, $g$ and
$f$ being the dimensionless BCS Matsubara Green functions of the uncoupled leads. 
Within this approximation the flow equations lead to energy-independent self-energies,
corresponding to an effective non-interacting model with renormalized parameters.
It is important to notice that this approximation exhibits also the limitation already
pointed out in the previous section as it imposes electron-hole symmetry which is not
satisfied in the magnetic phase. Nevertheless the approximation allows to identify a transition
to a phase with inversion of the Josephson current which is driven by an "overscreening" of
the induced pairing determined by $\Sigma_{\Delta}$. Fig. \ref{current-fRG} shows the
comparison of fRG results with those obtained with the NRG method.
The agreement is rather good for large $\Delta/\Gamma$ but it becomes poorer in the 
$\pi$-phase. We believe that the agreement could be improved allowing for a broken
symmetry state within the same fRG approach.

\subsection{Quantum Monte-Carlo}
\label{QMC}

The Josephson current in the S-QD-S system has also been analyzed using Quantum Monte Carlo (QMC) simulations
by Siano and Egger \cite{sian04}. The method used was the Hirsh-Fisher algorithm adapted to
this particular problem. They consider the deep Kondo regime $U/\Gamma \gg 1$ and
$\epsilon_0/\Gamma \ll -1$ and show that the results for the Josephson current  
exhibit a universal dependence with $T_K$ provided that $U/\Gamma > 5$. They identify 
the transition between the different phases at $\Delta/T_K \simeq 0.51, 0.875$ and $1.105$ for
$0-0'$, $0'-\pi'$ and $\pi'-\pi$ respectively \cite{erratum-siano}. Being a finite temperature calculation
the resulting current-phase relations do not exhibit sharp discontinuities in the intermediate
phases. This smooth behavior was criticized in Ref. \cite{choi05} pointing out that the QMC
results did not match the NRG ones of Ref. \cite{choi04} at finite temperatures, which was attributed
by Siano and Egger in their reply \cite{reply-siano} to a possible limited accuracy 
of the NRG calculation of Ref. \cite{choi04}. More recent 
NRG calculations of Ref. \cite{karr08} give a good agreement with QMC results at finite temperature
for $\phi$ values between $\pi/2$ and $\pi$, whereas QMC underestimates the current in the range
$0-\pi/2$. It is claimed in Ref. \cite{karr08} that the origin of the discrepancy lies in the
fact that the first excited state in this phase range is smaller or of the order of the
temperature values used in the calculations of \cite{sian04}. 

The QMC method has more recently been applied to analyze the spectral properties
of this model in Ref. \cite{luitz10}. The authors employ the so-called weak-coupling continuous-time
version of the method which is based on a perturbative expansion around the $U=0$ limit. They show
that the results for the spectral densities are in qualitative good agreement with the ones obtained
in the zero band-width approximation introduced in Ref. \cite{veci03}, which was discussed 
before in this section. 

\subsection{Experimental results}
\label{exp-SQDS-eq}

Several physical realizations of the S-QD-S system have been obtained in the last few
years by means of contacting carbon nanotubes (CNT), C60 molecules or semiconducting nanowires 
with superconducting electrodes (for a recent review see Ref. \cite{defran10}). In view of
the existence of this review on the experiments in this section we give only a brief summary of the
main findings and its relation to the theoretical work.

CNTs have provided so far the most promising setups for a direct test of the theoretical predictions
concerning the Josephson effect through a QD. The first experiments detecting a supercurrent
through a CNT-QD strongly coupled to the leads (i.e. $\Gamma \gg \Delta, U$) were performed
by Jarillo-Herrero et al \cite{jari06}. These experiments were basically performed in the resonant-tunneling
regime with a single-level spacing $\delta \epsilon \gg \Gamma$. The results indicated a
strong correlation between the critical current $I_c$  and the normal conductance $G_N$. However, the
product $I_c R_N$ deviated from a constant value due to the effect of the electromagnetic environment
suppressing the critical current more strongly in off-resonance conditions due to phase fluctuations.

\begin{figure}[h!]
\begin{center}
\includegraphics[scale=0.25]{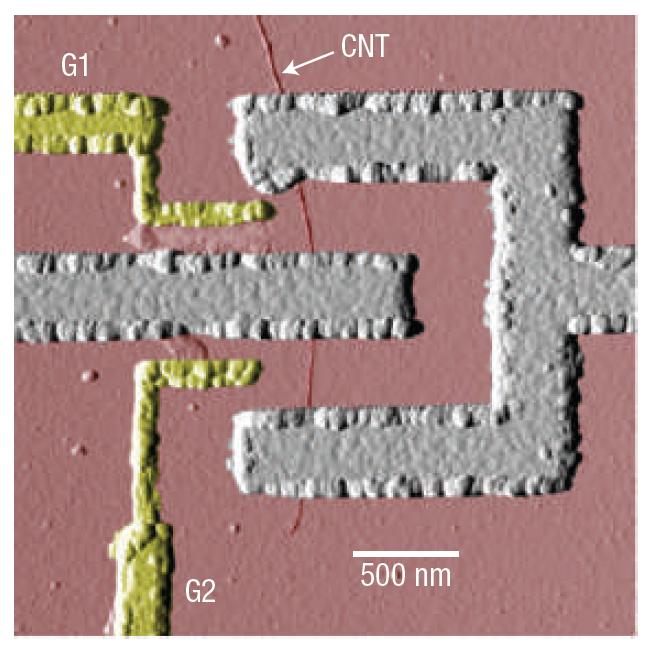}
\includegraphics[scale=0.25]{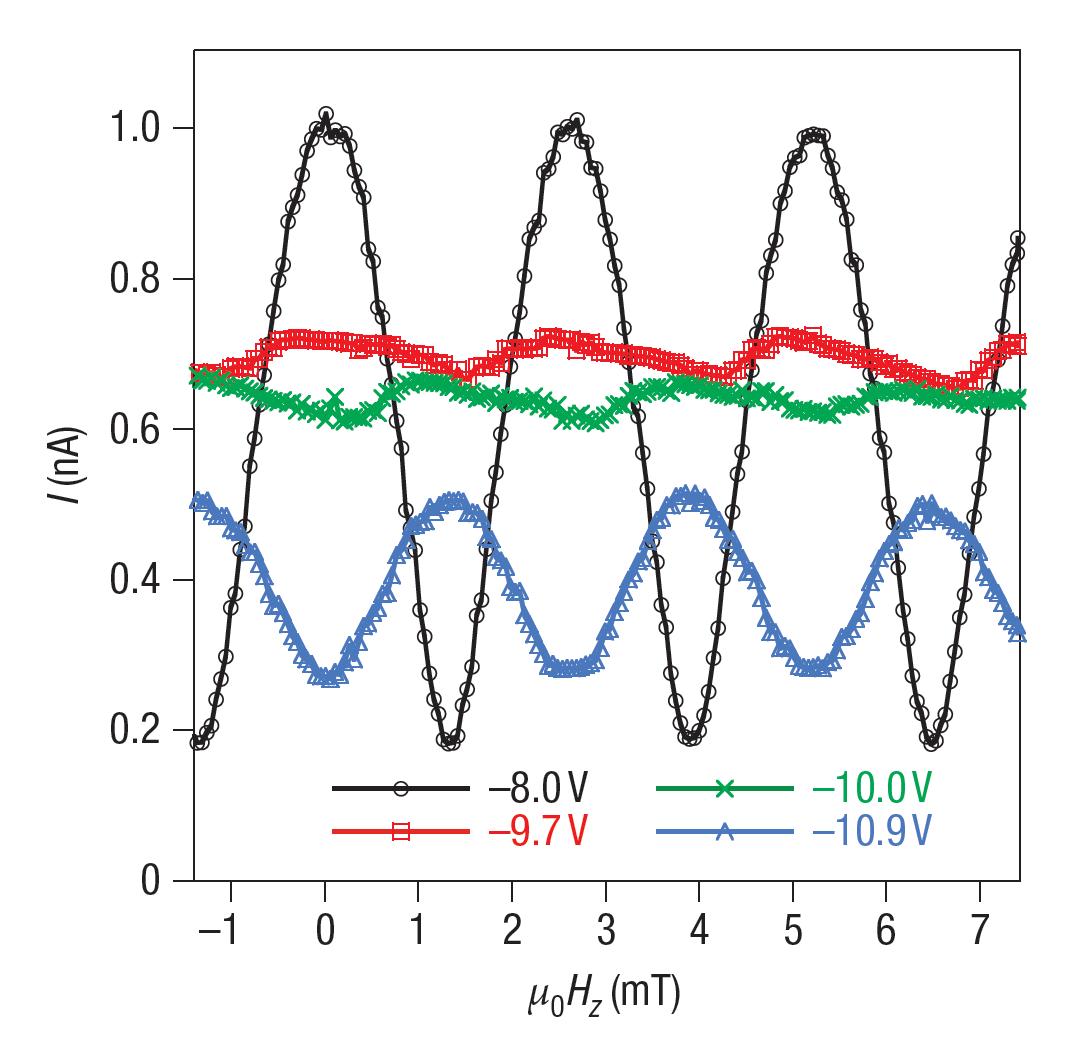}
\includegraphics[scale=0.2]{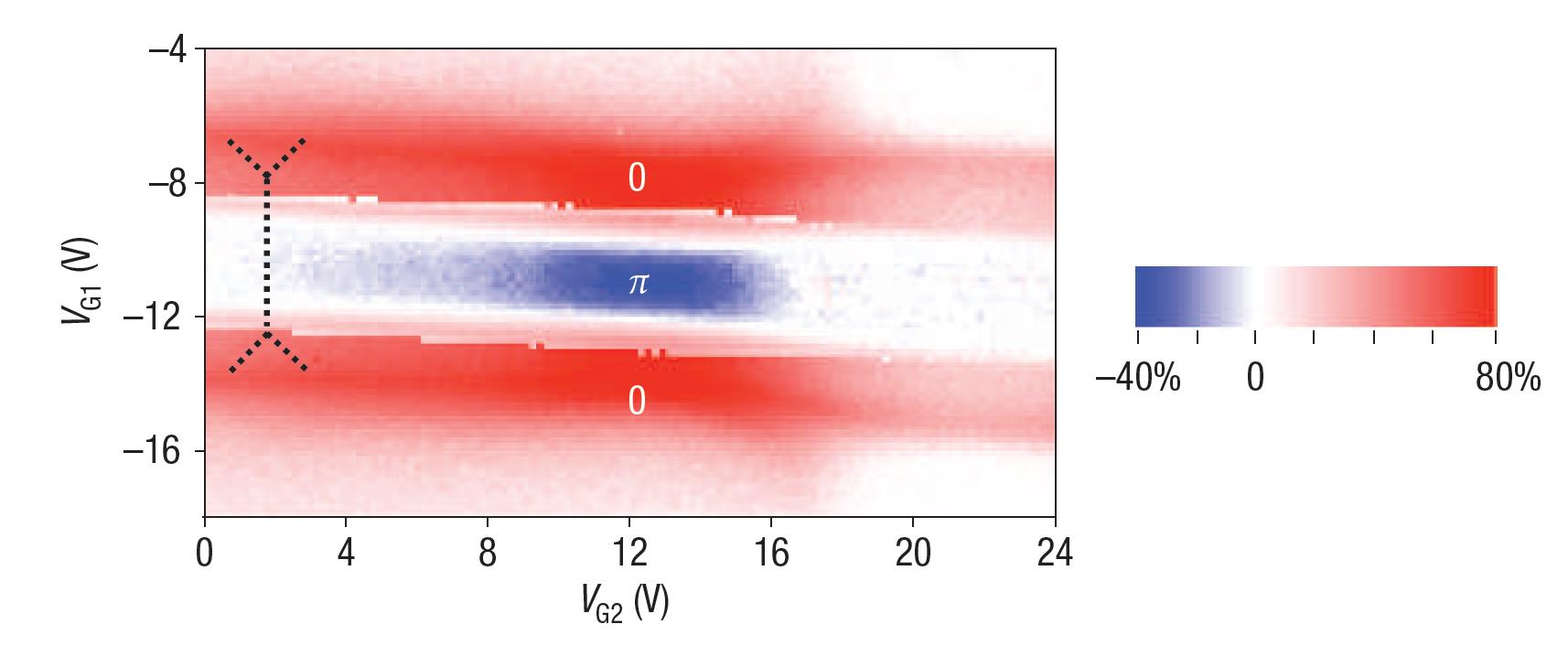}
\caption{Experimental setup used in Ref. \cite{cleu06} for analyzing the $0-\pi$ transition in CNT QDs (upper panel).
The middle panel shows the reversal of the oscillatory pattern of the current as a function of the magnetic
flux through the SQUID across the transition. The lower panel shows the correlation of the $\pi$-phase region
with the Kondo ridges in the normal state (indicated by the dashed line).
Reprinted by permission
from Macmillan Publishers Ltd: Nature Physics \cite{cleu06}, copyright (2006).}
\label{fig-nature-nano}
\end{center}
\end{figure}

In a subsequent work by the same group the first experimental evidence of $\pi$-junction behavior
in a S-QD-S system was obtained using semiconducting InAs nanowires with Al leads in a SQUID 
configuration \cite{dam06}. However, the observed features in this experiment could not be 
explained completely on the basis of a single-level model but rather a multilevel description
was necessary. In particular the authors showed that in this case the $\pi$-junction behavior
is not necessarily linked to the parity of the number of the electrons in the dot (this will be further
discussed in Sect. \ref{multi}). 
On the other hand, $\pi$-junction behavior have also been demonstrated in Ref. \cite{cleu06}
using CNT-QDs in a SQUID geometry. The corresponding experimental setup 
is depicted in Fig. \ref{fig-nature-nano}. In this SQUID configuration the transition to the
$\pi$-phase is directly demonstrated in the measured current-phase relation as a function
of one of the applied gate voltages (see Fig. \ref{fig-nature-nano}). Remarkably, this experiment
allows to correlate the appearance of the $\pi$-junction behavior with the presence of Kondo
correlations in the normal state. The dotted lines in the lower panel of Fig. \ref{fig-nature-nano} indicate the
Kondo ridge appearing in the normal state (which is recovered by applying a magnetic field).

Evidence of a $0-\pi$ transition has also been found in CNT-QD systems by analyzing the
current-voltage characteristic \cite{inge07}. In this work it was shown that the evolution of
the zero-bias conductance can be correlated to the behavior of the critical current
when transversing the $0-\pi$ boundary. As it corresponds to a non-equilibrium situation
this analysis will be discussed later in Sect. \ref{SQDS-neq}. A similar observation holds for
other experimental works \cite{veci04,grov07,eich09}
which will be discussed in that section.

\begin{figure}
\begin{center}
\includegraphics[scale=0.29]{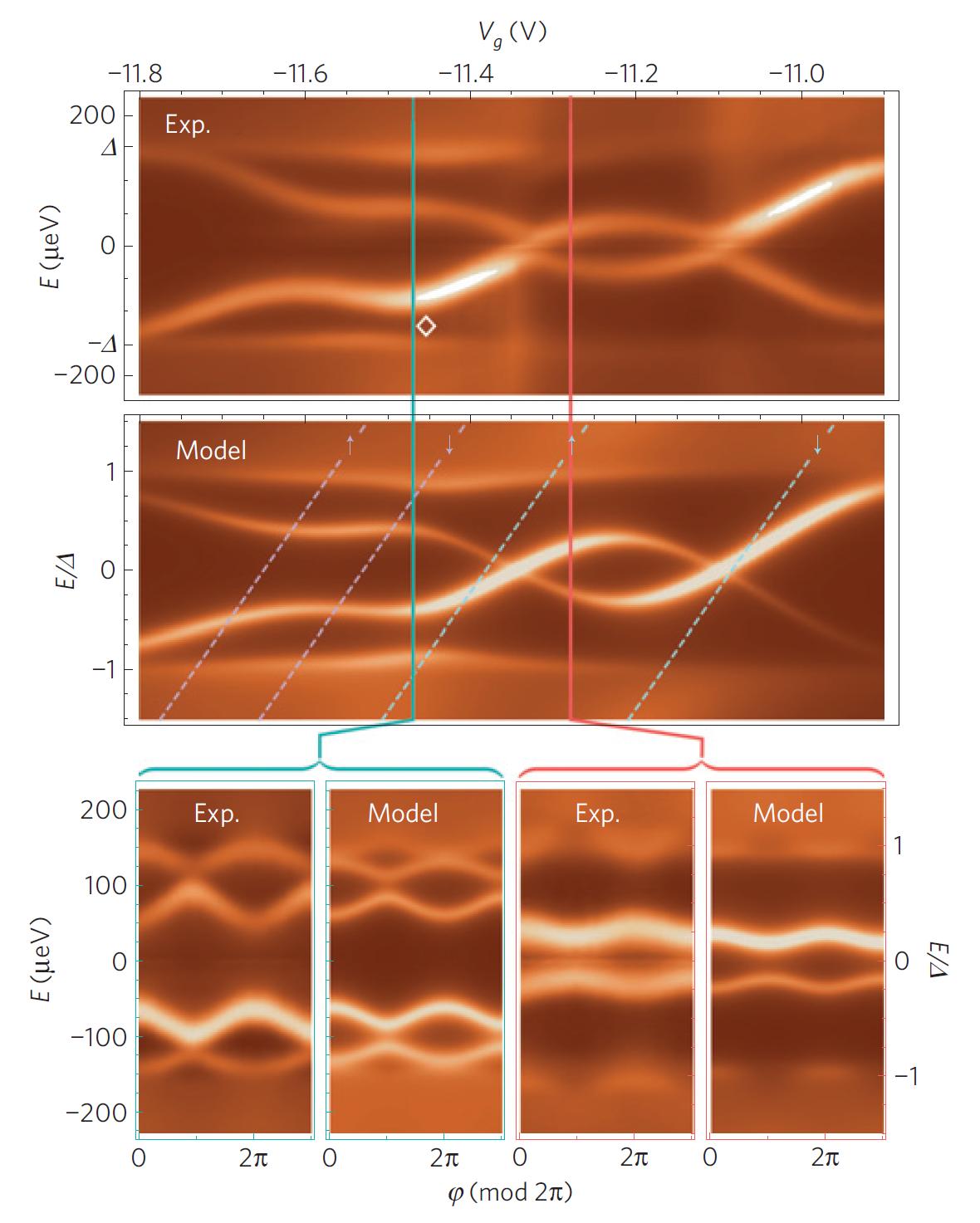}
\caption{Color scale plots of the local density of states in a CNT QD coupled to SC leads
as reported in Ref. \cite{pill10} as a function of gate voltage and phase difference. The figure
also shows the results obtained from a phenomenological model, as discussed in the text.
Reprinted by permission
from Macmillan Publishers Ltd: Nature Physics \cite{pill10}, copyright (2010).}
\label{pillet-10-fig4}
\end{center}
\end{figure}

Finally, it is worth pointing out recent experimental developments which have allowed to directly
measure the Andreev bound states spectrum of a CNT-QD coupled to superconducting leads 
in a SQUID configuration \cite{pill10}. In this experiment a weakly coupled lead was deposited at
the center of the CNT to allow for tunneling spectroscopy measurements. In this way
both the phase and gate voltage
variation of the Andreev bound states were determined. The results could be fitted satisfactorily
using a simplified phenomenological model corresponding to the superconducting Anderson model
within a mean field approximation, in which an exchange field is included to represent the magnetic
ground state. An example of the comparison between theory and experiment
is given in Fig. \ref{pillet-10-fig4}. This analysis showed that a double dot model was in
general necessary to fit the experimental results. Recent experimental work on graphene 
QDs coupled to SC leads \cite{dirks11} provided  also evidence on the crossing of ABs as a function of the gate
potential which is consistent with a magnetic ground state.

\section{Quantum dots with normal and superconducting leads}
\label{NQDS}

A single-level quantum dot coupled to a normal and a superconducting
lead provides a basic model system to study electron transport 
in the presence of Coulomb and pairing interactions. Compared to the S-QD-S situation, this case
has a simpler response in non-equilibrium conditions due to the absence of the ac Josephson
effect. For this reason this system has been widely analyzed theoretically. 

As in any N-S junction the low bias transport properties are dominated by Andreev 
processes. This mechanism is in general highly modified by resonant tunneling through
the localized levels in the dot. In addition, charging effects can
strongly suppress the Andreev reflection in certain parameter ranges. Furthermore there
is also an interesting interplay between Kondo and pairing correlations as in the case
of the S-QD-S system discussed above.

As an illustration of the general formalism we derive here the linear transport properties
of the non-interacting model. One can straightforwardly write the dot retarded Green function for
this case from expression (\ref{retarded-advanced-GF}) by setting $\Delta_L=0$. Then, from the
expression of the current in terms of Keldysh Green functions and using the corresponding
Dyson equation one can write

\begin{eqnarray}
I_L &=& \frac{e}{h} \sum_k \int d\omega \mbox{Tr} \left[ \tau_3 \left(V_{kL}\hat{g}^{+-}_{kL}V_{Lk} \hat{G}^{-+}_{00}   \right. \right. \nonumber\\
&&  \left. \left. -V_{kL} \hat{g}^{-+}_{kL} V_{Lk} \hat{G}^{+-}_{00} \right)\right] ,
\end{eqnarray}
where $\hat{g}_{kL}^{+-,-+}$ are the Keldysh Green functions of the uncoupled normal lead.
By further using $G^{+-,-+}_{00} = \sum_{\mu,k} G^r_{00} V_{\mu k} \hat{g}_{k\mu}^{+-,-+} V^*_{\mu k} \hat{G}^a_{00}$ and
taking the wide band approximation for the uncoupled leads one can obtain
the following expression for the temperature dependence linear conductance \cite{schw99,cuevas01}

\begin{equation}
G = \frac{16e^2}{h} \Gamma_N \Gamma_S \int d\omega \mbox{Im} \left(G^r_{12} G^a_{11} \right) \left(-\frac{\partial n_F}{\partial \omega}\right)
\end{equation}
which, at zero temperature reduces to the simple expression arising from the contribution of 
pure Andreev reflection processes

\begin{equation}
G = \frac{4e^2}{h} \frac{4\Gamma_N^2\Gamma_S^2}{\left(\epsilon_0^2 + \Gamma_N^2 + \Gamma_S^2\right)^2}
\end{equation}

As shown in Ref. \cite{beenakker92} this expression for the non-interacting case is equivalent to 
the formula $G= (4 e^2/h) \tau^2/(2 - \tau)^2$, with 
$\tau = 4 \Gamma_N\Gamma_S/(\epsilon_0^2 + (\Gamma_N+\Gamma_S)^2)$ being
the normal transmission through the dot at the Fermi energy. In contrast to the case of a NS quantum point contact
with essentially energy independent transmission,
in the dot case the Andreev processes become resonant at $\epsilon_0=0$ reaching the maximum value $G=4e^2/h$.

\subsection{Effect of interactions (linear regime)}
\label{NQDS-linear}

One of the first attempts to describe the effect of Coulomb interactions in the NDQS system was
presented by Fazio and Raimondi \cite{fazi98} using the equation of motion technique truncated to the
second order in the tunneling to the leads. They derived expressions for the mean current using the 
Keldysh formalism and extending the so-called Ng ansatz \cite{ng93} to the superconducting case.
The claim of an extended temperature range for the zero bias anomaly due to the Kondo resonance was
subsequently corrected in \cite{fazi99}.

The problem was addressed in Ref. \cite{kang98} by assuming that the relation
$G= (4e^2/h) \tau^2/(2 - \tau)^2$ of the non-interacting case still holds by substituting $\tau$ by the
normal transmission of the interacting case. Within this assumption that work suggested an
increase of the conductance in the Kondo regime by a factor of two with respect to the normal case. 
As shown by subsequent works which we discuss below, this enhancement is not always possible, the
general case is rather the opposite. 

The conductance of the interacting N-QD-S system was also analyzed in Ref. \cite{schw99} within the
infinite-U slave-boson mean field approximation. This approximation reduces the problem into an
effective Fermi liquid description with renormalized parameters $\tilde{\Gamma}_{N,S}$ and
$\tilde{\epsilon}_0$. Within this approximation both $\Gamma_{N,S}$ are renormalized equally and
therefore the condition for the maximum conductance is reached for the symmetric case as in the
normal state. As the authors acknowledge this result is valid only in the deep Kondo regime
$\Delta \ll T_K$, otherwise residual interactions would renormalize the left and right tunneling rates differently,
in particular $\Gamma_S$ coupling the dot with the superconductor would be significantly suppressed by
interactions.

The problem was subsequently addressed by Clerk et al. \cite{cler00} using the extension of the
NCA to the superconducting case already mentioned in Sect. \ref{SQDS-eq}. They analyze three different models:
the N-QD-S Anderson model and a single channel magnetic and a two-channel non-magnetic contact between normal 
and superconducting electrodes. We comment here only the results for the first model. They find an overall
decrease of the quasiparticle spectral density at the Fermi energy together with the appearance of
additional Kondo peaks at $\pm \Delta$. As a consequence, their conclusion was that there is no enhancement of the
linear conductance due to Andreev processes in contrast to the claim of previous works.  

Diagrammatic techniques for the finite-U N-QD-S Anderson model were applied in Ref. \cite{cuevas01}. Within this
approach the linear conductance can be expressed as the one corresponding to the non-interacting model
with asymmetrically renormalized parameters 

\begin{equation}
G = \frac{4e^2}{h} \frac{4\Gamma_N^2\tilde{\Gamma}_S^2}{\left(
\tilde{\epsilon}_0^2 + \Gamma_N^2 + \tilde{\Gamma}_S^2\right)^2}
\label{conductance-pert-ns}
\end{equation}
where $\tilde{\Gamma}_S = \Gamma_S - \Sigma_{12}(0)$ and $\tilde{\epsilon}_0 = \epsilon_0 - \Sigma_{11}(0)$,
$\Sigma_{\mu,\nu}$ being the dot self-energy elements in Nambu space. Although this result is valid in general within
a diagrammatic analysis, concrete results were obtained in this work by means of an interpolated second-order
approach. From Eq. (\ref{conductance-pert-ns}) it was found that even when the starting bare parameters
correspond to the symmetric case $\Gamma_N = \Gamma_S$, interactions would tend to reduce the conductance
by inducing an asymmetry, i.e. leading to $\tilde{\Gamma}_S \ne \Gamma_N$, as was suggested in 
Ref. \cite{schw99}. However, this equation also predicts the possibility that an adequate tuning of the
bare coupling parameters could yield an enhancement of the conductance up to the unitary limit ($4e^2/h$ in the
NS case) which would not correspond to the maximum conductance in the normal case.

\begin{figure}
\begin{center}
\includegraphics[scale=0.3]{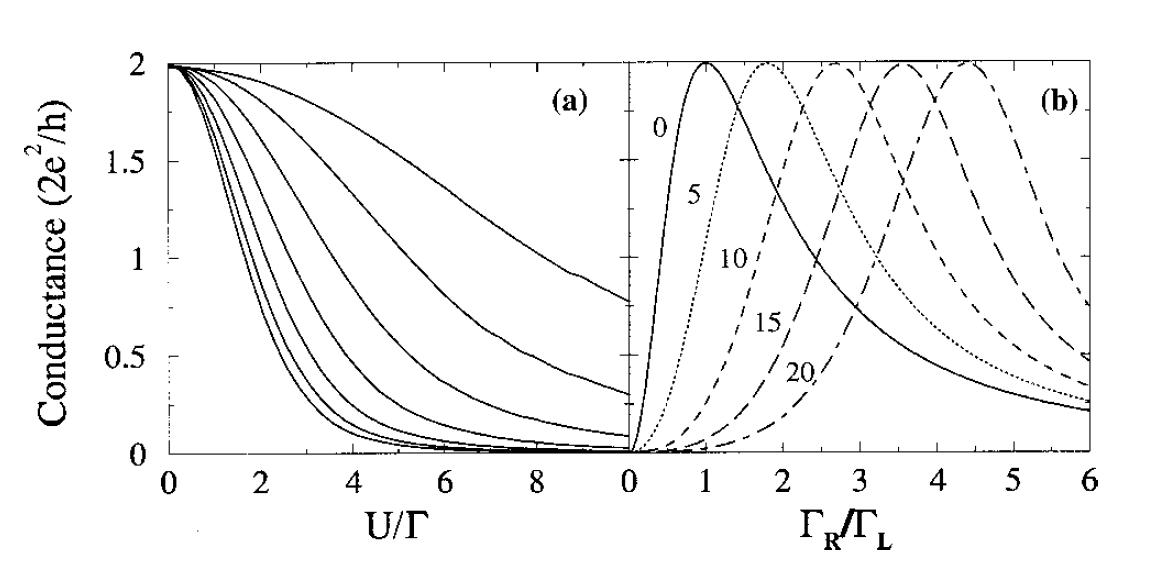}
\caption{Linear conductance for the N-QD-S system in the second-order self-energy approximation of Ref. \cite{cuevas01}. 
(a) symmetric case as a function of $U/\Gamma$ for different values of $\Gamma/\Delta= 0.125, 0.25, 0.5, 1.0, 2.0, 4.0$ and 8.0, from
top to bottom. (b) Same as (a) for asymmetric coupling to the leads $\Gamma_L \neq \Gamma_R$ and different values of $U/\Gamma_L$
(within the text we have set $\Gamma_L\equiv\Gamma_N$ and $\Gamma_R\equiv\Gamma_S$).
Reprinted figure with permission
from J.C. Cuevas {\it et al.}, Physical Review B {\bf 63}, 094515, 2001 \cite{cuevas01}.
Copyright (2001) by the American Physical Society.}
\label{cond-ns-cuevas}
\end{center}
\end{figure}

The approximation of Ref. \cite{cuevas01} is based on the evaluation of the second order diagrams, which due to
the proximity induced pairing in the dot are formally the same as those of Fig. \ref{diagrams-2nd-order} 
which were discussed for the
S-QD-S case. For the non-symmetric case the authors used an interpolative ansatz which recovers the correct behavior
in the $\Gamma/U \rightarrow 0$ (atomic) limit. The obtained behavior of the conductance as a function of $U/\Gamma$
for the symmetric and non-symmetric cases is illustrated Fig. \ref{cond-ns-cuevas}.

As can be observed, in the symmetric case the conductance drops steadily from the unitary limit as $U/\Gamma$ increases,
the scale of this decay being set by the parameter $\Gamma/\Delta$. On the other hand the right panel of Fig.
\ref{cond-ns-cuevas} illustrates that the unitary limit can be restored by an adequate tuning of the ratio
$\Gamma_N/\Gamma_S$.

The behavior of the linear conductance in the N-QD-S was also analyzed in \cite{kraw04,doma08} using the EOM technique
with different decoupling schemes. While in Ref. \cite{kraw04} it was obtained that the conductance due to 
Andreev processes was completely suppressed in the Kondo regime, an improved approximation for the EOM decoupling
procedure in Ref. \cite{doma08} showed that there is in fact a finite zero bias anomaly in the Andreev 
conductance although it is in general much smaller than the one in the normal case, its value depending
on the ratio $\Gamma_N/\Gamma_S$. This behavior is in qualitative agreement with the results of the diagrammatic
approach discussed before. However, the unitary limit is not reached within this approach. 

\begin{figure}
\begin{center}
\includegraphics[scale=0.7]{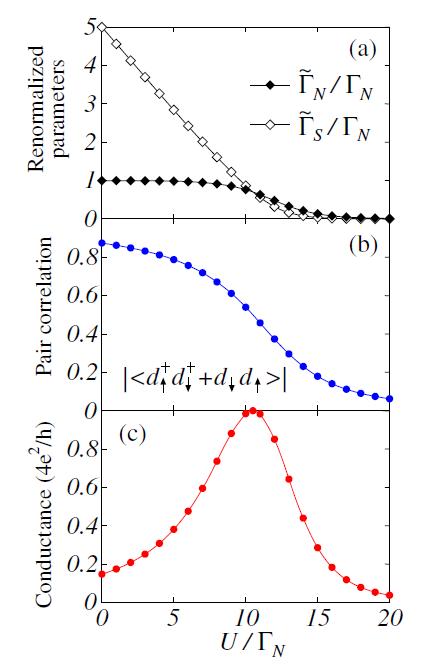}
\caption{Renormalized couplings to the leads $\tilde{\Gamma}_{N,S}$ (upper panel); Induced pairing amplitude (middle panel) and
linear conductance (lower panel) as a function of $U/\Gamma_N$ for the N-QD-S system obtained using NRG method in the 
$\Delta \rightarrow \infty$ limit. Reprinted figure with permission
from Y. Tanaka {\it et al.}, Journal of the Physical Society of Japan {\bf 76}, 074701, 2007 
\cite{tana07-2}. Copyright (2007) by the Physical Society of Japan.}
\label{fig-tanaka-3}
\end{center}
\end{figure}

More recently the linear conductance of the N-QD-S model has been studied using the NRG method \cite{tana07-2}. 
In this work the limit $\Delta \rightarrow \infty$ was taken from the start, which allows to map the 
problem into the case of a QD with a local pairing amplitude $\Delta_d = \Gamma_S$ coupled to a
single normal electrode (the $\Delta \rightarrow \infty$ limit for the S-QD-S case was discussed in Sect. \ref{SQDS-eq}). 
The NRG algorithm can be considerably simplified by this assumption because a simple Bogoliubov transformation
allows to get rid of the local pairing term leading to a problem which is formally equivalent to a 
normal Anderson model, which implies Fermi liquid behavior. As in the diagrammatic approach discussed before,
the main effect of the interactions is to renormalize the couplings $\Gamma_{N,S}$ and the dot level position.
Figure \ref{fig-tanaka-3} illustrates the evolution of the renormalized parameters $\tilde{\Gamma}_{N,S}$
as a function of $U$ for the half-filled case $\epsilon_0 = -U/2$ with 
initial parameters $\Gamma_S = 5 \Gamma_N$. As can be observed in the upper panel of Fig. \ref{fig-tanaka-3}, 
the main effect of increasing the interaction $U$ is to reduce the effective coupling to the superconductor, $\tilde{\Gamma}_S$,
while the effective coupling to the normal lead $\tilde{\Gamma}_N$ remains almost constant up to the region 
$U \sim 10 \Gamma_N$ when the Kondo effect is significant. Eventually the system reaches the condition
$\tilde{\Gamma}_N \simeq \tilde{\Gamma}_S$ and the conductance increases up to the unitary limit, as shown in the
lower inset of Fig. \ref{fig-tanaka-3}. This behavior is in good agreement with the prediction of the diagrammatic
theory of Ref. \cite{cuevas01}. On the other hand, the induced pairing amplitude (middle panel in Fig. \ref{fig-tanaka-3}),
exhibits the expected monotonous decrease for increasing intradot repulsion.

\begin{figure}
\begin{center}
\includegraphics[scale=0.3]{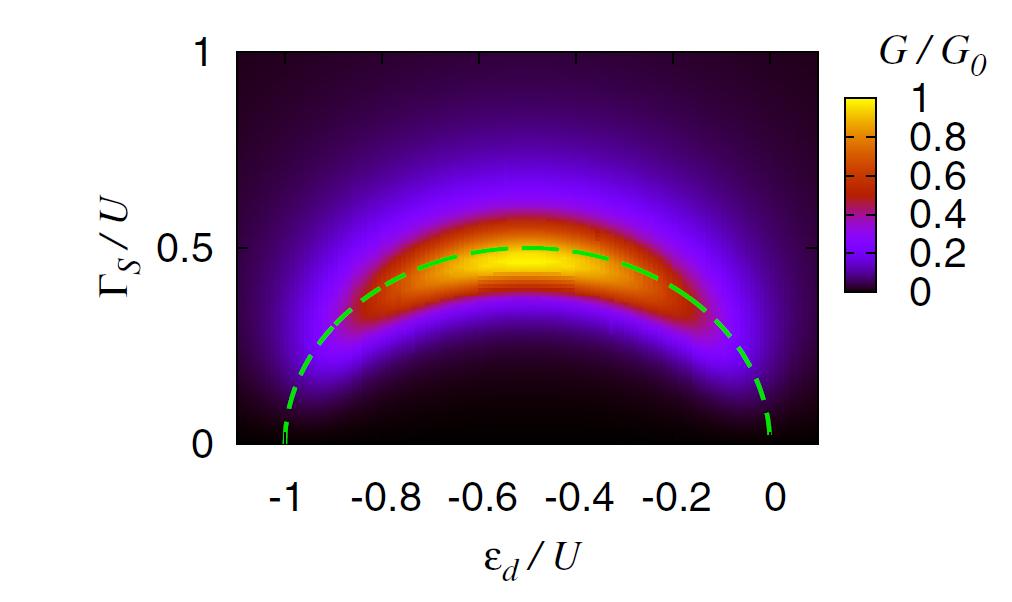}
\caption{Color-scale map of the linear conductance in the N-QD-S system in the $\Gamma_S/U$-$\epsilon_0/U$ plane obtained
in Ref. \cite{tana07-2} using the NRG method as in Fig. \ref{fig-tanaka-3}.
Reprinted figure with permission
from Y. Tanaka {\it et al.}, Journal of the Physical Society of Japan {\bf 76}, 074701, 2007
\cite{tana07-2}. Copyright (2007) by the Physical Society of Japan.}
\label{fig-tanaka-6}
\end{center}
\end{figure}

The behavior of the linear conductance outside the half-filled case obtained from these NRG calculations is illustrated 
in Fig. \ref{fig-tanaka-6}. In this color-scale map it can be clearly observed that the unitary limit is reached
mainly along the curve $((\epsilon_0/U + 1/2)^2 + (\Gamma_S/U)^2)^{1/2}= 1/2$, which corresponds to the single-doublet
transition in the $\Gamma_N \rightarrow 0$ limit. When $\Gamma_S/U < 0.5$ (corresponding to the doublet state in the
$\Gamma_N \rightarrow 0$ limit within the dashed curve in Fig. \ref{fig-tanaka-6}), the conductance as a function of $\epsilon_0$
exhibits a double peaked structure, whereas for $\Gamma_S/U > 0.5$ only a single peak is found which can be
correlated to the superconducting singlet ground state of the system in the $\Gamma_N \rightarrow 0$ limit.

\subsection{Non-linear regime} 
\label{NQDS-nonlinear}

The non-linear regime in the N-QD-S system has received so far much less attention and 
there are still aspects, specially those related to the Kondo effect which are not sufficiently understood.
This regime has been analyzed using the EOM technique employing different decoupling schemes
in Refs. \cite{fazi98,sun01,kraw04} and \cite{doma08}. The approximation used in Ref. \cite{fazi98} has
been already described in the context of the linear regime. On the other hand, the approximation used in Ref. \cite{sun01} 
consisted in truncating the EOM equations at the level of the two particle Green functions by substituting the
leads fermionic operators by their average values. Within this decoupling the authors find the appearance of Kondo features
in the dot spectral density. In addition to the usual features at $\omega = \pm eV$ for the regime $\Delta > \Gamma_S >\Gamma_N$,
they also find excess Kondo like features at $\omega = \pm(2\epsilon_0 + U - eV)$. These features have been explained
as arising from co-tunneling processes involving Andreev tunneling from the QD-S interface and normal tunneling from
the N-QD interface. 

\begin{figure}
\begin{center}
\includegraphics[scale=0.4]{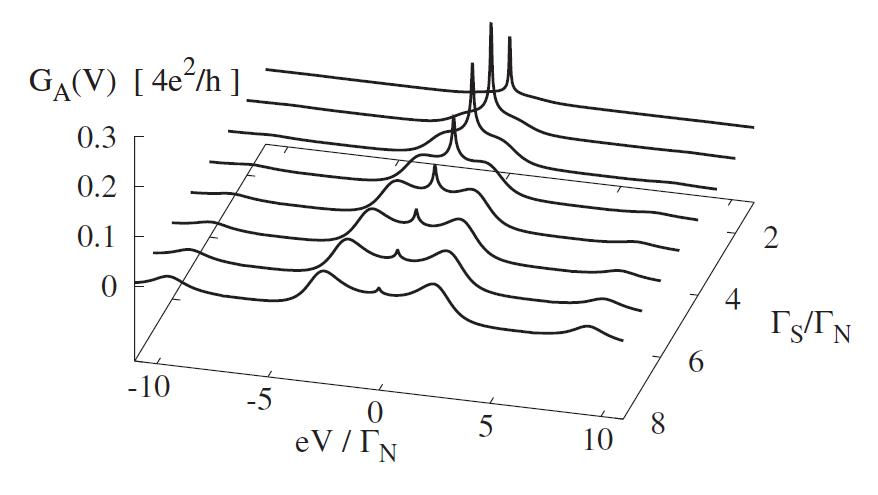}
\caption{Non-linear conductance for the N-QD-S system obtained using the EOM method in the decoupling scheme of
Ref. \cite{doma08} in the $\Delta \rightarrow \infty$ limit as a function of $eV/\Gamma_N$ and $\Gamma_S/\Gamma_N$. Reprinted figure with permission
from T. Domanski and A. Donabidowicz, Physical Review B {\bf 78}, 073105, 2008 \cite{doma08}.
Copyright (2008) by the American Physical Society.}
\label{fig-doma-6}
\end{center}
\end{figure}

In Refs. \cite{kraw04,doma08} already commented for the linear regime, the finite voltage case was also analyzed.
The non-linear conductance obtained for the $\Delta \rightarrow \infty$ limit
in Ref. \cite{doma08} both as a function of the bias voltage and the asymmetry
in the coupling parameters $\Gamma_S/\Gamma_N$ is shown in Fig. \ref{fig-doma-6}. The parameters of this case
correspond to the Kondo regime of the normal state $\epsilon_0 = -1.5 \Gamma_N$ and $U = 10 \Gamma_N$. 
This figure exhibits the expected features like the qualitative evolution of the zero bias anomaly already
discussed in the previous section and the splitting of the dot resonances due to the proximity effect. 
Additional peaks at $eV \sim \pm U$ can be observed corresponding to the population of the higher charge
states.

\begin{figure}
\begin{center}
\includegraphics[scale=0.35]{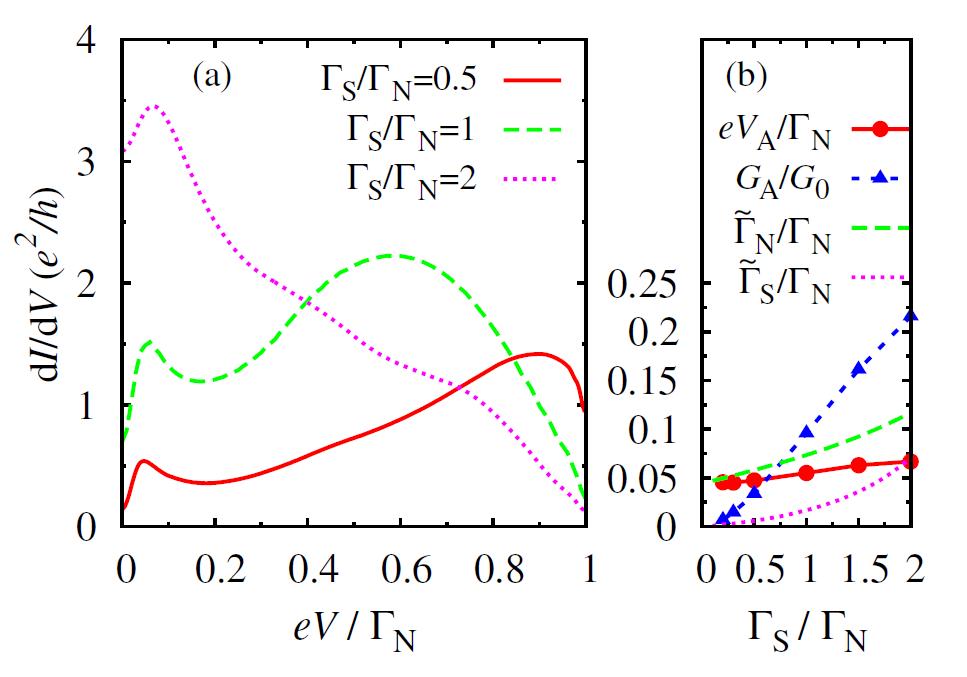}
\caption{Non-linear conductance for the N-QD-S system using the interpolative self-energy approach of Ref. \cite{yama10}.
The results correspond to the symmetric case with $\Gamma_N=\Delta$ and different values of $\Gamma_S$ and $U$. Reprinted figure with permission
from Y. Yamada {\it et al.}, Journal of the Physical Society of Japan {\bf 79}, 043705, 2010
\cite{yama10}. Copyright (2010) by the Physical Society of Japan.}
\label{fig-yama-2}
\end{center}
\end{figure}

The non-linear case has been analyzed more recently in Ref. \cite{yama10} by extending the interpolative
self-energy approach of Ref. \cite{cuevas01} to the non-equilibrium case. The authors consider the case
$\Delta = \Gamma_N$ and $\epsilon_0 = -U/2$ for different values of $\Gamma_S$ and $U$.
Their main findings are illustrated in Fig. \ref{fig-yama-2} corresponding to the
non-linear conductance for $U=20\Gamma_N$. They observe a Kondo peak 
which is displaced from zero bias, whose height increases with increasing $\Gamma_S$ while its position is 
only weakly modified. When reducing $\Gamma_S$ a second peak develops which shift progressively towards the
gap edge $eV/\Gamma_N = 1$. 

\begin{figure}
\begin{center}
\includegraphics[scale=0.3]{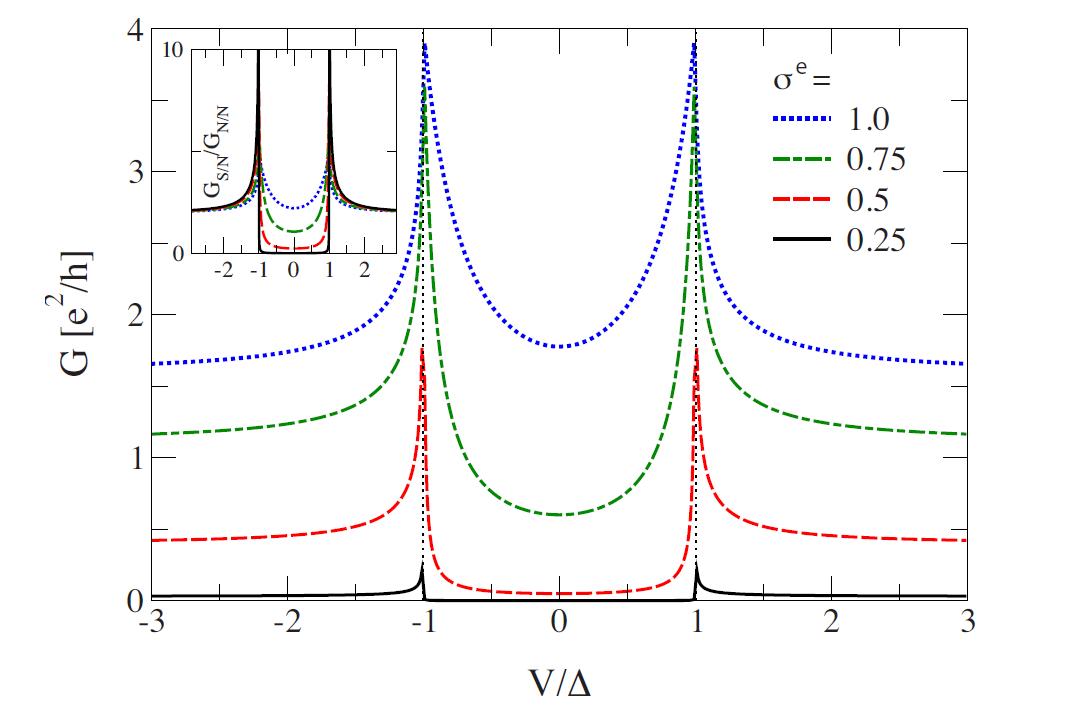}
\includegraphics[scale=0.3]{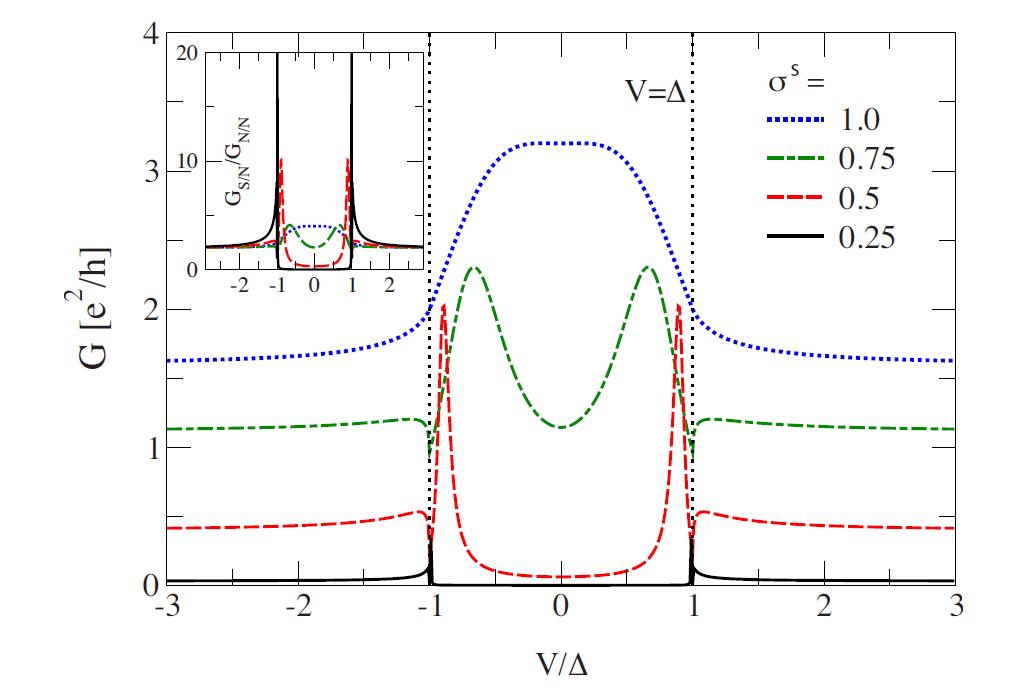}
\caption{Non-linear conductance for the N-QD-S system in the effective cotunneling model of Ref. \cite{koerting10}
for even (upper panel) and odd (lower panel) cases. The parameters $\sigma^e$ and $\sigma^s$ control the normal
transmission through the dot in the even and odd cases respectively. The inset shows the differential conductance
normalized to the one in the normal case. Reprinted figure with permission
from V. Koerting {\it et al.}, Physical Review B {\bf 82}, 245108, 2010 \cite{koerting10}.
Copyright (2010) by the American Physical Society.}
\label{fig-koerting-4}
\end{center}
\end{figure}

We should also mention the work of Ref. \cite{koerting10} in which the non-equilibrium
transport through a N-QD-S system is studied within an effective cotunneling model. Within this approach the
self-energy is calculated to leading order in the cotunneling amplitude from which 
the nonlinear cotunneling conductance can be obtained. By neglecting charge fluctuations in the dot 
two different regimes are found corresponding to the case of even and odd number of electrons. For the even
case the system becomes equivalent to an effective S/N junction with the subgap transport due to
Andreev reflection processes. On the other hand, for the odd case they find that the net spin within the
dot leads to the appearance of subgap resonances 
giving rise to a peak-dip structure in the differential
conductance. The typical conductance curves that are found for both cases are shown in Fig. 
\ref{fig-koerting-4}. As can be observed, in the even case (upper panel in Fig. \ref{fig-koerting-4}) 
the behavior
of the conductance is similar to a conventional NS junction with an effective transmission set by
the second-order cotunneling amplitude. In the odd case (lower panel) the double peak structure within the
subgap region evolves into a single zero bias peak as the cotunneling amplitude increases. 

To conclude this section it appears that our present knowledge of the non-equilibrium N-QD-S
system is still limited and further research would be desirable, particularly to understand the 
behavior of Kondo features at finite applied voltages and the crossover between 
the different parameter regimes so far analyzed. In this respect we refer the interested reader
to a recent work \cite{yamada11} that has been published after submitting this review.

\subsection{Experimental results}
\label{NQDS-exp}

\begin{figure}
\begin{center}
\includegraphics[scale=0.5]{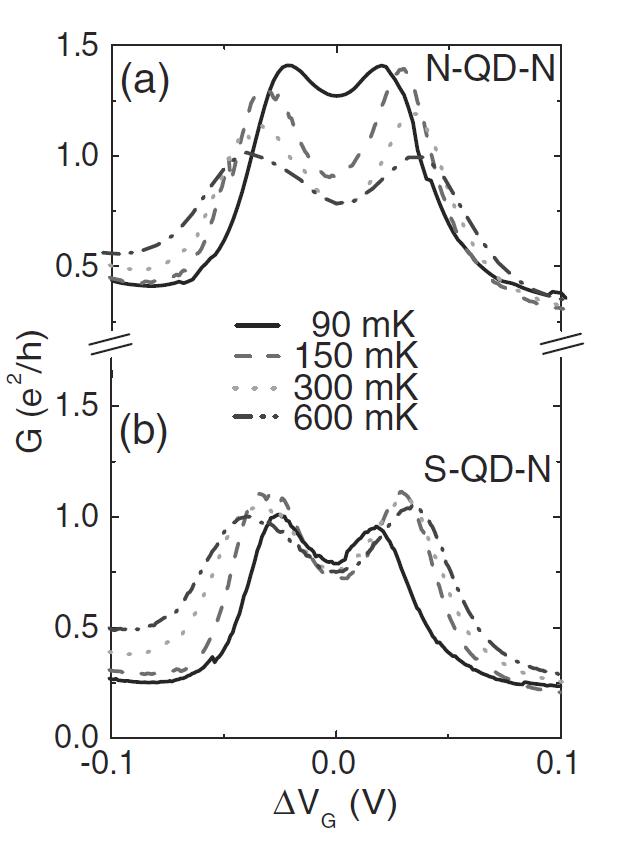}
\caption{Comparison between the linear conductance in the NQDN and SQDN systems as a function 
of the gate voltage for different temperatures
in the experimental realization of Ref. \cite{grab04}. Reprinted figure with permission
from M.R. Graber {\it et al.}, Nanotechnology {\bf 15}, S479, 2004 \cite{grab04}.
Copyright (2004) by IOP Publishing Ltd.}
\label{fig-graber-3}
\end{center}
\end{figure}

Unlike the case of the S-QD-S system, 
only a few works have addressed the issue of the transport properties of N-QD-S systems experimentally.
This is probably due to the technical difficulties associated to the fabrication of such hybrid
systems. The first experimental realization of this configuration was presented by Gr\"aber et al. 
\cite{grab04} using a multiwall CNT as a QD coupled to Au (normal) and Al/Au (superconducting) leads
at each side. They first analyzed the normal case by applying a magnetic field of $25 mT$, clearly
observing Kondo features in the linear conductance, as shown in the upper panel of Fig. \ref{fig-graber-3}.
At the lowest temperature of $90 mK$ the normal conductance reached values $\sim 1.5 e^2/h$, indicating
good and rather symmetric coupling to the leads. When one of the leads become superconducting
it was observed that the temperature dependence characteristic of the Kondo regime was very much suppressed,
as shown in the lower panel of Fig. \ref{fig-graber-3}. This behavior is in qualitative agreement with the
theoretical results of Refs. \cite{cuevas01,tana07-2} for the case of a nearly symmetrically coupled dot.

\begin{figure}
\begin{center}
\includegraphics[scale=0.22]{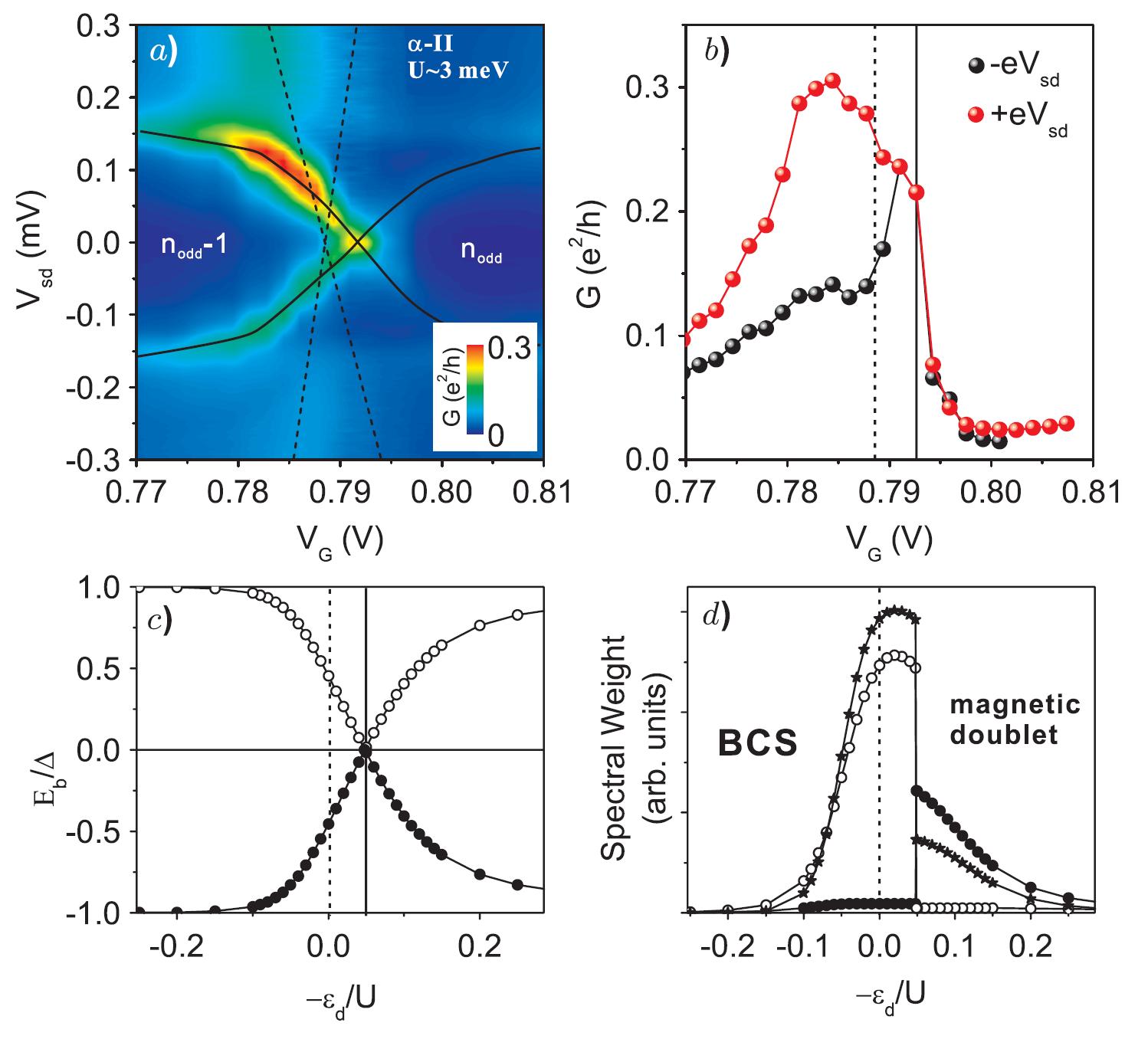}
\caption{(a) Color scale map of the differential conductance as a function of the source-drain and gate voltages
in the experimental realization of the N-QD-S system of Ref. \cite{deac10}. (b) plot of the peak conductance following
the subgap resonances indicated by the full lines in (a).  Panels (c) and (d) show the Andreev levels position and
weight respectively obtained using the NRG method for $U/\Delta=20$ and $\Gamma_S/\Delta=0.7$.
Reprinted figure with permission
from R.S. Deacon {\it et al.}, Physical Review Letters {\bf 104}, 076805, 2010 \cite{deac10}.
Copyright (2010) by the American Physical Society.}
\label{fig-deacon-3-a}
\end{center}
\end{figure}

A different experimental realization of the N-QD-S system was presented in Refs. \cite{deac10,deac10b}.
These authors used a self-assembled InAs QD with diameters of the order of $\sim 100 nm$ coupled to
a Ti/Au (N lead) and a Ti/Al (S lead). In the first of these works the authors
focused on devices with large coupling asymmetry $\Gamma_S \gg \Gamma_N$ in which the Kondo effect is
suppressed by the strong proximity effect. In this limiting situation the normal lead is basically providing
a means to probe spectroscopically the Andreev spectra of the QD-S system. 
The experiment provided evidence of the transition between the singlet and the doublet ground state for the
QD-S system when the number of electrons changed from even to odd. As shown in Fig. \ref{fig-deacon-3-a} this
was reflected in the behavior of the Andreev states within the gap exhibiting a crossing point
together with a large drop in the conductance. These experimental results were in good qualitative
agreement with NRG calculations for the QD-S system.

\begin{figure}
\begin{center}
\includegraphics[scale=0.25]{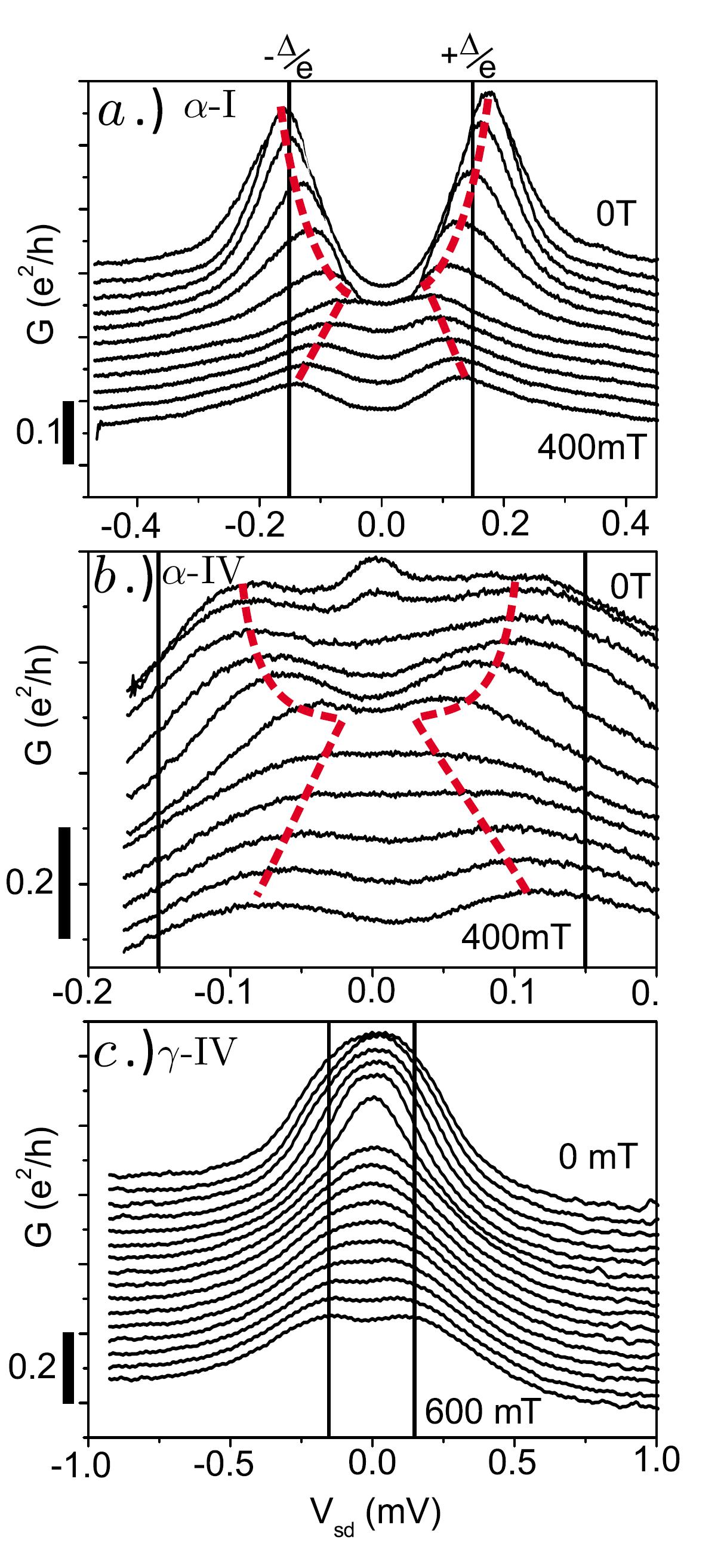}
\caption{Differential conductance at the center of odd occupation regions for three different samples
of the experimental realization of the N-QD-S system of Ref. \cite{deac10b} as a function of the applied
magnetic field. The three cases correspond to different values of the $\Gamma_S/\Gamma_N$ parameter:
0.045 (upper panel) and 8.0 (middle and lower panels). 
Reprinted figure with permission
from R.S. Deacon {\it et al.}, Physical Review B {\bf 81}, 121308, 2010 \cite{deac10b}.
Copyright (2010) by the American Physical Society.}
\label{fig-deacon-3-b}
\end{center}
\end{figure}

In a subsequent work by this group \cite{deac10b} the same experimental realization but 
with varying coupling asymmetry was analyzed. The main results of this work are shown in Fig. 
\ref{fig-deacon-3-b} corresponding to asymmetries $\Gamma_S/\Gamma_N = 0.045$ (upper panel) and $8.0$ (two lower
panels). For the first case with sufficiently large $\Gamma_N$ one would expect the formation of a Kondo
resonance due to the good coupling of the QD with the normal electrode. However, the conductance
which is mediated by Andreev processes is suppressed inside the gap due to the very small coupling to
the superconductor. The cases with asymmetries of the order of 8.0 were not in the extreme situation
of the previous work (with $\Gamma_S/\Gamma_N \sim 50$) and thus did exhibit Kondo features as
can be observed in the lower panels of Figs. \ref{fig-deacon-3-b}.

\section{Voltage biased S-QD-S systems}
\label{SQDS-neq}

Including a finite bias voltage between the superconducting
electrodes in the S-QD-S system poses an additional difficulty in the theory
due to the intrinsic time-dependence of the ac Josephson effect. 
Even in the non-interacting case the inclusion of MAR processes
up to infinite order constitutes a challenging problem which
in general requires a numerical analysis. 

\begin{figure}
\begin{center}
\includegraphics[scale=0.35]{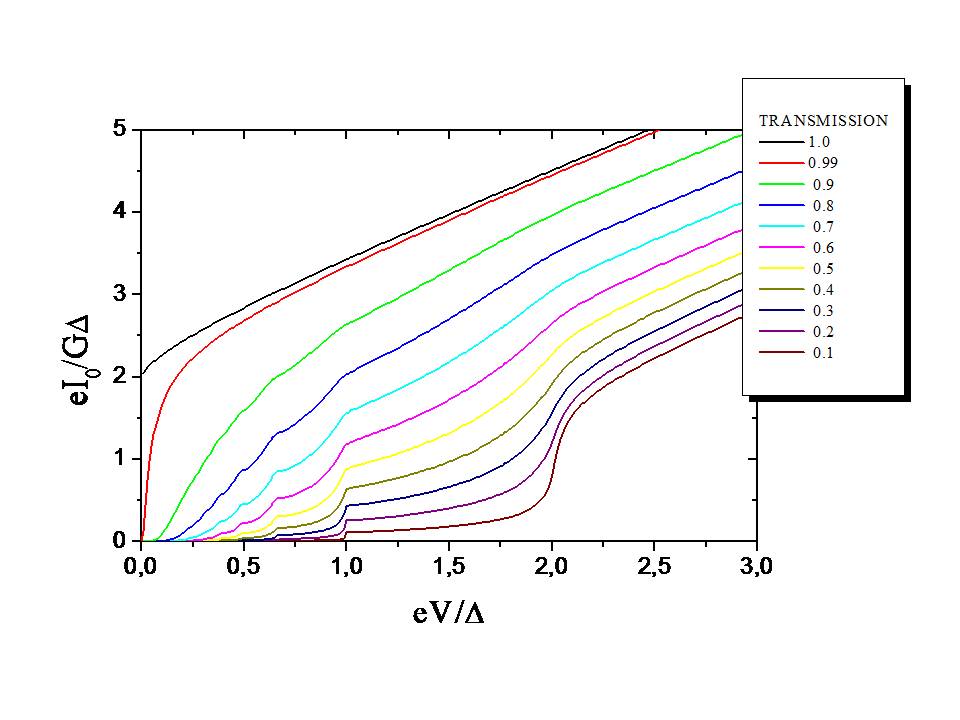}
\caption{dc I-V characteristic for a superconducting single channel contact for different values of the
normal transmission.}
\label{subgap-contact}
\end{center}
\end{figure}

Although the mechanism of MAR processes for explaining the subgap structure 
in superconducting junctions was introduced in the early '80s \cite{kbt,obtk},
it was not until the mid '90s that a full quantitative theory of
the I-V characteristics in a superconducting
contact of arbitrary transparency was developed using either a scattering approach 
\cite{bratus95,averin95} or a non-equilibrium Green function approach \cite{cuevas96}.
This theoretical progress allowed a very accurate description of experimental
results for atomic size contacts \cite{scheer97,scheer98}. For the discussion in the present section
it could be useful to remind the main features of the I-V characteristics of a one channel contact.
Fig. \ref{subgap-contact} shows the evolution of the dc current as a function of the contact 
transmission obtained using the theory of Ref. \cite{cuevas96}. As can be observed, at sufficiently 
low transmission the current exhibits
a subgap structure with jumps at $V=2\Delta/n$, corresponding to the threshold voltage for an $n$-order
MAR process. As the transmission is increased the subgap structure is progressively smeared out and eventually
at $\tau=1$ the behavior of the I-V curve is almost linear except in the limit $V\rightarrow0$ where it
saturates to a finite value $2e\Delta/h$ \cite{averin95,cuevas96}. 

The case of a non-interacting resonant level coupled to superconducting electrodes was first
analyzed in Ref. \cite{levy97}.
We discuss briefly here the Green function formalism for this case which has the advantage of
allowing to include the effect of interactions in a second step. The main
point in this formalism is to realize that even when the Green functions
$\breve{G}(t,t')$ depends on the two time arguments, 
in the case of a constant voltage bias the dependence on the mean time $(t+t')/2$ can only correspond to
the harmonics of the fundamental frequency $eV/\hbar$ \cite{cuevas96}. This allows to express all 
quantities in terms of the components $\breve{G}_{nm}(\omega)$ corresponding to a double Fourier
transformation \cite{arnold} of $\breve{G}(t,t')$ defined as \cite{martin-rodero99}

\begin{equation}
\breve{G}_{nm}(\omega) = \int dt \int dt' e^{-iV(nt-mt')} e^{i\omega(t-t')} \breve{G}(t,t')
\end{equation}

The Fourier components $\breve{G}_{nm}$ obey an algebraic Dyson equation in the discrete
space defined by the harmonic indices which can be solved using a standard recursive algorithm.  
A compact expression of these equations for the dot case, given in Ref. \cite{dell08}, 
is

\begin{equation}
(\breve{G_{00}})^{-1}_{nm} = \left(\omega_n - \epsilon_0 \sigma_z\right) \tau_z - \sum_{\mu=L,R}
\Gamma_{\mu} \sigma_z \tau_z \breve{g}_{nm}(\omega) \sigma_z \tau_z
\end{equation}
where $\omega_n = \omega + nV$, while $\sigma_i$ and $\tau_i$ correspond to Pauli matrices in the Nambu and 
Keldysh space respectively and

\begin{equation}
\breve{g}_{nm} = \left( \begin{array}{cc} \delta_{nm} \breve{X}(\omega_n \mp V/2) & \delta_{n,m\mp1} 
\breve{Y}(\omega_n \mp V/2) \\
\delta_{n,m \pm 1} \breve{Y}(\omega_n \pm V/2) & \delta_{nm} \breve{X}(\omega_n \pm V/2) \end{array} \right)
\end{equation}
where the matrix $\breve{X}(\omega) = -\omega \breve{Y}/\Delta$ in Keldysh space are given by

\begin{equation}
\breve{X}(\omega) = \left\{\begin{array}{lr} -\frac{\omega}{\sqrt{\omega^2 - \Delta^2}}\tau_z & |\omega| > \Delta \\
\frac{i|\omega|}{\sqrt{\Delta^2 - \omega^2}} \left(
\begin{array}{cc} 2n_F(\omega) - 1 &  2n_F(\omega) \\ 
2n_F(-\omega) & 2n_F(\omega)- 1 \end{array} \right) & |\omega| < \Delta \end{array} \right. 
\end{equation}

\begin{figure}
\begin{center}
\includegraphics[scale=0.35]{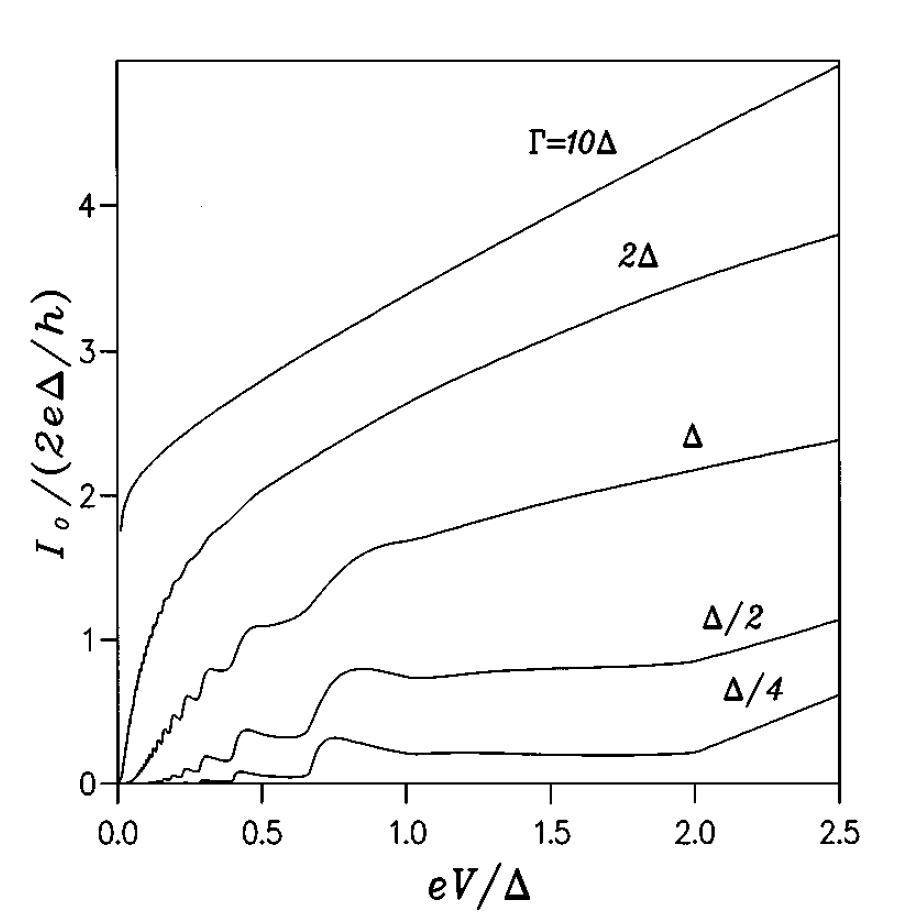}
\caption{dc I-V characteristic for the non-interacting S-QD-S system with $\epsilon_0=0$ and different values
of $\Gamma=\Gamma_L=\Gamma_R$. Reprinted figure with permission
from A. Levy Yeyati {\it et al.}, Physical Review B {\bf 55}, 6137, 1997 \cite{levy97}.
Copyright (1997) by the American Physical Society.}
\label{subgap-dot-symmetric}
\end{center}
\end{figure}

\begin{figure}
\begin{center}
\includegraphics[scale=0.5]{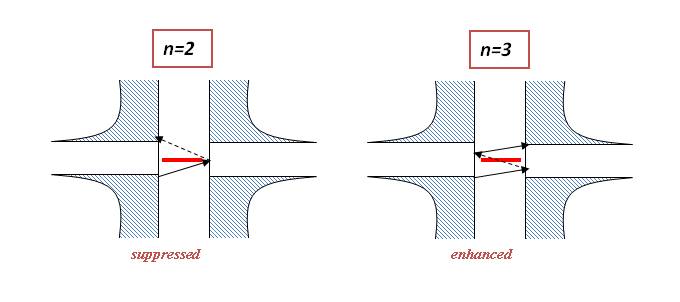}
\caption{Schematic representation of MAR processes of order $n=2$ and $n=3$. The central horizontal line represents the
level position, assumed to be located at the Fermi level.}
\label{even-odd-mar}
\end{center}
\end{figure}

The presence of a discrete resonant level between the superconducting leads can strongly modify the
I-V characteristics with respect to the quantum point contact case. This is illustrated in Fig. \ref{subgap-dot-symmetric}
which corresponds to a resonant level located at zero energy with decreasing tunneling rates to the leads.
As the figure shows, in the limit of large $\Gamma$ the I-V curves tend to that of a perfect transmitting contact. 
In the opposite limit $\Gamma \ll \Delta$ there appears a pronounced subgap structure. In contrast to the contact
case, the current jumps associated to the threshold of MAR processes appear only for the condition $V=2\Delta/n$ 
with $n$ being an odd integer, while the features at $2e\Delta/n$ with even $n$ are suppressed. This can be understood
qualitatively from the schematic pictures of Fig. \ref{even-odd-mar}. They represent the $n=2$ and $n=3$ MAR 
processes with arrows indicating propagation of electrons (full lines) or holes (broken lines). In the $n=2$ case
the MAR ``trajectory" in energy space does not cross the resonant level while in the $n=3$ case the resonant condition
is fulfilled. 

\begin{figure}
\begin{center}
\includegraphics[scale=0.25]{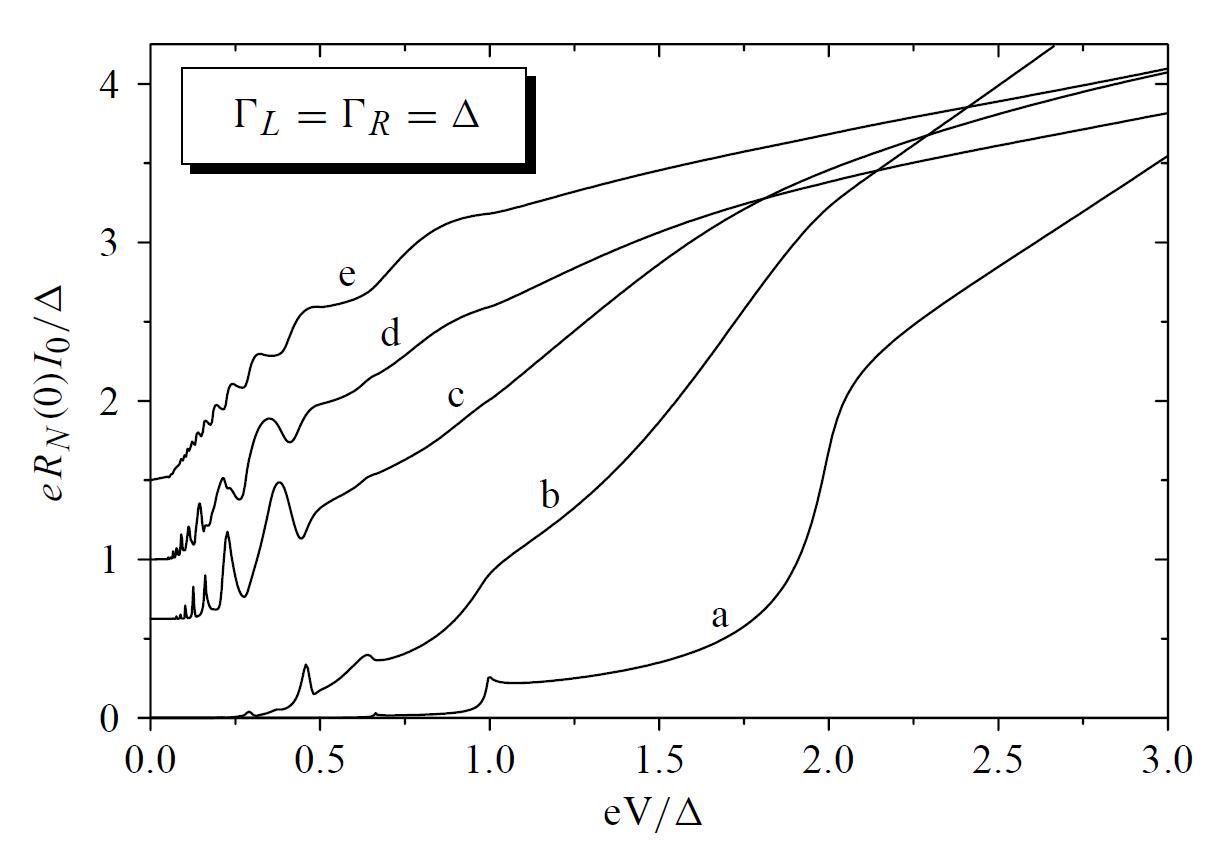}
\caption{dc I-V characteristic for the non-interacting S-QD-S system with $\Gamma_L=\Gamma_R=\Delta$ and different
dot level positions: 5 (a), 2 (b), 1 (c), 0.5 (d) and 0 (e) in units of $\Delta$. Reprinted figure 
with permission from A. Mart\'{\i}n-Rodero {\it et al.}, Superlattices and Microstructures {\bf 25}, 
925, 1999 \cite{martinrodero99}. Copyright (1999) by Elsevier.}
\label{subgap-dot-nonsymmetric}
\end{center}
\end{figure}

\begin{figure}[htb!]
\begin{center}
\includegraphics[scale=0.25]{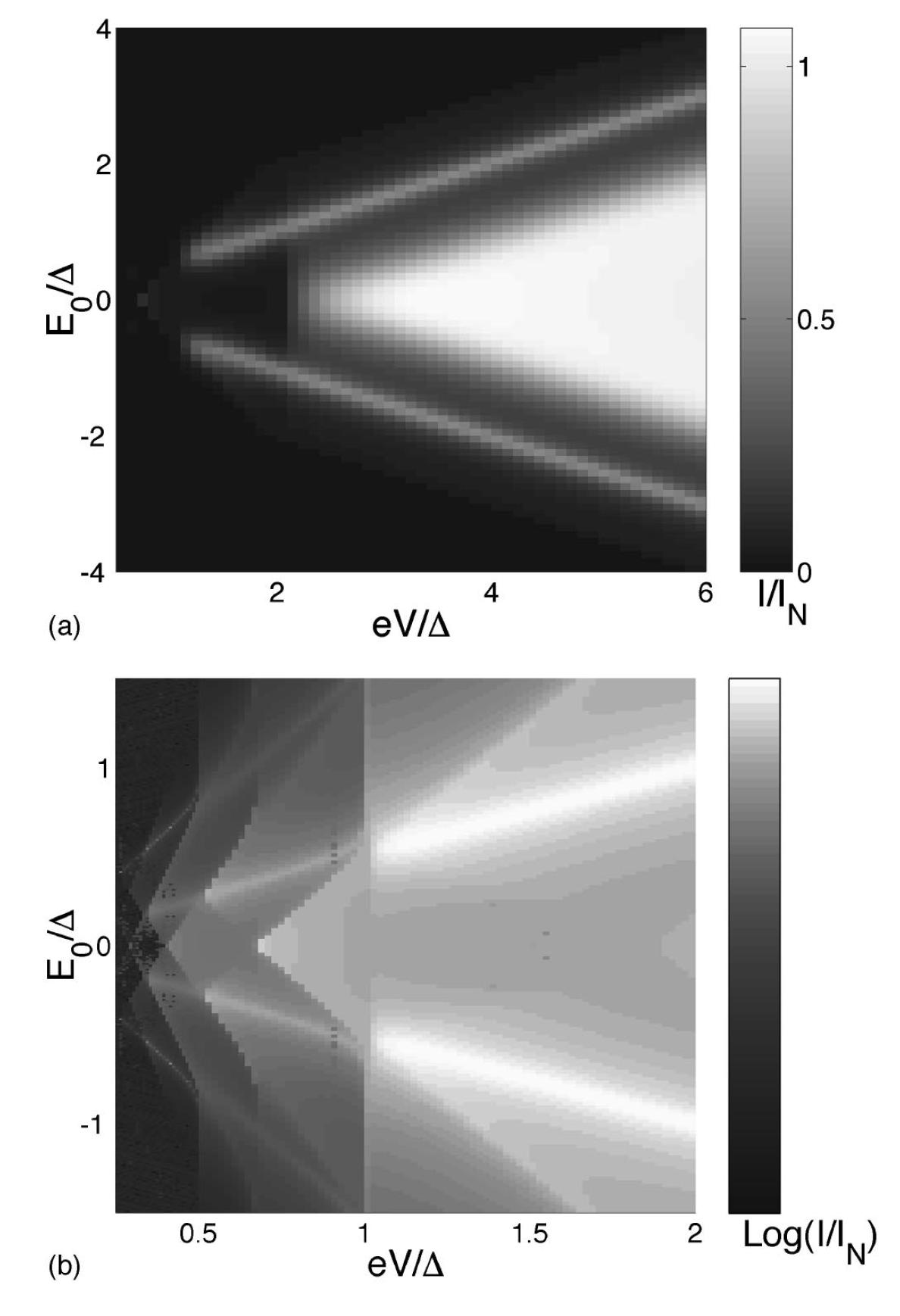}
\caption{Color scale map of the the dc current in the $\epsilon_0$-$eV$ plane for the non-interacting
S-QD-S system. The upper panel corresponds to $eV > \Delta$ and $\Gamma = 0.2 \Delta$ while the lower
panel corresponds to $eV < 2\Delta$ and $\Gamma = 0.05 \Delta$. 
Reprinted figure with permission
from G. Johansson {\it et al.}, Physical Review B {\bf 60}, 1382, 1999 \cite{joha99}.
Copyright (1999) by the American Physical Society.}
\label{subgap-shumeiko}
\end{center}
\end{figure}

\begin{figure}[htb!]
\begin{center}
\includegraphics[scale=0.25]{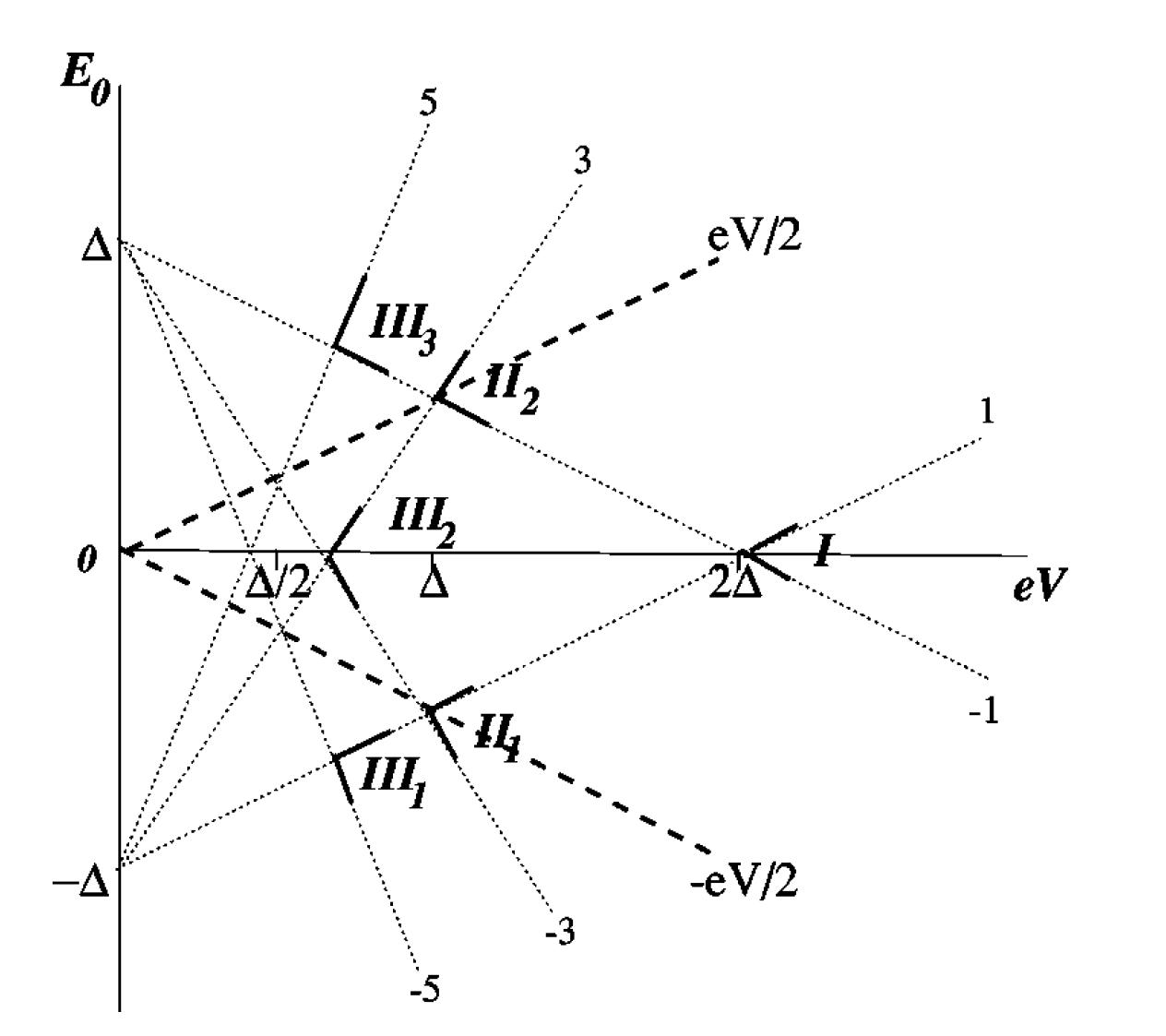}
\caption{Lines indicating the boundaries for the resonant regions of the single, double and triple 
quasiparticle currents in the $\epsilon_0$-$eV$ plane for the non-interacting S-QD-S system. 
Reprinted figure with permission
from G. Johansson {\it et al.}, Physical Review B {\bf 60}, 1382, 1999 \cite{joha99}.
Copyright (1999) by the American Physical Society.}
\label{subgap-scheme-shumeiko}
\end{center}
\end{figure}

The subgap features are quite sensitive to the level position. Fig. \ref{subgap-dot-nonsymmetric}
illustrates
the evolution of the I-V characteristics as a function of the level position $\epsilon_0$ for the case
$\Gamma_L=\Gamma_R=\Delta$. As can be observed, when the level is far from the gap region the behavior of
a weakly transmitting contact is recovered, while in the case where the level approaches the gap,
the subgap features become more pronounced and correspond to resonant conditions which depend both on 
$\Delta$ and $\epsilon_0$. In this complex situation a more clear picture of the overall behavior 
was presented in Ref. \cite{joha99}. Figs. \ref{subgap-shumeiko} show the intensity plot of the current
in the $\epsilon_0-V$ plane. The upper panel illustrates the behavior of the current for $eV > \Delta$
for $\Gamma = 0.2 \Delta$, showing clearly the onset of single quasiparticle current for $eV > 2\Delta$
at $\epsilon_0=0$. For $\epsilon_0 \neq 0$ this current is only significant in a wedge-like zone
limited by $\epsilon_0 = \pm (V-2\Delta)/2$. It can also be noticed in addition the presence of
resonant peaks at $V/2 = \pm \epsilon_0$ which are reminiscent of the resonant condition for the normal case.
The lower panel shows the intensity map in the region $eV < 2\Delta$ for $\Gamma=0.05\Delta$. This 
illustrates the onset of higher order MAR processes, which also appear to be limited into wedge-like
regions bounded by the condition $\epsilon_0 = \pm (\Delta - neV/2)$ with odd $n$. The schematic figure
\ref{subgap-scheme-shumeiko} indicates the different resonant regions for the single, double and 
triple quasi-particle currents. 

In subsequent works further analysis of the non-interacting S-QD-S case out of equilibrium
was presented \cite{sun02,jonc09}. While in Ref. \cite{sun02} the Hamiltonian approach was used
to analyze the out of equilibrium dot spectral density and the ac components of the current,
in Ref. \cite{jonc09} the effect of dephasing simulated by a third normal reservoir coupled
to the dot has been studied. This work will be further commented in Sect. \ref{multi}. 

\subsection{Effect of Coulomb interactions}
\label{SQDS-neq-int}

The inclusion of intradot interactions in the out of equilibrium S-QD-S system introduces
an additional difficulty in an already challenging theoretical problem, as shown in the previous 
section. So far the attempts have been restricted to some limiting cases which have been treated
using approximate methods. One of these special limiting situations which was first analyzed 
was the case of a quantum dot in the strong Coulomb blockade regime \cite{whan96,levy97}. These
works were motivated by the experimental results of Ref. \cite{ralph95} for transport through
small metallic nanoparticles coupled to superconducting leads. In this strong blockade regime 
multiple quasiparticle processes are suppressed and the current is basically due to single
quasiparticle tunneling. 

\begin{figure}
\begin{center}
\includegraphics[scale=0.3]{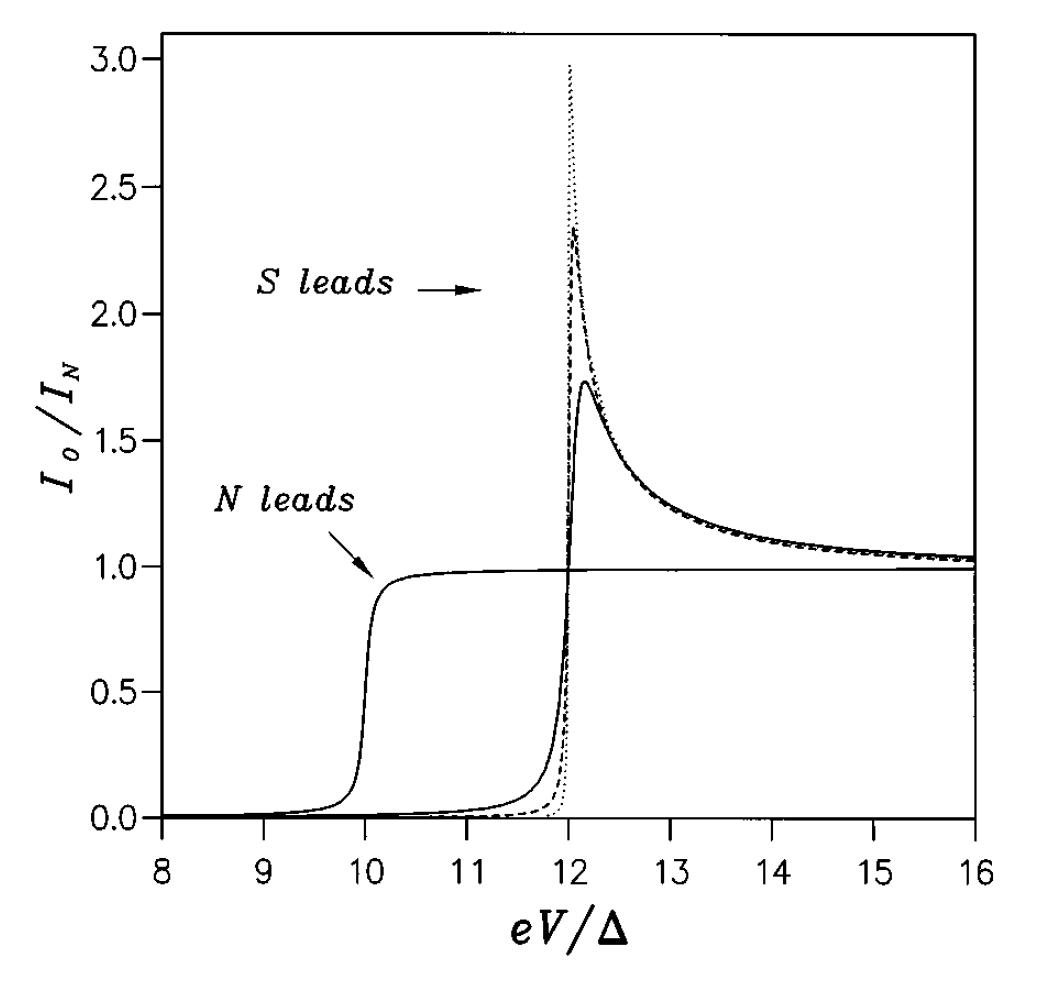}
\caption{dc current-voltage characteristic for a S-QD-S system in the strong Coulomb blockade regime
for different values of $\Gamma_L=$ $5\times 10^{-3}\Delta$ (full line), $10^{-3}\Delta$ (dashed line) and
$2\times 10^{-4}\Delta$ (dotted line), with $\epsilon=5\Delta$. Reprinted figure with permission
from A. Levy Yeyati {\it et al.}, Physical Review B {\bf 55}, 6137, 1997 \cite{levy97}.
Copyright (1997) by the American Physical Society.}
\label{fig-2-us97}
\end{center}
\end{figure}

In Ref. \cite{whan96} the current was calculated in this regime by means of a master equation
approach assuming a sequential tunneling regime. The single-particle tunneling rates were
calculated using the Fermi golden-rule. A slightly different method was used in Ref.
\cite{levy97} where resonant tunneling through an effective one-electron level describing the
dot in the limit $U \rightarrow \infty$ was considered. The corresponding expression for the
tunneling current was given by

\begin{eqnarray}
I(V) &=& \frac{4e}{h} \int d\omega \frac{\Gamma_L(\omega) \Gamma_R(\omega)}{(\omega - \epsilon)^2 +
\left[\Gamma_L(\omega)+\Gamma_R(\omega)\right]^2} \nonumber\\
&& \times\left[n_F(\omega-eV/2) - n_F(\omega+eV/2)\right] ,
\end{eqnarray}
where $\epsilon$ denotes the effective level and 

\[ \Gamma_{L,R}(\omega) = (\Gamma/2) \mbox{Re}\left[ |\omega\pm V/2|/
\sqrt{(\omega \pm V/2)^2 - \Delta^2} \right]. \]

The corresponding result for different values of $\Gamma$ are shown in
Fig. \ref{fig-2-us97}. This result differs from the simple sequential tunneling picture, which would
predict $I(V) \sim \Gamma_L(\epsilon)\Gamma_R(\epsilon)/(\Gamma_L(\epsilon)+\Gamma_R(\epsilon))$, 
exhibiting
an intrinsic broadening of the BCS-like feature, in agreement with the experimental observation \cite{ralph95}.
A similar expression was obtained in Ref. \cite{kang98} using the equation of motion approach in the
atomic limit which produces a correction factor in the current, 
$\sim \sum_{\sigma} (1 - <n_{0\sigma}>)$ due to the strong Coulomb interaction.

In order to analyze the interplay of MAR and Kondo correlations it is necessary to rely on other
approaches. In Ref. \cite{avis03} the current-voltage of this system was obtained using the slave
boson mean field approximation already discussed in Sect. \ref{SQDS-eq}. In the infinite-U version of the method
used in Ref. \cite{avis03} the problem becomes equivalent to an effective non-interacting model
with renormalized parameters $\tilde{\Gamma}$ and $\tilde{\epsilon}_0$, as indicated in subsection 
\ref{mean-field}. The
authors analyze the
evolution of the I-V curves and the shot-noise as a function of the parameter $T_K/\Delta$. As
is physically expected, the behavior is similar to that of a perfectly transparent contact when
$T_K/\Delta \gg 1$ developing a clear subgap structure in the opposite limit.

\begin{figure}
\begin{center}
\includegraphics[scale=0.3]{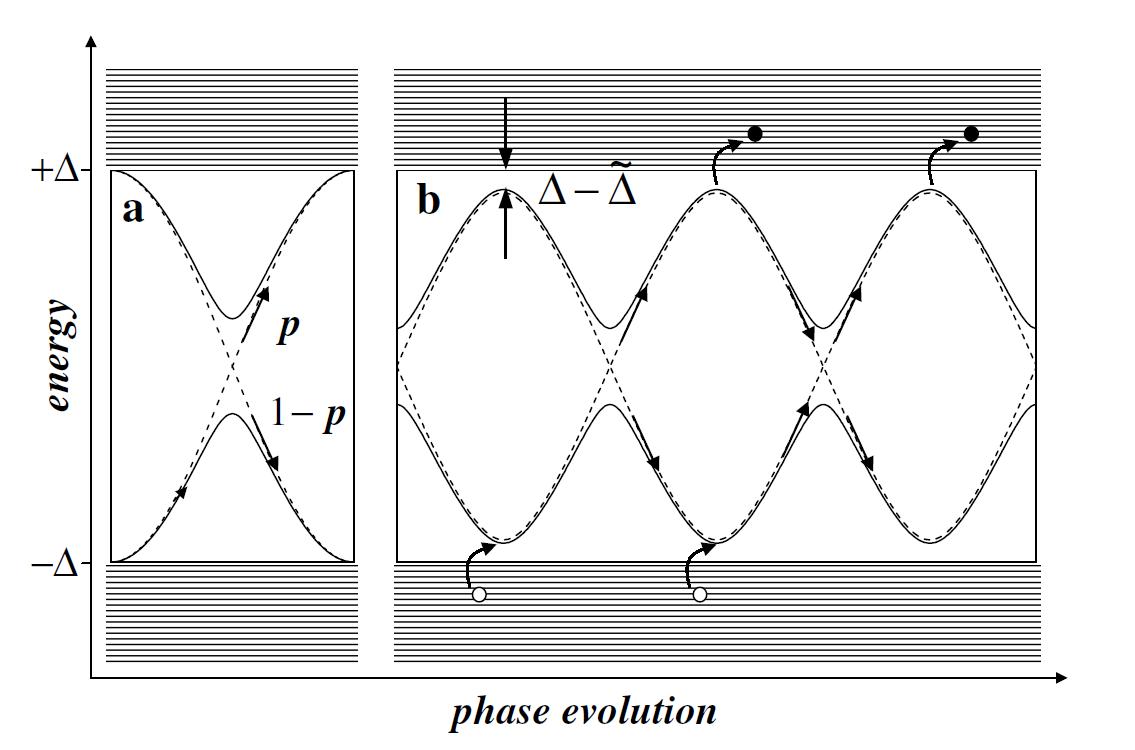}
\caption{Schematic representation of the ABS dynamics explaining the low bias dc current in the case 
of a single channel contact (panel a) and for the S-QD-S system in the Kondo regime (panel b). 
Reprinted figure with permission
from A. Levy Yeyati {\it et al.}, Physical Review Letters {\bf 91}, 266802, 2003 \cite{levy03}.
Copyright (2003) by the American Physical Society.}
\label{ABS-LZ}
\end{center}
\end{figure}

\begin{figure}
\begin{center}
\includegraphics[scale=0.25]{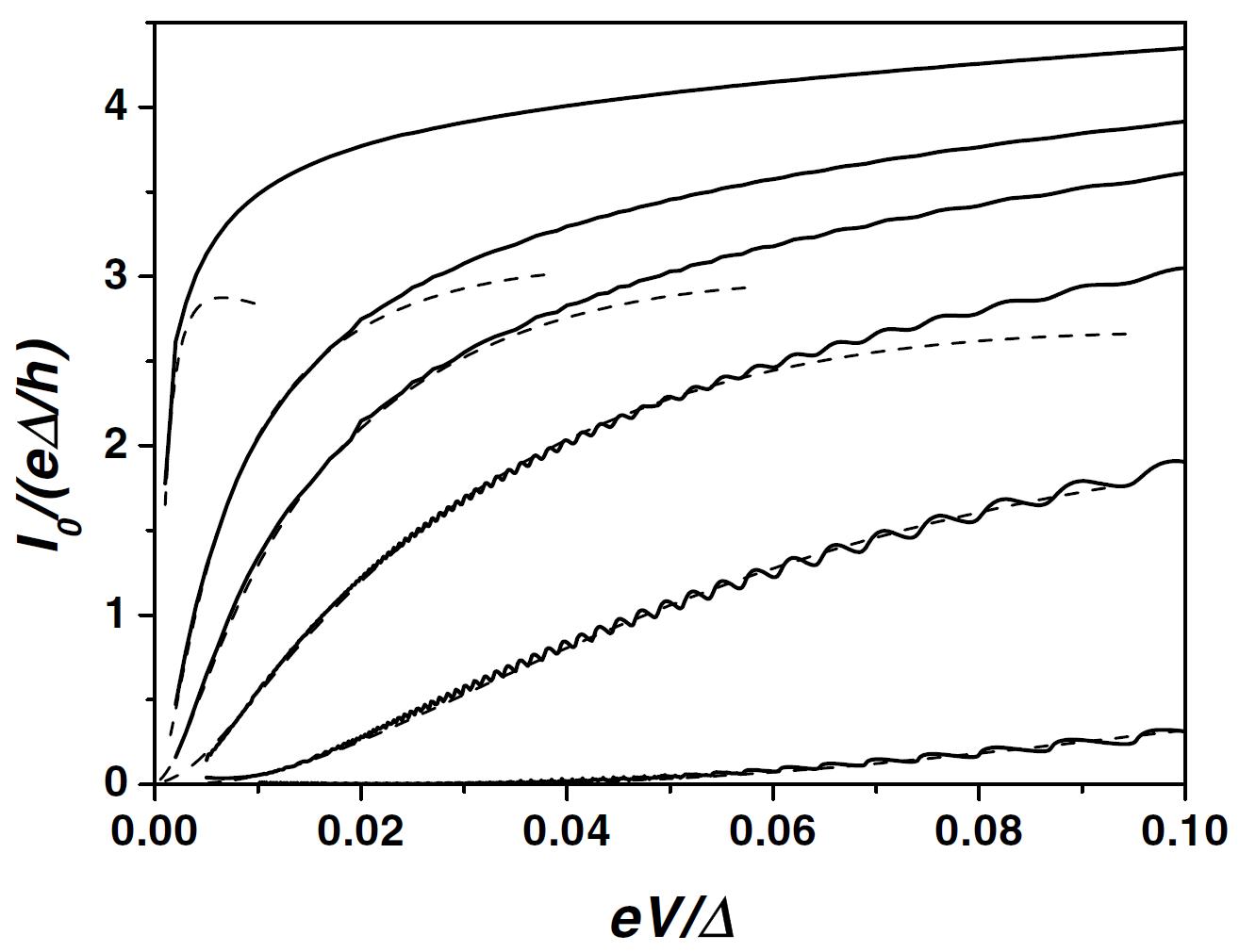}
\caption{Comparison of the low bias dc current in the S-QD-S system in the Kondo regime obtained from the 
ABS dynamics (dashed lines) and by the full numerical calculation (full line) for different values
of the effective coupling to the leads $\tilde{\Gamma}_L = \tilde{\Gamma}_R = 1, 2, 3, 4, 5$ 
and $10$ in units of $\Delta$, from bottom to top. Reprinted figure with permission
from A. Levy Yeyati {\it et al.}, Physical Review Letters {\bf 91}, 266802, 2003 \cite{levy03}.
Copyright (2003) by the American Physical Society.}
\label{low-bias-ABS-LZ}
\end{center}
\end{figure}

A similar approach was used in Ref. \cite{levy03} in order to analyze the low bias transport 
properties of a S-QD-S in the Kondo regime $T_K \gg \Delta$. It was shown that these properties
can be understood in terms of the dynamics of the subgap Andreev states. In this limit the
ABSs satisfy the equation corresponding to the non-interacting case, ie. Eq. (\ref{ABSequation}),
with renormalized parameters $\tilde{\Gamma}_{L,R}$ and $\tilde{\epsilon}_0$ instead of $\Gamma_{L,R}$
and $\epsilon_0$. The low bias quasiparticle current through the system arises from
transitions between the continuum occupied and empty states below and above the superconducting gap
which occur by means of Landau-Zener like processes involving the ABSs. This is illustrated
in Fig. \ref{ABS-LZ} first for the case of a quantum point contact (left panel) and then
for the S-QD-S case (right panel). In the last case it is necessary to have a transition between
the ABSs and the continuum in addition to the Landau-Zener transition between the lower and upper ABSs. 
The results for the low bias dc current which are obtain from this analysis, shown in Fig. 
\ref{low-bias-ABS-LZ}, are in good agreement with the results of a full numerical calculation
including MAR processes up to infinite order.
The analysis based on the dynamics of the ABSs was also used in Ref. \cite{veci04} for a comparison
of experimental results in the low bias regime. In contrast to Ref. \cite{levy03}, in this work
a phenomenological damping rate $\eta$ was introduced in order to fit the experimental data 
obtained for multiwall CNTs connected to Au/Al leads.
                
\begin{figure}
\begin{center}
\includegraphics[scale=0.25]{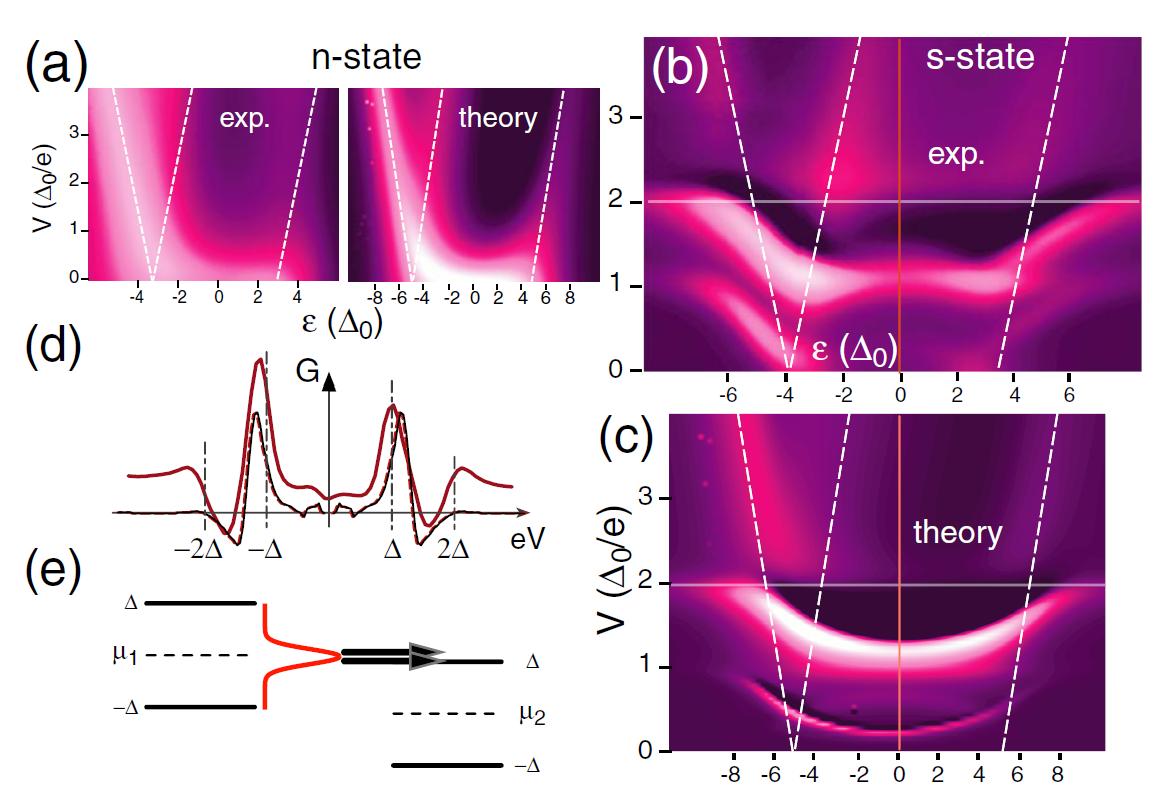}
\caption{Experimental and theoretical results for the differential conductance of the
S-QD-S system in an odd valley. Panel (a) corresponds to the normal state (experimental and
theoretical results), while (b) and (c) show the corresponding results in the superconducting state.
Panel (d) shows the conductance vs voltage bias along the line indicated in red in panels (b) and (c).
The theoretical curves corresponds to $\Gamma_L=\Delta$, $\Gamma_R=0.03\Delta$ and $U=10\Delta$.
Reprinted figure with permission
from A. Eichler {\it et al.}, Physical Review Letters {\bf 99}, 126602, 2007 \cite{eich07}.
Copyright (2007) by the American Physical Society.}
\label{eichler}
\end{center}
\end{figure}

A step beyond the infinite-$U$ SBMF approach was taken in Ref. \cite{eich07} where the finite-U
SBMF method was used to determine the I-V characteristics of a S-QD-S system. This approach
allowed to describe the observed differences in the subgap structure between situations with
even and odd number of electrons in the dot for a SWCNT QD coupled to Al/Ti electrodes.
In Ref. \cite{eich07} the effective parameters of the finite-U SBMF approach were obtained
for the leads in the normal state and assumed to remain unmodified in the superconducting case.
The results of this work for the subgap structure are shown in Fig. \ref{eichler} where the 
comparison with the experimental data is given. The figure corresponds to an {\it odd} valley
exhibiting clear Kondo features in the normal state. The more intriguing feature of the 
differential conductance in the superconducting case was the presence of a pronounced structure
for $eV \sim \Delta$, which cannot be explained by the non-interacting theory. This feature
was attributed to the large asymmetry $\Gamma_L/\Gamma_R \sim 30$ which produces a Kondo 
resonance pinned at the chemical potential of the left lead. As is schematically depicted in
Fig. \ref{eichler}(e) this resonance would produce and enhancement of the current for
$eV \sim \Delta$. 

\begin{figure}
\begin{center}
\includegraphics[scale=0.2]{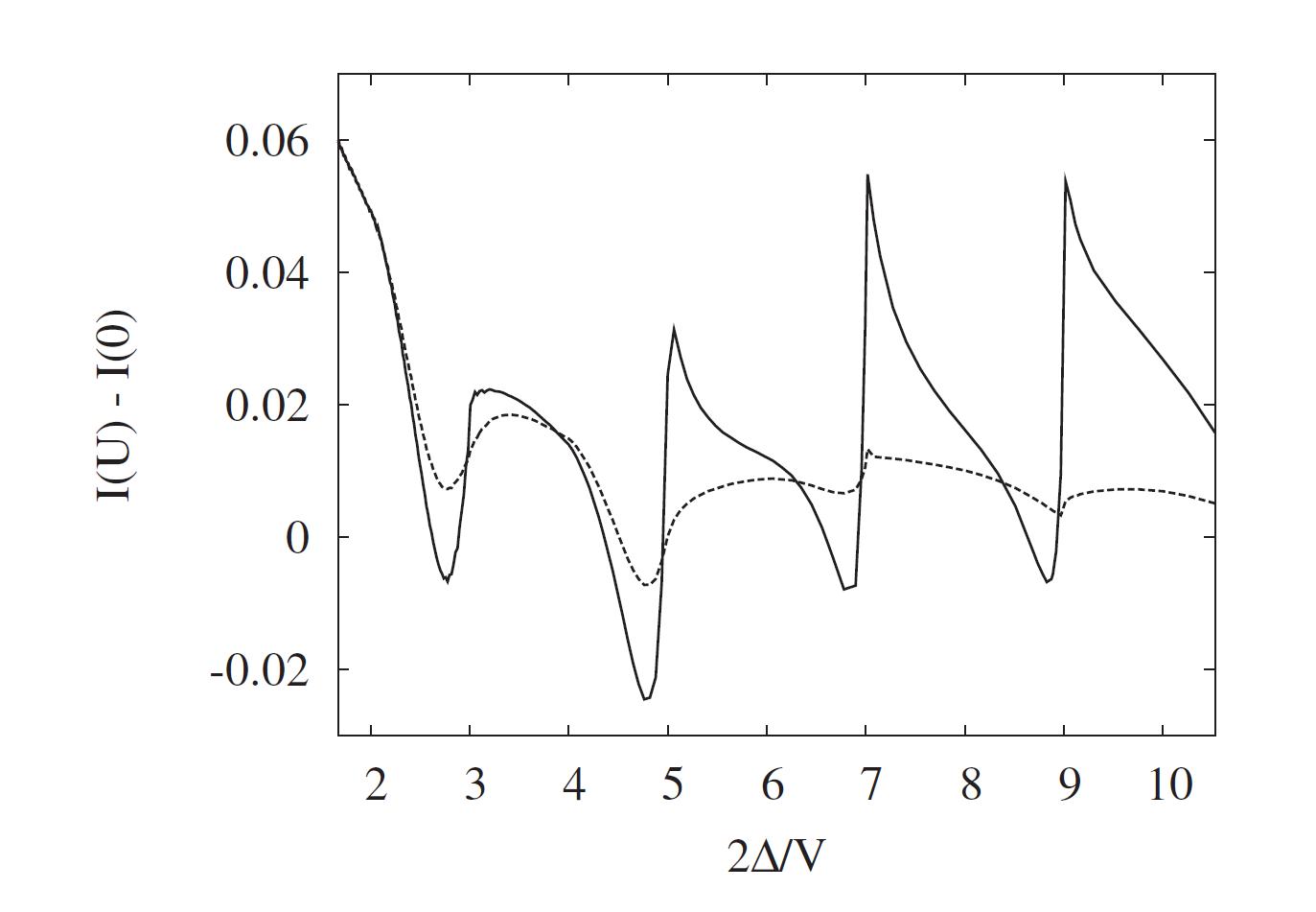}
\caption{Variation of the dc current in the symmetric S-QD-S system with respect to the non-interacting case
obtained in Ref. \cite{dell08} using the second-order self-energy approach for 
$U/\Gamma=\Gamma/\Delta=0.5$. The dashed curve gives the 
self-consistent first order result. Reprinted figure with permission
from L. Dell'Anna {\it et al.}, Physical Review B {\bf 77}, 104525, 2008 \cite{dell08}.
Copyright (2008) by the American Physical Society.}
\label{dell-fig3}
\end{center}
\end{figure}

The effect of interactions in the subgap structure has also been considered in Ref. \cite{dell08}
by means of a perturbative approach in which the dot self-energy was calculated up to second
order in $U$. The method thus includes the diagrams already discussed in Sect. \ref{SQDS-eq} but extended
to the non-equilibrium situation. To avoid heavily time-consuming computation of the  multiple frequency integrals, 
the authors calculate the diagrams in time representation and then Fourier transform the final result. 
They consider the weak interaction regime $U/\Gamma < 1$ thus avoiding the regime of
$\pi$-junction behavior. The most remarkable result is the observation of an enhancement of
the current due to the interactions, which is more pronounced for voltages approaching the odd
MAR onset conditions $2\Delta/(2r+1)$. This is illustrated in Fig. \ref{dell-fig3} for the
case $\epsilon_0 = 0$ where the difference $I(U)-I(0)$ as a function of $2\Delta/V$ is
represented. The enhancement is observed both in the self-consistent first order approximation
and when including the second-order diagrams. This current enhancement is reminiscent yet different
from the ``antiblockade" behavior due to dynamical Coulomb blockade effects on MAR transport, 
as discussed in Ref. \cite{us2005}.
It is worth mentioning that MAR transport through a resonant level coupled
to a localized phonon mode was studied in Ref. \cite{zazunov06} using second order perturbation
theory in the electron-phonon coupling.

\subsection{Summary of Experimental results}
\label{SQDS-neq-exp}

\begin{figure}
\begin{center}
\includegraphics[scale=0.4]{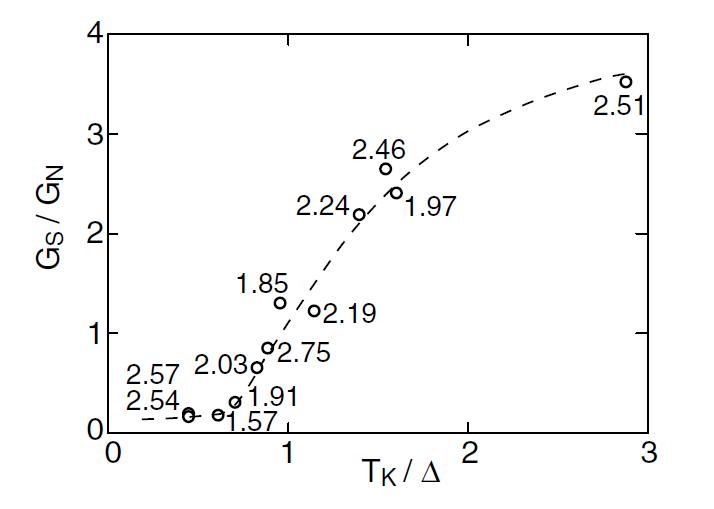}
\caption{Linear conductance in the Kondo regime for the experimental realization of the S-QD-S system of 
Ref. \cite{buit02} normalized to its value in normal state as a function of $T_K/\Delta$.
Reprinted figure with permission
from M.R. Buitelaar {\it et al.}, Physical Review Letters {\bf 89}, 256801, 2002 \cite{buit02}.
Copyright (2002) by the American Physical Society.}
\label{buit02-fig-4}
\end{center}
\end{figure}

\begin{figure}
\begin{center}
\includegraphics[scale=0.23]{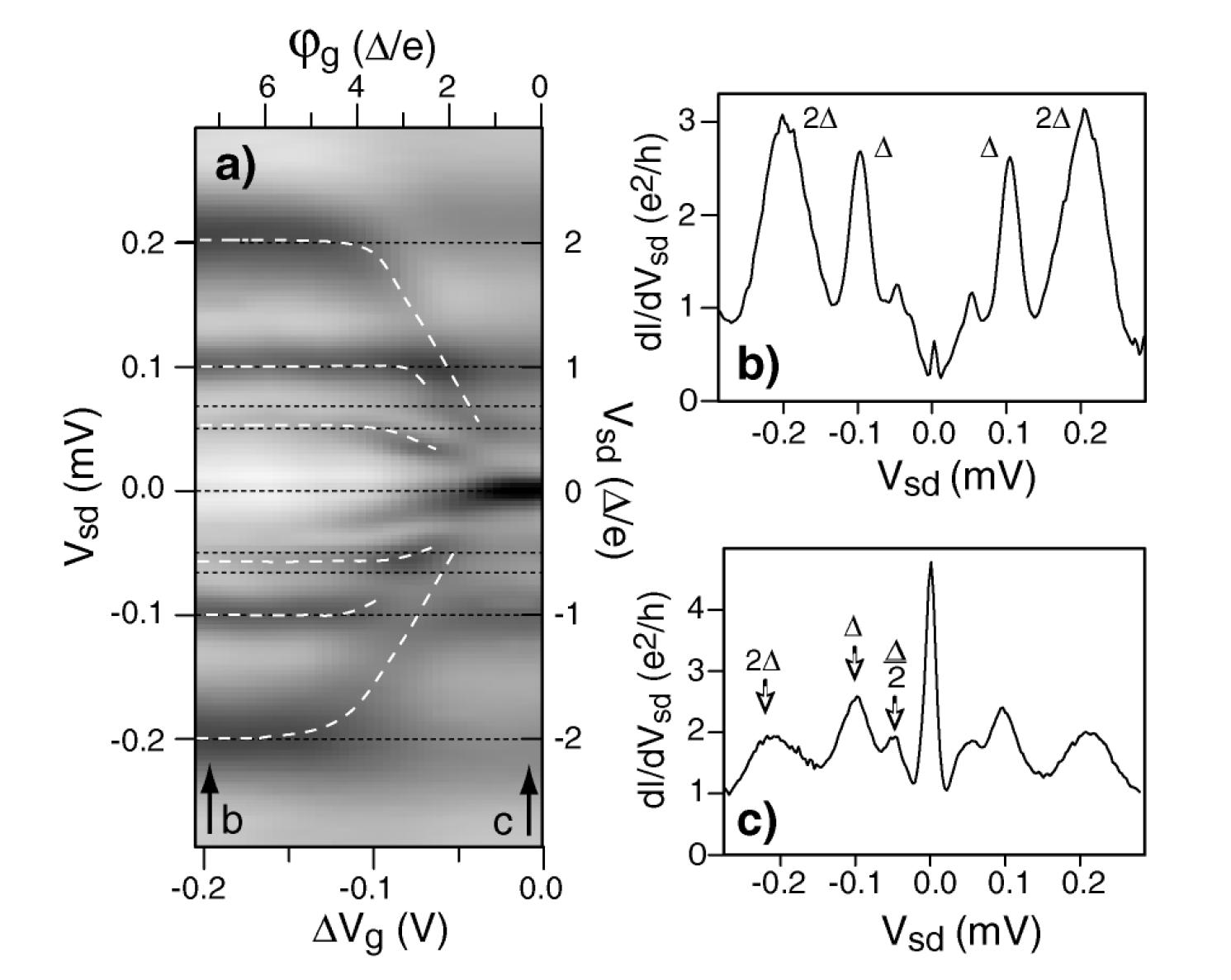}
\caption{Results for the differential conductance in the experimental realization of the
S-QD-S system of Ref. \cite{buit03}. Panel (a) color map for this quantity in the bias voltage-gate voltage plane.
The dashed lines indicate the evolution of the MAR resonances. Panels (b) and (c) show plot of the conductance
as a function of bias for the lines indicated by the arrows in panel (a). Reprinted figure 
with permission
from M.R. Buitelaar {\it et al.}, Physical Review Letters {\bf 91}, 057005, 2003 \cite{buit02}.
Copyright (2003) by the American Physical Society.}
\label{buit03-fig-3}
\end{center}
\end{figure}

The already commented work by Ralph et al. \cite{ralph95} can be considered one of the first
realizations of a S-QD-S system in which the current-voltage characteristic was measured.
This case corresponded, however, to the strong blockade regime in which the subgap structure
is absent. It was not until 2002 that experiments on CNTs coupled to Al leads \cite{buit02,buit03}
allowed a clear observation of the subgap features. Ref. \cite{buit02} mainly focused in the
linear conductance which can exhibit either an enhancement or a suppression with 
respect to the normal case depending on the ratio $T_K/\Delta$. The results of this work are 
summarized in Fig. \ref{buit02-fig-4}. On the other hand in Ref. \cite{buit03} the authors 
analyzed the MAR induced subgap structure for the same type of systems in more detail. 
As shown in Fig. \ref{buit03-fig-3} clear peaks in the differential conductance are observed
at the positions $eV \sim 2\Delta$, $\Delta$ and $\Delta/2$. The intensity of these peaks
evolves as a function of the dot level position (controlled by the gate voltage $V_g$). 
Contrary to the theoretical expectations the peak at $\Delta$ is still visible at
resonance whereas the expected feature from the non-interacting model at $2\Delta/3$ 
is not observed. The authors of Ref. \cite{buit03} suggest that the discrepancy can
be attributed to the effect of interactions not included in their theoretical analysis.
These results could be analyzed in the light of the already commented arguments of
Ref. \cite{eich07} which attributed the pronounced $\Delta$ peak to the combined
effect of coupling asymmetry and Kondo effect.    

The competition between Kondo effect and superconductivity was also observed in 
Ref. \cite{buiz07} in which self-assembled InAs quantum dots contacted with Al leads
were analyzed. A magnetic field was used to control the size of the superconducting
gap parameter and the linear conductance as a function of $\Delta/k_B T_K$ was measured.
The results exhibited a rather universal behavior as a function of this parameter.
However, in contrast to Ref. \cite{buit02} the ratio $G_S/G_N$ did not exceed unity for
$\Delta/k_B T_K <1$ while decreasing as expected for $\Delta/k_B T_K >1$.
Although the authors attributed this difference to a stronger Coulomb repulsion 
which would in their case heavily damp the MAR processes, one would expect that
this effect would be already included when scaling $\Delta$ in units of $T_K$. 
The absence of conductance enhancement for large $T_K$ could be also pointing out
to an ingredient in this system not included in the simplest Anderson model like
spin-orbit interactions.

\section{Beyond the single level model: multidot, multilevel and multiterminal systems} 
\label{multi}

In recent years there has been an increasing interest in more complex situations
which cannot be described by the simplest single level Anderson model. These 
include situations where transport occurs through more than a single dot or
where several quantum channels in a single dot are involved. In addition, 
there is also great interest in analyzing the transport properties of 
hybrid quantum dot systems coupled to several superconducting and/or normal
electrodes in a multiterminal configuration. These configurations could allow
to explore non-local electronic transport, in particular the possibility of 
creating entangled electron pairs by means of crossed or non-local Andreev
processes. To describe these
developments we organize this section as follows: in the first subsection
we discuss the case of Josephson transport through double and multiple dot systems,
in the second one we consider this effect for a multilevel dot,
and finally we consider multiple dot systems including both normal and
superconducting electrodes as well as in a multiterminal configuration.

\subsection{Josephson effect through multidot systems}
\label{multi-josephson}

\begin{figure}
\begin{center}
\includegraphics[scale=0.4]{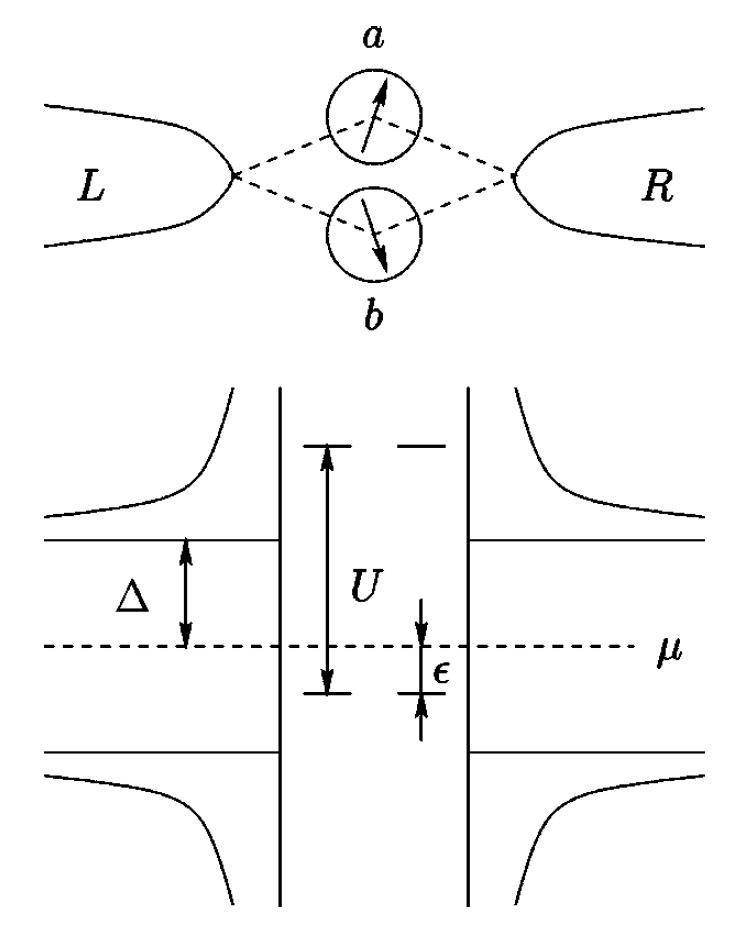}
\caption{Schematic representation of a double dot system coupled in parallel to two
superconducting electrodes (upper panel) considered in Ref. \cite{choi00}. The lower panel describe the leads spectral density and
the parameters of the double dot Anderson model used in this reference. 
Reprinted figure with permission
from M.S. Choi {\it et al.}, Physical Review B {\bf 62}, 13569, 2000 \cite{choi00}.
Copyright (2000) by the American Physical Society.}
\label{choi00-fig1}
\end{center}
\end{figure}

Transport through double quantum dots connected either in parallel or in
series to superconducting electrodes has been extensively analyzed in the
literature. Most of the theoretical works describe this situation by
using a single-level Anderson model to represent each dot and
introducing extra terms describing the coupling to the leads. 
Choi et al. considered in Ref. \cite{choi00} the case of
two dots connected in parallel to superconducting leads 
as depicted in Fig. \ref{choi00-fig1}. By analyzing
the problem to the fourth order in the tunneling to the leads they 
derived an effective Hamiltonian coupling the localized spins in both
dots. In the regime $0<-\epsilon_0 << \Delta << U$ this Hamiltonian
adopts the form

\begin{equation}
H_{eff} \simeq J \left( 1 + \cos{\varphi} \right) \left[ {\bf S}_a.{\bf S}_b - \frac{1}{4}\right],
\end{equation}
where $J$ is an exchange coupling between the localized spins in dots $a$ and $b$; and
$\varphi$ is defined as

\begin{equation}
\varphi = \phi_L - \phi_R - \frac{\pi}{\Phi_0} \int \left(dl_a+dl_b\right).{\bf A} ,
\end{equation}
where the last term corresponds to the phase accumulated on each path of the loop
due to the magnetic field.
These results indicate that the Josephson current through such a device would be sensitive
to the total spin of the double dot. In order to probe the spin state of the double dot 
system the authors propose to incorporate it into a SQUID geometry in which an additional
tunnel junction is included in one of its arms. Further elaboration on similar ideas were
presented in Ref. \cite{hur06} for a triple dot between superconducting leads. This system
is shown to behave under certain conditions as a {\it mesoscopic pendulum} where the
singlets injected through a pair of dots oscillate between two different configurations
like in the resonating valence bond model.

\begin{figure}
\begin{center}
\includegraphics[scale=0.25]{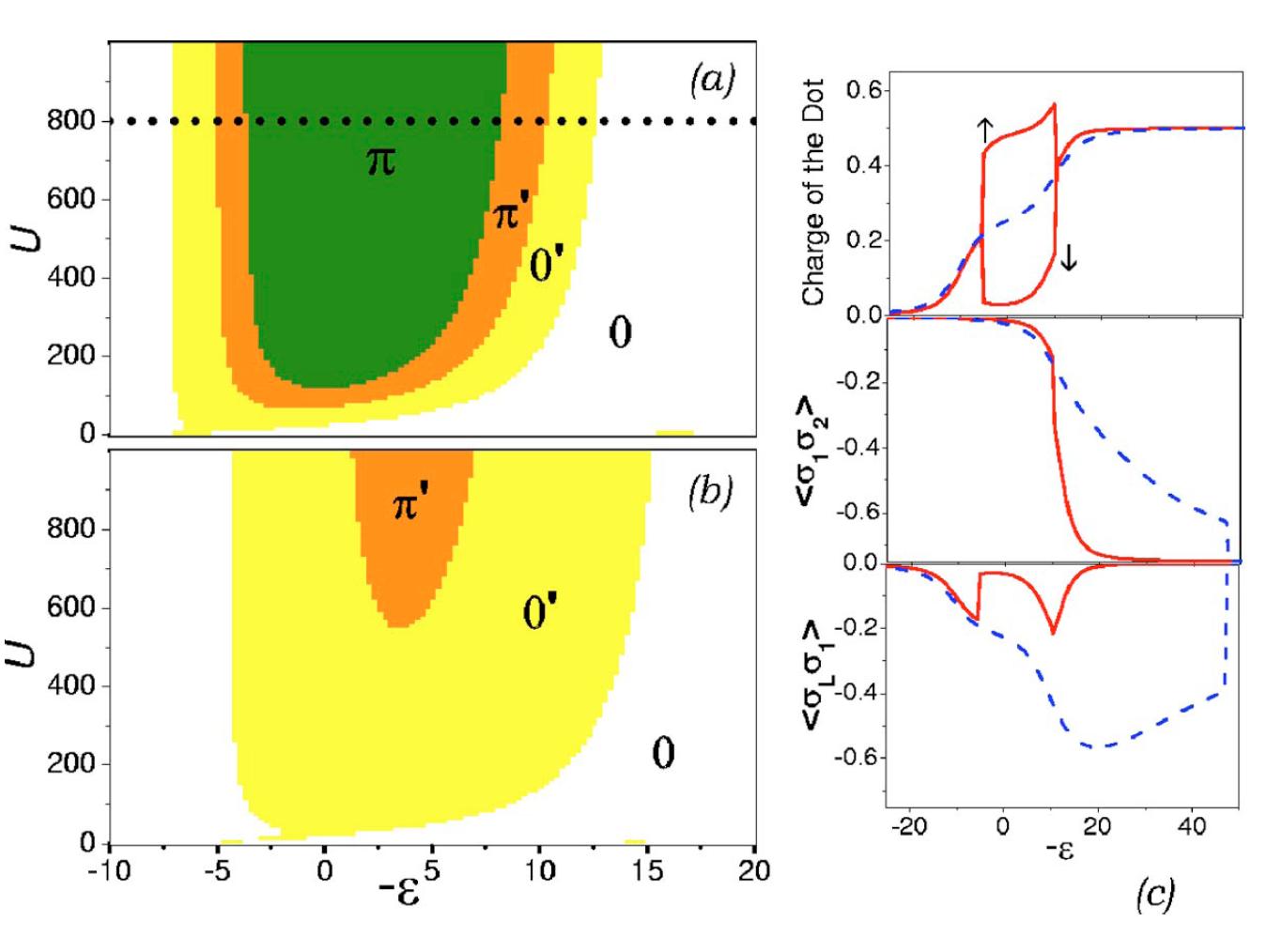}
\caption{$U-\epsilon$ phase diagrams, dot population and spin correlation functions for the series double dot
model considered in Ref. \cite{berg06} obtained using the zero band-width approximation to describe the 
superconducting leads. The parameters are $t_{12}=10\Delta$, $t_L=t_R=2\Delta$ (a) and $t_L=t_R=2.5\Delta$ (b).
The plots in panel (c) are taken along the dotted line corresponding to $U=800\Delta$ shown in panel (a). Reprinted figure with permission
from F.S. Bergeret {\it et al.}, Physical Review B {\bf 74}, 132505, 2006 \cite{berg06}.
Copyright (2006) by the American Physical Society.}
\label{bergeret-06-fig2}
\end{center}
\end{figure}

\begin{figure}
\begin{center}
\includegraphics[scale=0.25]{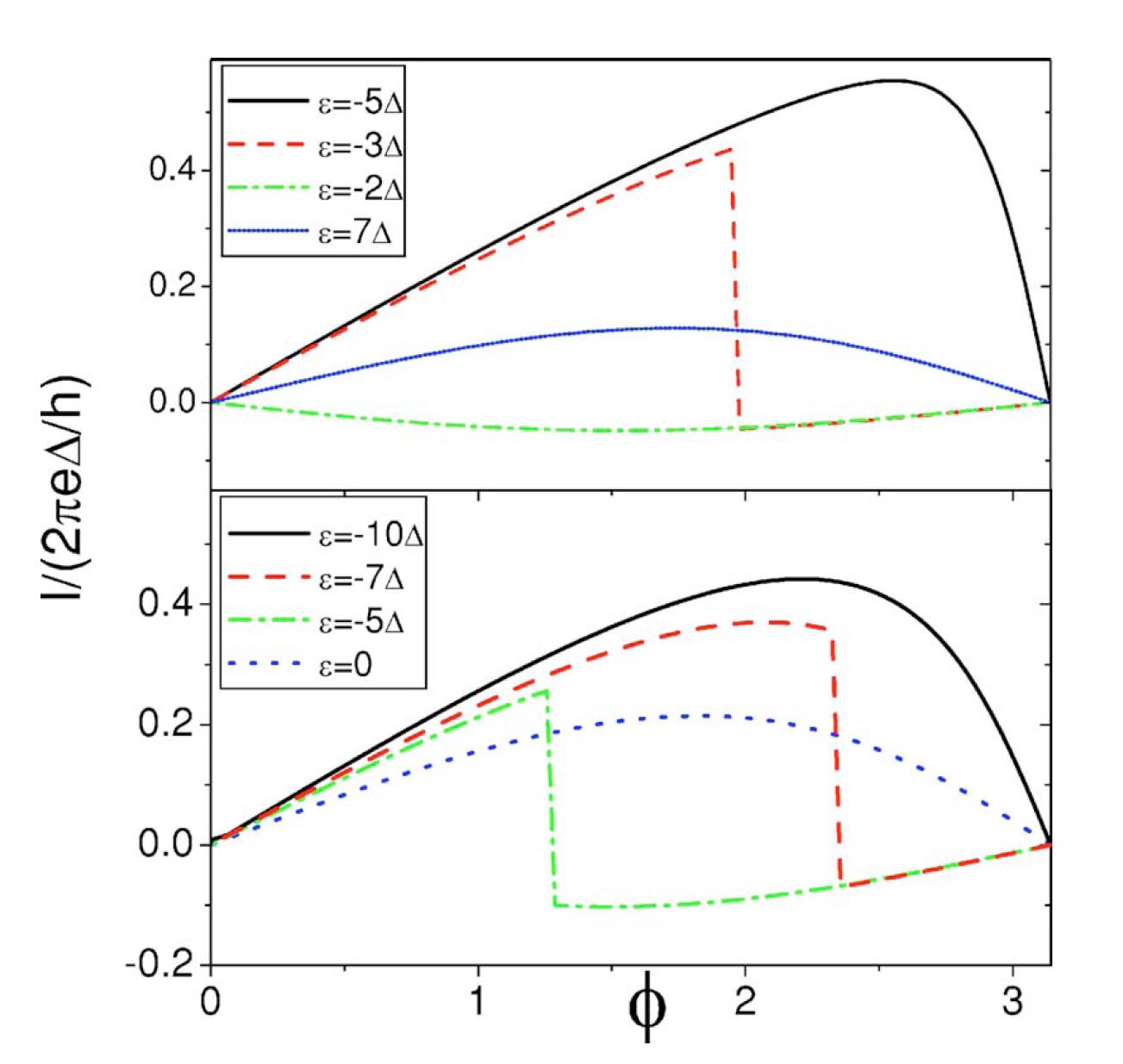}
\caption{Current-phase relation for the model considered in Ref. \cite{berg06} obtained using the finite-U SBMF method
with parameters $U=800\Delta$, $t_{12}=10\Delta$, $\Gamma_L=\Gamma_R=2.25\Delta$ (upper panel) and $\Gamma_L=\Gamma_R=4\Delta$ (lower panel). Reprinted figure with permission
from F.S. Bergeret {\it et al.}, Physical Review B {\bf 74}, 132505, 2006 \cite{berg06}.
Copyright (2006) by the American Physical Society.}
\label{bergeret-06-fig3}
\end{center}
\end{figure}

The case of two QDs in series connected to superconducting leads was first analyzed
in Ref. \cite{zhu02}. The authors start by diagonalizing exactly the isolated double dot Hamiltonian
including a term describing the interdot repulsion $V$ and then introducing the external
leads by a Dyson-type equation. This approach should be valid in the limit of vanishing 
coupling to the leads and is equivalent to the one commented in Sect. \ref{SQDS-eq} for the single dot
case. For the case $U > V > t$, where $t$ is the interdot hopping parameter, and for $k_BT \gg
\Gamma$ they find that the system exhibits a 0 type current-phase relation except when a finite Zeeman splitting
is included. The absence of a $\pi$-junction behavior for any dot filling is probably due to the
extremely small value of $\Gamma$ compared to $k_BT$. This will be further discussed below.

A similar situation was later considered in Ref. \cite{berg06}, where a double dot model
was analyzed using both small cluster numerical diagonalizations (discussed before in Subsect.
\ref{diagonalization} for the single dot case) together with the finite-U SBMF technique. In addition to
the transition to the $\pi$-phase this work aimed to investigate the interplay between different
possible magnetic correlations including Kondo and anti-ferromagnetic coupling between the
localized spins within each dot. Part of the results are shown in Fig. \ref{bergeret-06-fig2}
where the phase-diagram in the $U$ vs $-\epsilon$ plane is given for the case $t_{12} \gg \Delta$,
$t_{12}$ being the interdot hopping parameter.
In the range of parameters of Fig. \ref{bergeret-06-fig2} the system exhibits a $\pi$-phase
region associated to the transition between the empty and singly occupied double dot. 
When increasing the dot population to the level of one electron per dot antiferromagnetic
correlations dominate and the $\pi$-phase is no longer stable. This is further illustrated
in panel (c) of Fig. \ref{bergeret-06-fig2} where different spin-spin correlation functions
are shown both for the normal and the superconducting case. The results obtained using the
small cluster diagonalizations were confirmed by the finite-U SBMF calculations. In particular
these last calculations also show the gradual disappearance of the full $\pi$-phase when 
increasing the coupling to the leads, as illustrated in Fig. \ref{bergeret-06-fig3}.
This work also provided a possible scenario for explaining the experimental results of
Ref. \cite{kasu05} for fullerene dimers containing Gd magnetic atoms and suggested the
possibility to control the magnetic configuration of these atoms by means of the Josephson
current.

\begin{figure}
\begin{center}
\includegraphics[scale=0.2]{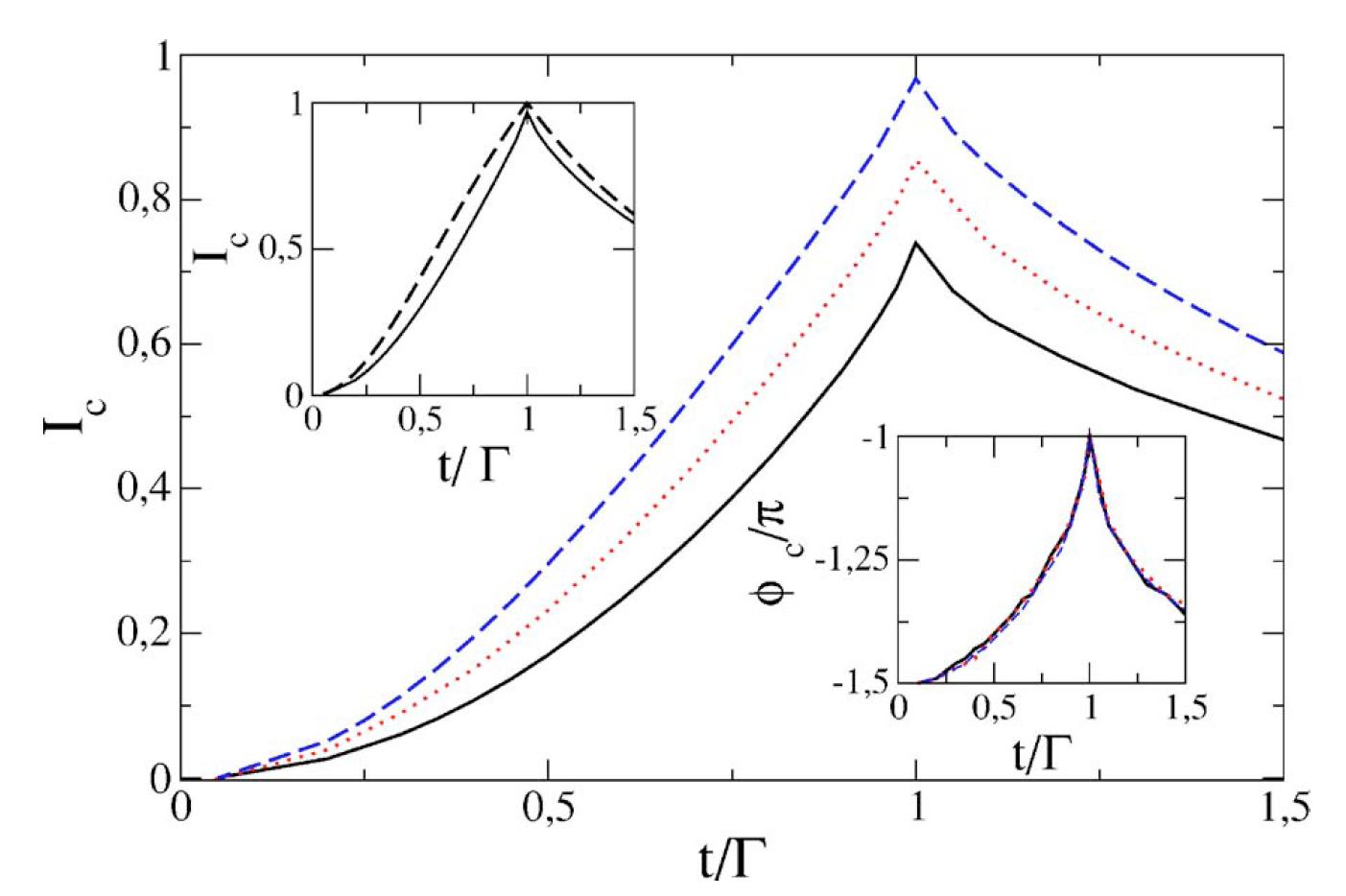}
\caption{Josephson critical current as a function of $t/\Gamma$ for the series double dot
model analyzed in Ref. \cite{lope07} using the infinite-U SBMF method for $\Delta/T_K = 0.1,
0.25$ and 0.5 (from bottom to top). The upper inset shows the comparison of the numerical
result for $\Delta/T_K=0.1$ (full line) with the prediction from a non-interacting model
with an effective transmission (dashed line). The lower inset shows the critical phase
at which the maximum current occurs. Reprinted figure with permission
from R. L\'opez {\it et al.}, Physical Review B {\bf 75}, 045132, 2007 \cite{lope07}.
Copyright (2007) by the American Physical Society.}
\label{lopez-07-fig4}
\end{center}
\end{figure}

The SBMF approach in the infinite-U version was applied to series and parallel double
quantum dots in Ref. \cite{lope07}. As in the case of a single QD this method cannot account
for the appearance of a $\pi$-junction phase. The results are nevertheless relevant for
the regime $T_K \gg \Delta$ where Kondo correlations dominate over pairing. While in the parallel 
case it is found that the Josephson critical current, $I_c$, decreases monotonically with the 
interdot hopping parameter $t$, in the series case a non-monotonous behavior is found.
This is illustrated in Fig. \ref{lopez-07-fig4} where $I_c$ exhibits a maximum 
at $t/\Gamma \sim 1$. The authors interpret the change in behavior of $I_c$ as a transition from
a regime characterized by two independent Kondo singlets involving each dot and the corresponding
lead to the formation of bonding and antibonding Kondo resonances. 

\begin{figure}
\begin{center}
\includegraphics[scale=0.2]{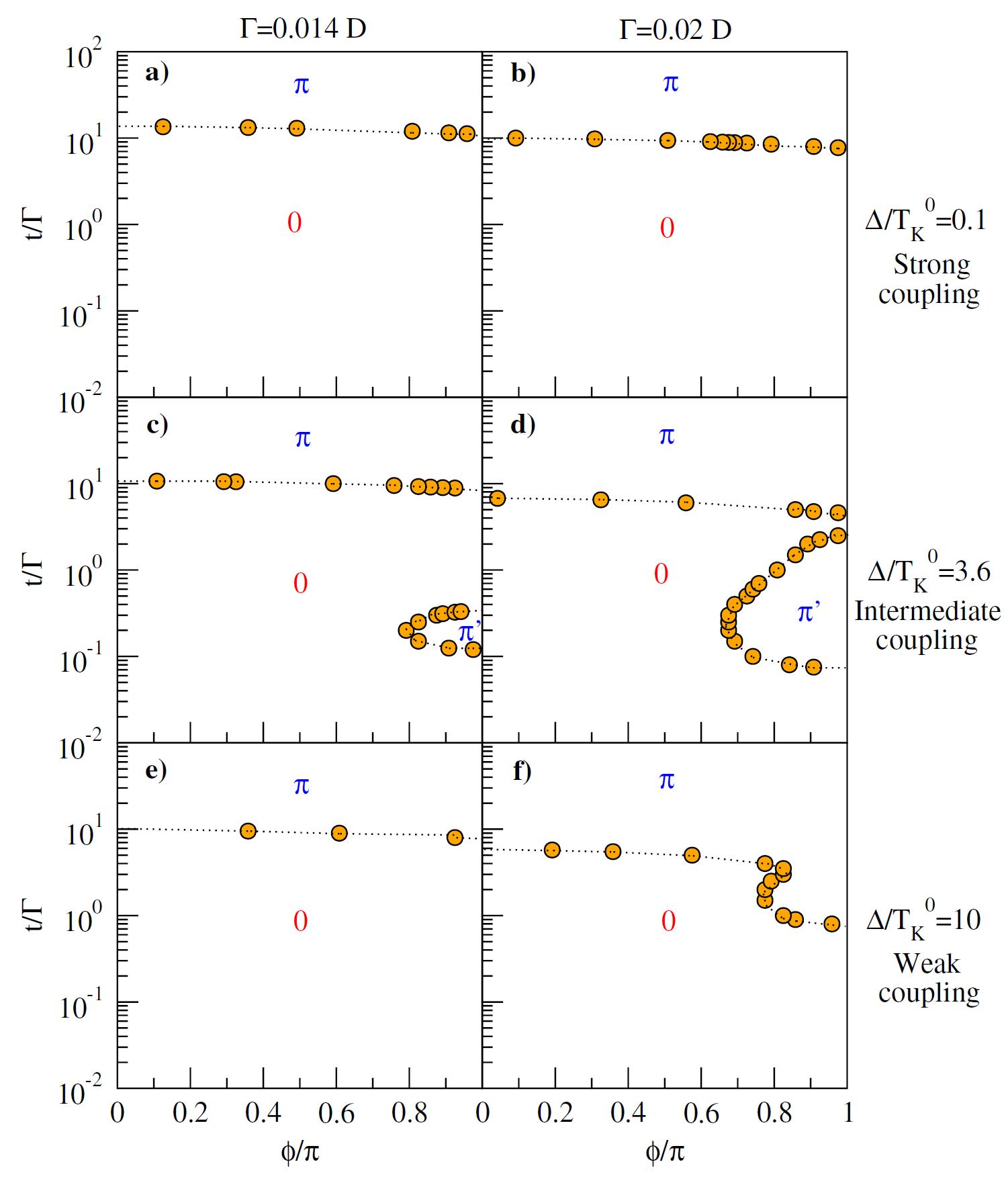}
\caption{Phase boundaries between the 0 and $\pi$ states in the series double dot model
of Ref. \cite{zitko10} obtained using the NRG method for $\Gamma = 0.014 D$ and
$\Gamma = 0.02 D$, where $D$ is the leads band-width. Reprinted figure with permission
from R. Zitko {\it et al.}, Physical Review Letters {\bf 105}, 116803, 2010 \cite{zitko10}.
Copyright (2010) by the American Physical Society.} 
\label{zitko-10-fig2}
\end{center}
\end{figure}

Quite recently, the series double dot system coupled to SC leads has been analyzed using the
NRG method \cite{zitko10}. The authors considered the regime $U \rightarrow \infty$ and
$-\epsilon \gg \Gamma$, which would correspond to the deep Kondo regime in a normal single QD.
In this range of parameters they find a rich phase diagram as a function of the interdot
coupling $t$ and the ratio $T_K/\Delta$. Some of their results are illustrated in Fig. \ref{zitko-10-fig2}
showing the regions corresponding to $0$, $\pi$ and $\pi'$ phases in the $t/\Gamma$ vs $\phi$
plane for different values of $T_K/\Delta$. One can notice the abrupt transition between 
$0$ and $\pi$ phases for $t \sim 10\Gamma$ which can be associated to a change in the DQD
population from an even to an odd number of electrons. An additional remarkable feature is
the appearance of a $\pi'$ "island" close to $\phi = \pi$ and for $t \sim \Gamma$ in the
intermediate coupling regime $T_K \sim \Delta$.

\begin{figure}
\begin{center}
\includegraphics[scale=0.3]{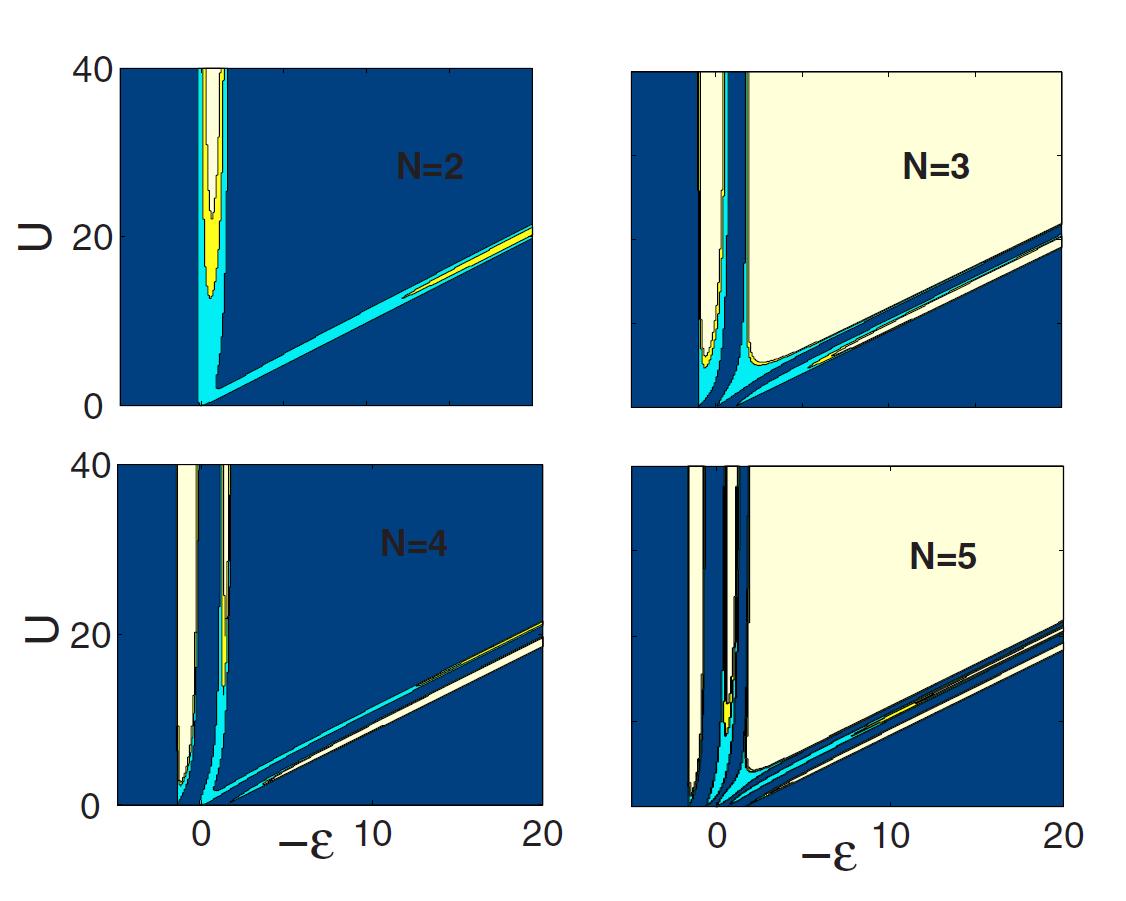}
\caption{$\epsilon-U$ phase diagrams for the quantum dot array coupled to superconductors
considered in Ref. \cite{berg07} with increasing number of dots $N=2,3,4$ and 5. The results
were obtained using the zero band-width approximation described in Subsect. \ref{diagonalization}
with parameters $t_L=t_R=t=\Delta$. Reprinted figure with permission
from F.S. Bergeret {\it et al.}, Physical Review B {\bf 76}, 174510, 2007 \cite{berg07}.
Copyright (2007) by the American Physical Society.}
\label{bergeret-07-fig2}
\end{center}
\end{figure}

\begin{figure}[htb!]
\begin{center}
\includegraphics[scale=0.3]{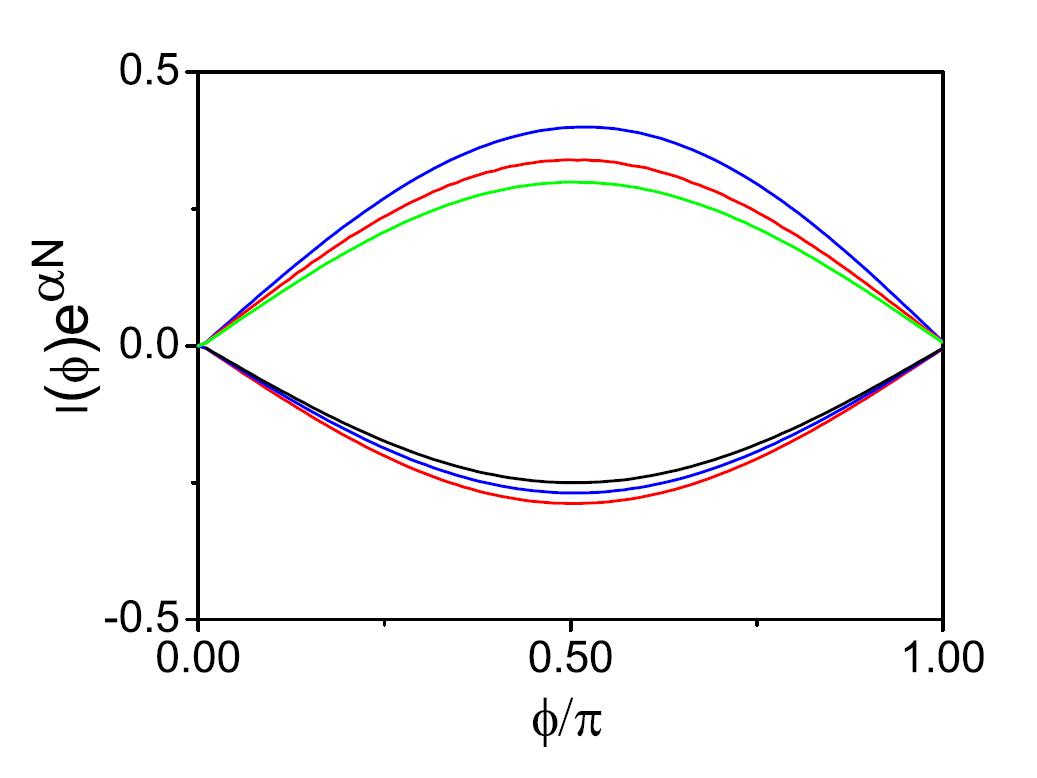}
\caption{Josephson current-phase relation for the quantum dot array model analyzed in Ref. \cite{berg07}
with $U=10\Delta$ at half-filling. The curves in the positive half plane correspond to $N=2,4$ and 6
(from top to bottom) while those taking negative values correspond to $N=1,3$ and 5 (from bottom to top).
The current is scaled by and exponential factor $\exp{\alpha N}$ with $\alpha \simeq 1.8$ in units of $e\Delta/h$. Reprinted figure with permission
from F.S. Bergeret {\it et al.}, Physical Review B {\bf 76}, 174510, 2007 \cite{berg07}.
Copyright (2007) by the American Physical Society.}
\label{bergeret-07-fig7}
\end{center}
\end{figure}

The evolution of the Josephson effect as a function of the number of dots connected in series
was studied in Ref. \cite{berg07}. The model considered in this work with all dot levels 
fixed at a same vale $\epsilon$, with the same local Coulomb repulsion $U$ and with dots
connected by a fixed hopping parameter $t$, is equivalent to a finite Hubbard chain connected
to two superconducting leads. The ground state properties of this model were obtained 
using the zero band-width limit description of the leads discussed in Subsect. \ref{diagonalization} and 
employing the Lanczos algorithm. Fig. \ref{bergeret-07-fig2} illustrates the evolution of the
phase diagram as $N$, the number of dots in the chain, is increased. One can clearly distinguish
the case of even and odd $N$. In the last case the diagram is similar to the single dot case
with a central $\pi$-phase region corresponding to the half-filled case. On the contrary, for
even $N$ the $\pi$-phase is absent around half-filling due to the dominance of antiferromagnetic
correlations between spins in neighboring dots. One can also notice the appearance of additional
narrower regions of $\pi$-phase character corresponding to fillings with odd number of electrons
in the dots region. The authors also analyzed the current-phase relation as a function of $N$
for the half-filled case. Fig. \ref{bergeret-07-fig7} shows that the critical current 
scales as $e^{-\alpha N}$ with a different sign depending on the parity of $N$. This behavior is
consistent with the prediction of field theoretical calculations for a 1D Luttinger liquid with
repulsive interactions where the fixed point corresponds to the absence of Josephson coupling
in the limit of an infinite long chain \cite{affleck00}.  

\subsection{Multilevel quantum dots}
\label{multilevel}

So far only the single-level Anderson model has been considered for describing
an individual quantum dot. A proper description of
actual physical realizations of quantum dots could require to consider
a multilevel generalization of this model. This has been already pointed out
in connection with the experiments of Ref. \cite{dam06} on InAs nanowires, whose
results where qualitatively accounted for using a multilevel model in which
two orbitals with different parity were involved. 

\begin{figure}
\begin{center}
\includegraphics[scale=0.35]{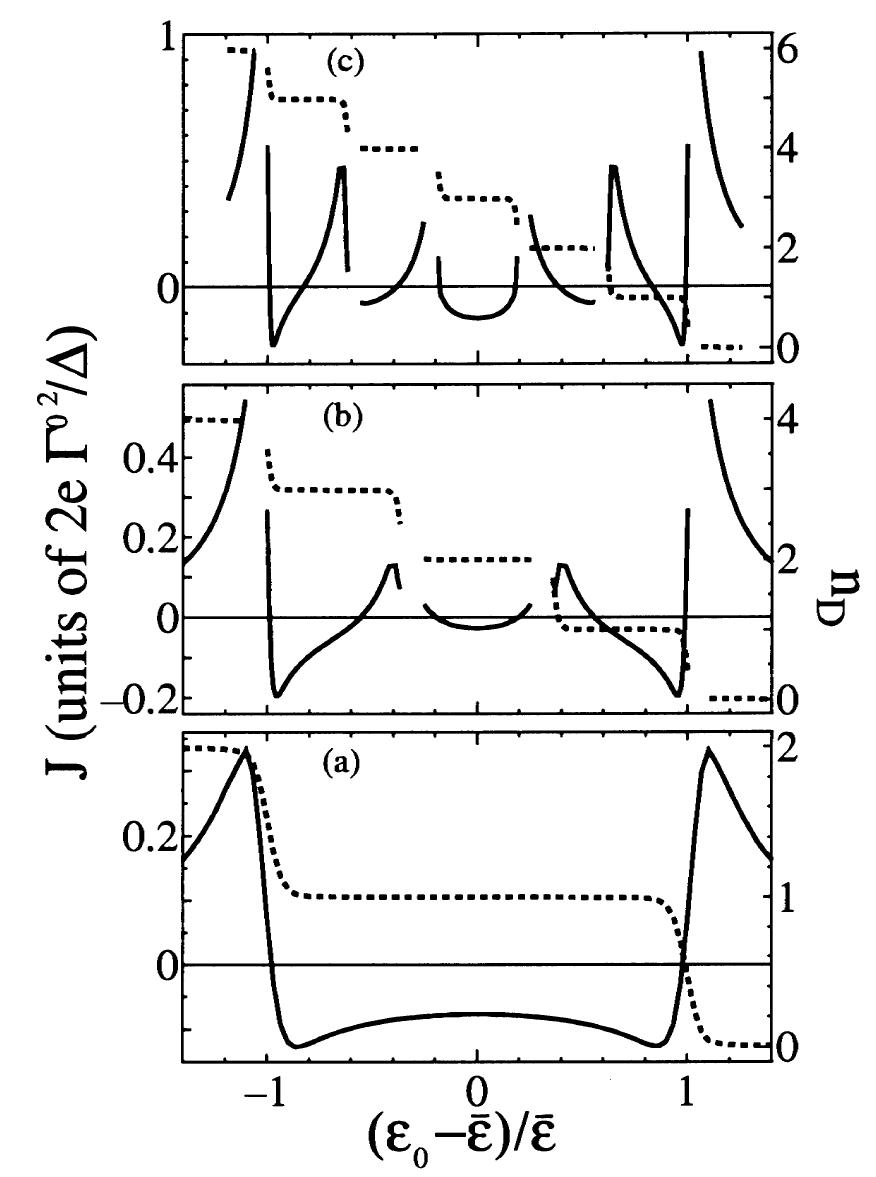}
\caption{Josephson current at $\phi=\pi/2$ in the multilevel S-QD-S model of Ref. \cite{shimizu98} 
as a function of the level position for the cases $n=1$ (a), $n=2$ (b) and $n=3$ (c), where $n$
denotes the level degeneracy. The dotted lines indicate the evolution of the dot total charge $n_d$.
Reprinted figure with permission
from Y. Shimizu {\it et al.}, Journal of the Physical Society of Japan {\bf 67}, 1525, 1998
\cite{shimizu98}. Copyright (1998) by the Physical Society of Japan.}
\label{shimizu-98-fig1}
\end{center}
\end{figure}

Multilevel effects in the Josephson current through a QD were first addressed
in Ref. \cite{shimizu98} by means of the Hartree-Fock approximation. The authors
showed that when non-diagonal processes involving different dot levels are
relevant the system can behave as a $\pi$-junction even in the absence of
a magnetic ground state. This is illustrated in Fig. \ref{shimizu-98-fig1}
where the Josephson current at $\phi=\pi/2$ is plotted as a function of 
the dot levels position, $\epsilon_0$, for the cases with 1, 2 and 3 levels
and including diagonal and non-diagonal couplings to the leads with the same
value $\Gamma$. One can clearly notice that whereas in the lower panel
(single level) the $\pi$-phase is only present for odd number of electrons
($n_d=1$) in the 2 and 3 level cases (panels b and c) the $\pi$ behavior is
also present for even occupancy plateaus. 

Similar ideas were discussed in Ref. \cite{rozhkov01}
for a multilevel situation with nearly degenerate levels ($\delta \epsilon \ll \Delta$)
connected at
the same two points to the leads and using the cotunneling approach.
The authors pointed out that the $\pi$-junction behavior is linked to two-particle
processes in which one of the electrons proceeds through an occupied state and the
other through an empty one.  

Further analysis of this multilevel case was presented in Ref. \cite{lee10} using 
both perturbation theory (cotunneling approach) and NRG calculations. They considered
a two-level situation coupled in parallel to single channel leads and including
an exchange term $J$ between the electron spins in each dot level. In the normal case
this exchange term for $J<0$ would drive a singlet-triplet transition.
Two different situations
are distinguished: a case in which the two orbitals have the same "parity" (i.e. 
${\mbox sign}(t_{1L}*t_{1R}) = {\mbox sign}(t_{2L}*t_{2R})$, where $t_{j\alpha}$ are the hopping from the $j$ 
level to the lead $\alpha$) and a case in which the parities are different.
In the first case and for $t_{1L}/t_{1R}=t_{2L}/t_{2R}$ the dot levels are only
coupled to the symmetric combination of the two leads, yielding an effective
one channel problem. In the second case the problem is equivalent to a two channel
two impurity model with exchange coupling.

\begin{figure}
\begin{center}
\includegraphics[scale=0.3]{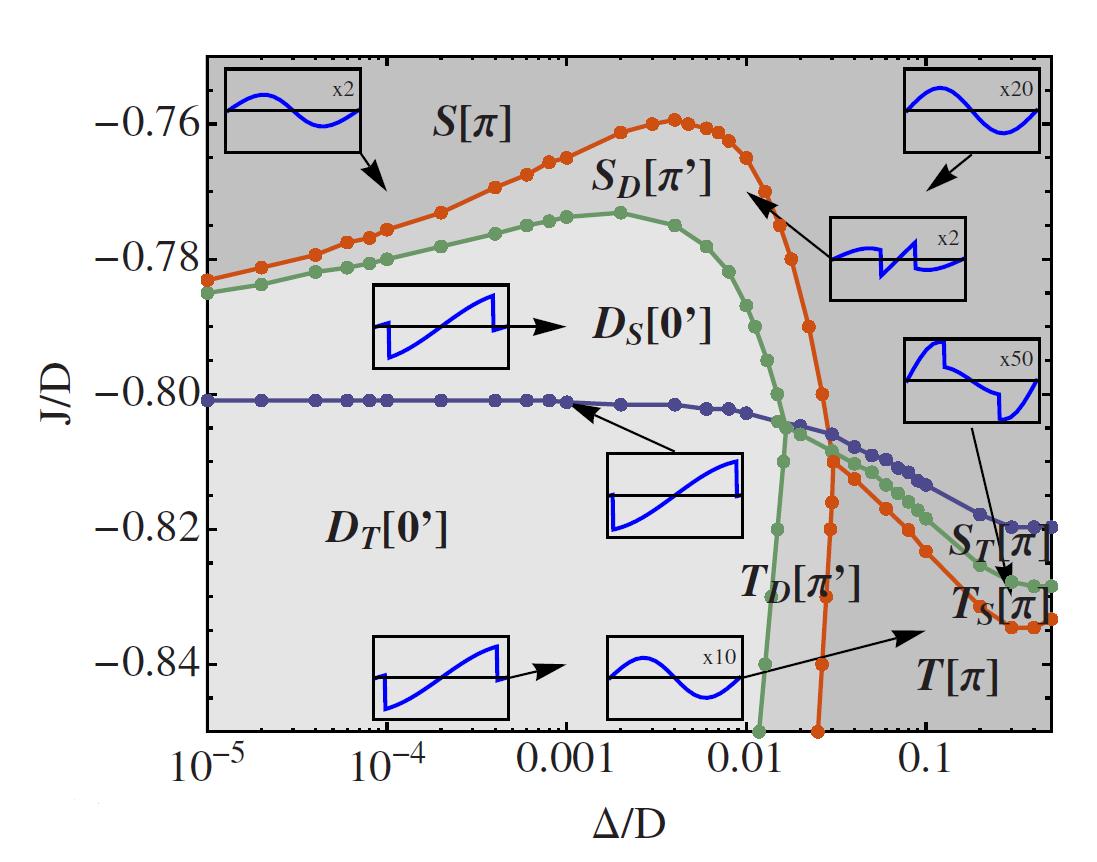}
\caption{Phase diagram in the $J-\Delta$ plane for the two level dot model of Ref. \cite{lee10} 
corresponding to an effective one channel situation with symmetric coupling to the leads.
The notation for the different phases is indicated in the main text.
Reprinted figure with permission
from M. Lee {\it et al.}, Physical Review B {\bf 81}, 155114, 2010 \cite{lee10}.
Copyright (2010) by the American Physical Society.}
\label{lee-10-fig10a}
\end{center}
\end{figure}

\begin{figure}
\begin{center}
\includegraphics[scale=0.3]{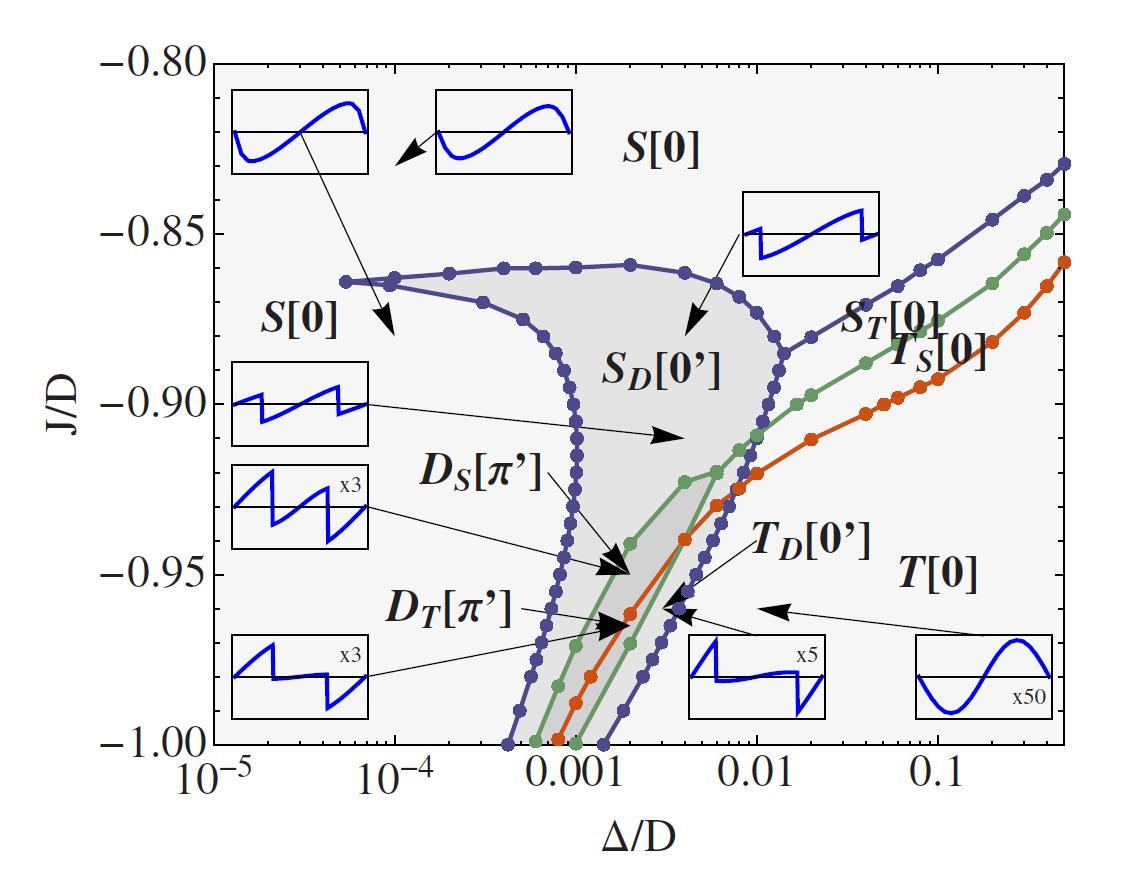}
\caption{Same as Fig. \ref{lee-10-fig10a} for the case corresponding to an effective two-channel
two-impurity model. Reprinted figure with permission
from M. Lee {\it et al.}, Physical Review B {\bf 81}, 155114, 2010 \cite{lee10}.
Copyright (2010) by the American Physical Society.}
\label{lee-10-fig14a}
\end{center}
\end{figure}

The system exhibits a very rich phase diagram depending on the several
model parameters. For illustration we show in Fig. \ref{lee-10-fig10a} and \ref{lee-10-fig14a}
the obtained phase diagrams in the $\Delta,J$ plane for the single channel and the two channel 
cases respectively for symmetric coupling to the leads. The different phases are denoted
by a capital letter indicating the spin of the dominant ground state, which can be either
$S$, $D$, or $T$ for spin 0, 1/2 or 1. In the case of mixed phases where the ground 
state changes with $\phi$ a subindex is included indicating the character of the
metastable state. There is finally a label which can be either 0 or $\pi$ indicating
the character of the current-phase relation in the dominant phase. The different types
of current-phase relations are shown as insets in Figs. \ref{lee-10-fig10a} and \ref{lee-10-fig14a}.

\begin{figure}[htb!]
\begin{center}
\includegraphics[scale=0.3]{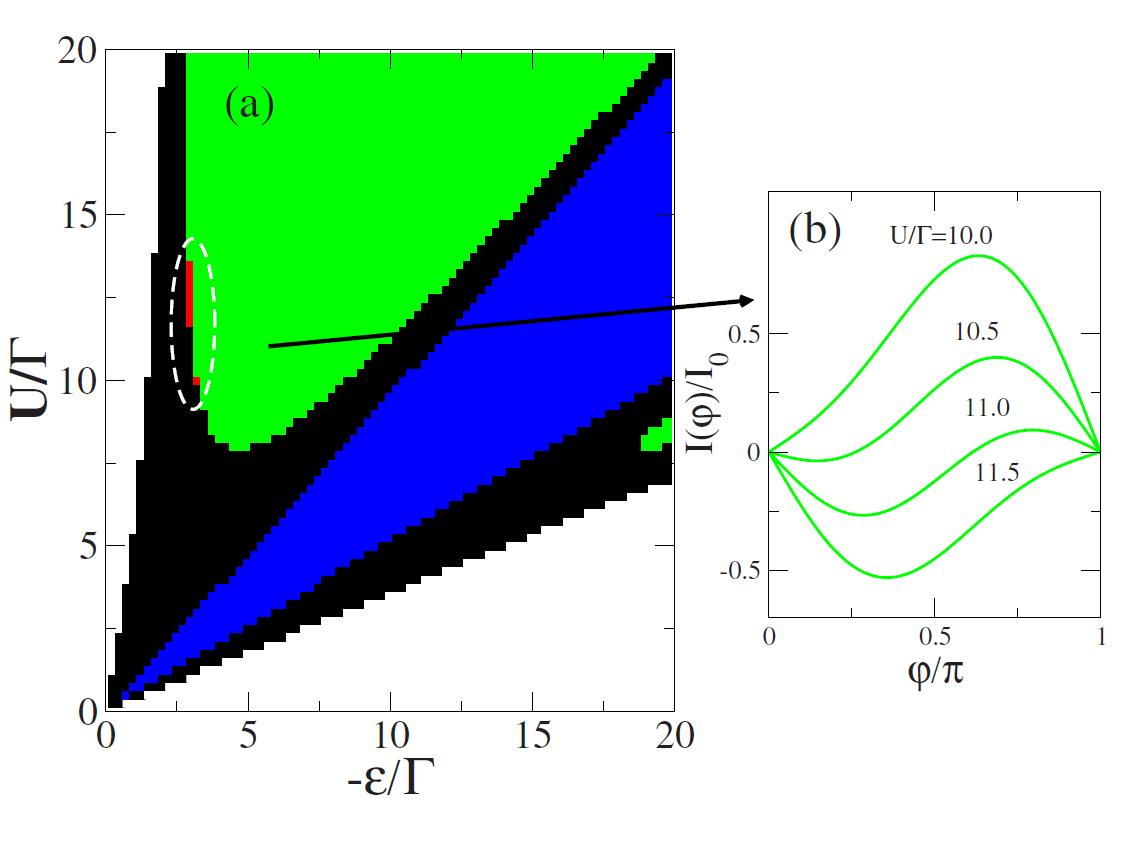}
\caption{Phase diagram in the $U-\epsilon$ plane for the SU(4) Anderson model with superconducting leads
analyzed in Ref. \cite{zazu10} obtained in the ZBW approximation for $\Delta=10\Gamma$. White regions correspond
to the $(S,T)=(0,0)$ and green to $(S,T)=(1/2,1/2)$. The black and blue regions correspond to mixed phases
with $(S,T)=(0,0)$ at $\phi=0$ and $(S,T)=(1/2,1/2$ or $(S,T)=(1,0)-(0,1)$ at $\phi=\pi$ respectively.
The right panel shows the evolution of the current-phase relation for $\epsilon/\Gamma = -5$ and several
values of $U/\Gamma$. The small region in red inside the dashed white line corresponds to a $\pi'$ mixed phase. Reprinted figure with permission
from A. Zazunov {\it et al.}, Physical Review B {\bf 81}, 012502, 2010 \cite{zazu10}.
Copyright (2010) by the American Physical Society.}
\label{zazunov-10-fig3}
\end{center}
\end{figure}

Another multilevel situation which has been recently analyzed is the case of a
four-fold degenerate level coupled to multichannel leads \cite{zazu10}. This situation 
has been
found in experiments on high quality CNT quantum dots with normal leads showing
clear signatures of four-fold degeneracy both in the Coulomb blockade and in the
Kondo regimes \cite{jarillo05,makrovski07}. The SU(4) Kondo effect has been analyzed
in the normal case by several authors like \cite{yeyati99}, \cite{choi05-SU4} and \cite{lim06}.
In Ref. \cite{zazu10} Zazunov et al. have studied the Josephson effect
in this case by considering a generalized SU(4) Anderson model with superconducting leads.
They obtained analytical results in two opposite regimes corresponding to the deep Kondo
limit $T_K \gg \Delta$ and the cotunneling limit. In the first case they obtained
a current-phase relation which corresponds to the superposition of effective non-interacting
channels with renormalized transmission $\tau = 1/2$, therefore deviating from the SU(2)
case where the unitary limit can be reached. In the cotunneling case
with $U \rightarrow \infty$ a 0-$\pi$ transition at $\epsilon_0=0$ was obtained,
as in the SU(2) case (see Ref. \cite{glazman89}) but with a different ratio
of the critical currents $I_c(-|\epsilon_0|)/I_c(|\epsilon_0|)=-1/4$.
The reduction of this ratio by a factor 2 with respect to the SU(2) case can be readily
understood by considering the number of processes leading to Cooper pair transfer
through the dot.

The authors also analyzed numerically the phase diagram of the model in the
regime $\Delta \gg \Gamma$. A first insight is obtained by taking the limit $\Delta \rightarrow \infty$
in which case the relevant Hilbert space is reduced to the $2^4$ dot states. 
Conservation of the total spin, $S$, and orbital pseudo-spin, $T$, allows to further decouple
this Hilbert space into three different sectors: $(S,T)=(0,0)$, $(S,T)=(1/2,1/2)$ and
$(S,T)=(1,0)$ or $(0,1)$ (these last two are degenerate in the SU(4) case).
The main features of the phase diagram obtained in this limit were shown to be 
preserved in the finite $\Delta \gg \Gamma$ regime, which was analyzed by means of
the zero band-width model for the leads. The phase diagram, illustrated in Fig. 
\ref{zazunov-10-fig3}, is essentially the same as in the $\Delta \rightarrow \infty$ limit except
for two properties: first, the appearance of tiny $\pi'$ type mixed
phase (indicated by the red regions in Fig. \ref{zazunov-10-fig3}) and second by the
change in the character of the current-phase relation of the $(S,T)=(1/2,1/2)$ with increasing
$U$. The panel on the right shows that this relation
evolve with $U/\Delta$: while for $\Delta \gg U$ it is 
typically of 0 type, for $U >\Delta$ it becomes of $\pi$ type. 

In closing this subsection we mention the recent appearance of a work by Lim et al.
analyzing the effect of including spin-orbit interactions within this model \cite{lim11},
which can be relevant for small radius CTNs \cite{kuemmeth}.
 
\subsection{Multidot-multiterminal systems with normal and superconducting leads}
\label{multiterminal}

\subsubsection{Josephson effect through a quantum dot in a three terminal configuration}

\begin{figure}
\begin{center}
\includegraphics[scale=0.25]{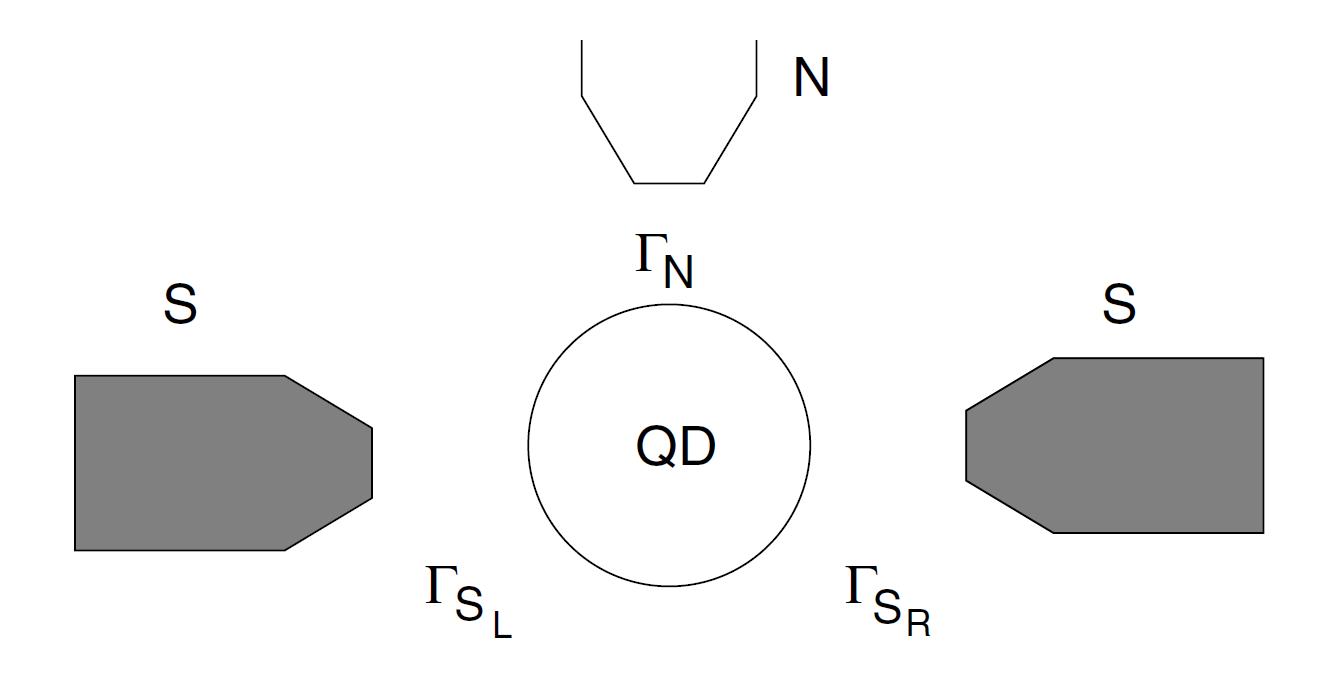}
\includegraphics[scale=0.27]{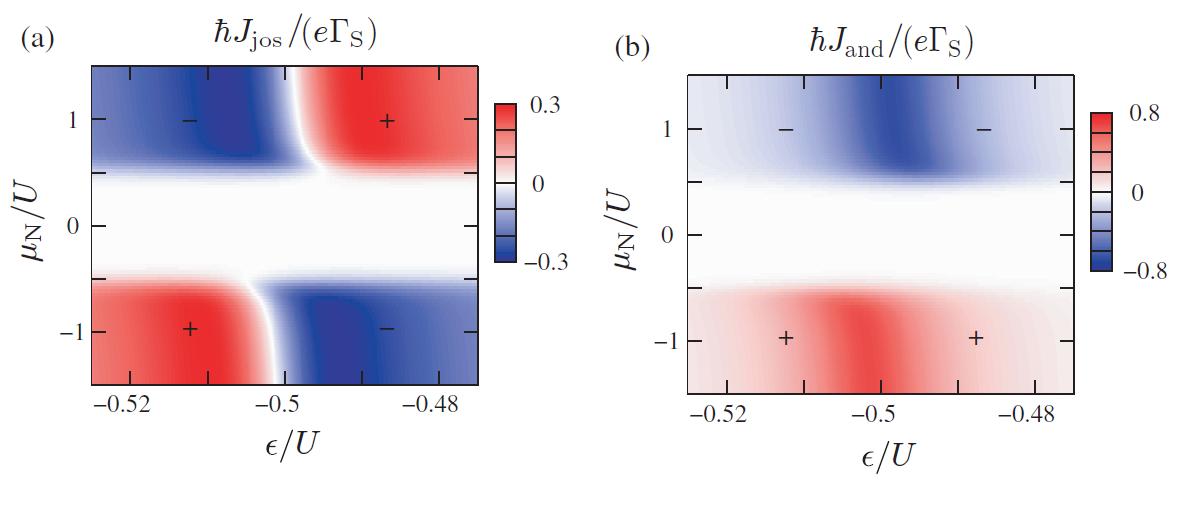}
\caption{Color map of the Josephson critical current (lower left panel) and Andreev current (lower right panel) 
for the QD dot coupled to two superconducting leads and an additional voltage biased normal one
in the $\mu_N/U-\epsilon/U$ plane. The upper panel gives a schematic representation of the setup considered in this work. Reprinted figure with permission
from M. Pala {\it et al.}, New Journal of Physics {\bf 9}, 278, 2007 \cite{pala07}.
Copyright (2007) by IOP Publishing Ltd.}
\label{pala-07-fig2}
\end{center}
\end{figure}

The Josephson effect through a single dot coupled to two superconductors and to a third normal
lead has been analyzed by several authors. This configuration is schematically depicted in the
upper panel of Fig. \ref{pala-07-fig2}.
An interesting non-equilibrium enhancement of
the Josephson effect was predicted in Ref. \cite{pala07}. The authors applied the real time
diagrammatic approach commented in Subsect. \ref{diagrammatic}  with the tunneling rates to the leads calculated
to the first order in $\Gamma_{S,N}$. They found that a significant Josephson current can
be induced by the voltage bias applied to the normal leads even in the situation 
$\Gamma_S < k_B T$ where the equilibrium Josephson current would be negligible.

Fig. \ref{pala-07-fig2} shows a color map of the Josephson critical current as a function of both the
level position and the chemical potential on the normal lead, $\mu_N$. One can notice the presence
of a white region for $|\mu_N| \sim < U/2$ where the Josephson current is negligible. Outside this
region it becomes of the order of $\sim e\Gamma_S/\hbar$ and exhibits a transition from 0 to $\pi$
behavior depending on the level position. The origin of this peculiar behavior can be traced to
the enhancement of the proximity effect pairing amplitude on the dot due to the non-equilibrium
population which increases the double population probability that is strongly suppressed at $\mu_N=0$
due to the charging energy. In a subsequent publication by the same group \cite{governale08} the
authors considered the same effect in the limit $\Delta \rightarrow \infty$ which allows to account
for the Josephson effect to all orders in $\Gamma_S$. The results obtained are qualitatively similar
and allow to identify the lines separating different regions with the Andreev bound states of the
S-QD-S system.

\begin{figure}
\begin{center}
\includegraphics[scale=0.25]{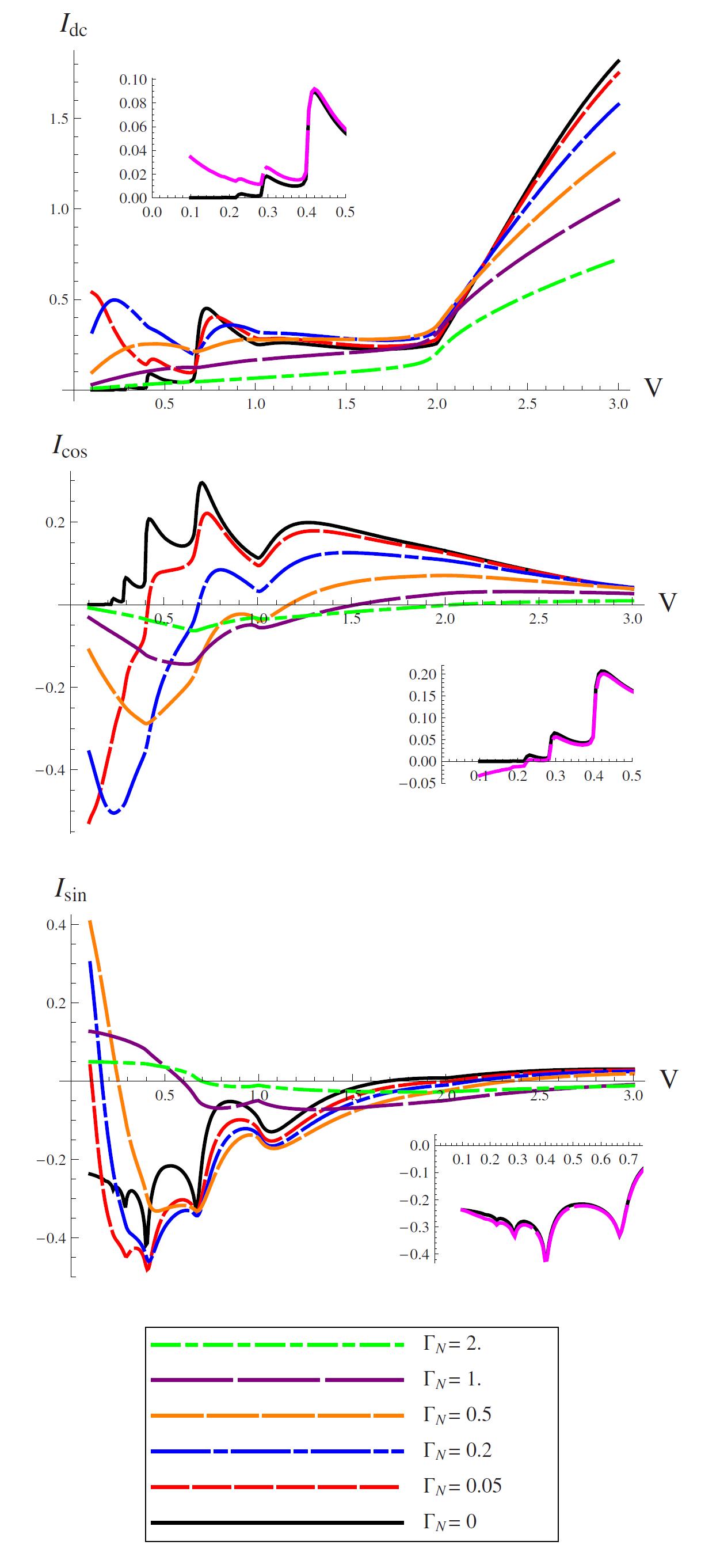}
\caption{dc and the two first harmonics of the current-voltage characteristic of a non-interacting
single level S-QD-S system with an additional normal lead connected to the dot with increasing
coupling $\Gamma_N$ analyzed in Ref \cite{jonckheere09}. The other parameters are $\epsilon_0$ 
and $\Gamma_S=0.2\Delta$. The insets show the comparison
of the dc current and the corresponding harmonics for $\Gamma_N=0$ (black line) and $\Gamma_N=0.02\Delta$ (red line). Reprinted figure with permission
from T. Jonckheere {\it et al.}, Physical Review B {\bf 80}, 184510, 2009 \cite{jonckheere09}.
Copyright (2009) by the American Physical Society.}
\label{jonckheere-09-fig4}
\end{center}
\end{figure}

It should be also mentioned within this context the work by Jonckheere et al. \cite{jonckheere09}
in which the effect of a third normal lead on the ac Josephson effect in a non-interacting
voltage biased S-QD-S system was analyzed. The main idea of this work was to study the transition
from the coherent MAR regime to the incoherent limit controlled by the coupling $\Gamma_N$ to
the normal lead. They show that while the dc Josephson current exhibits a monotonous decrease
with increasing $\Gamma_N$ the behavior of the dc quasiparticle current and its first ac
harmonics have a much more involved evolution, which is illustrated in Fig. \ref{jonckheere-09-fig4}.

\subsubsection{Andreev transport through double quantum dots}

\begin{figure}
\begin{center}
\includegraphics[scale=0.4]{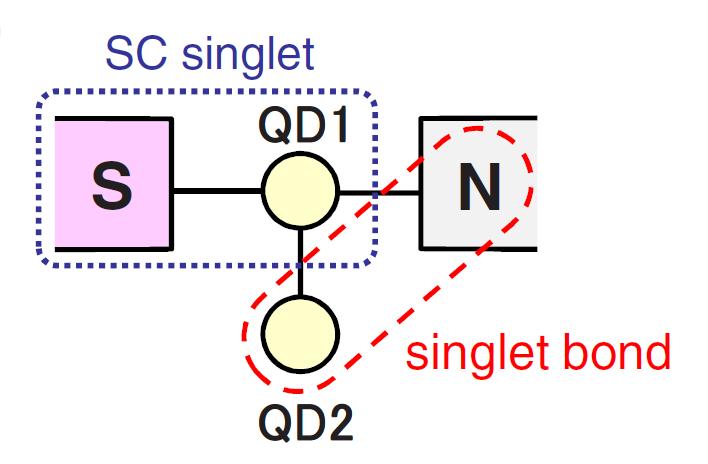}
\includegraphics[scale=0.2]{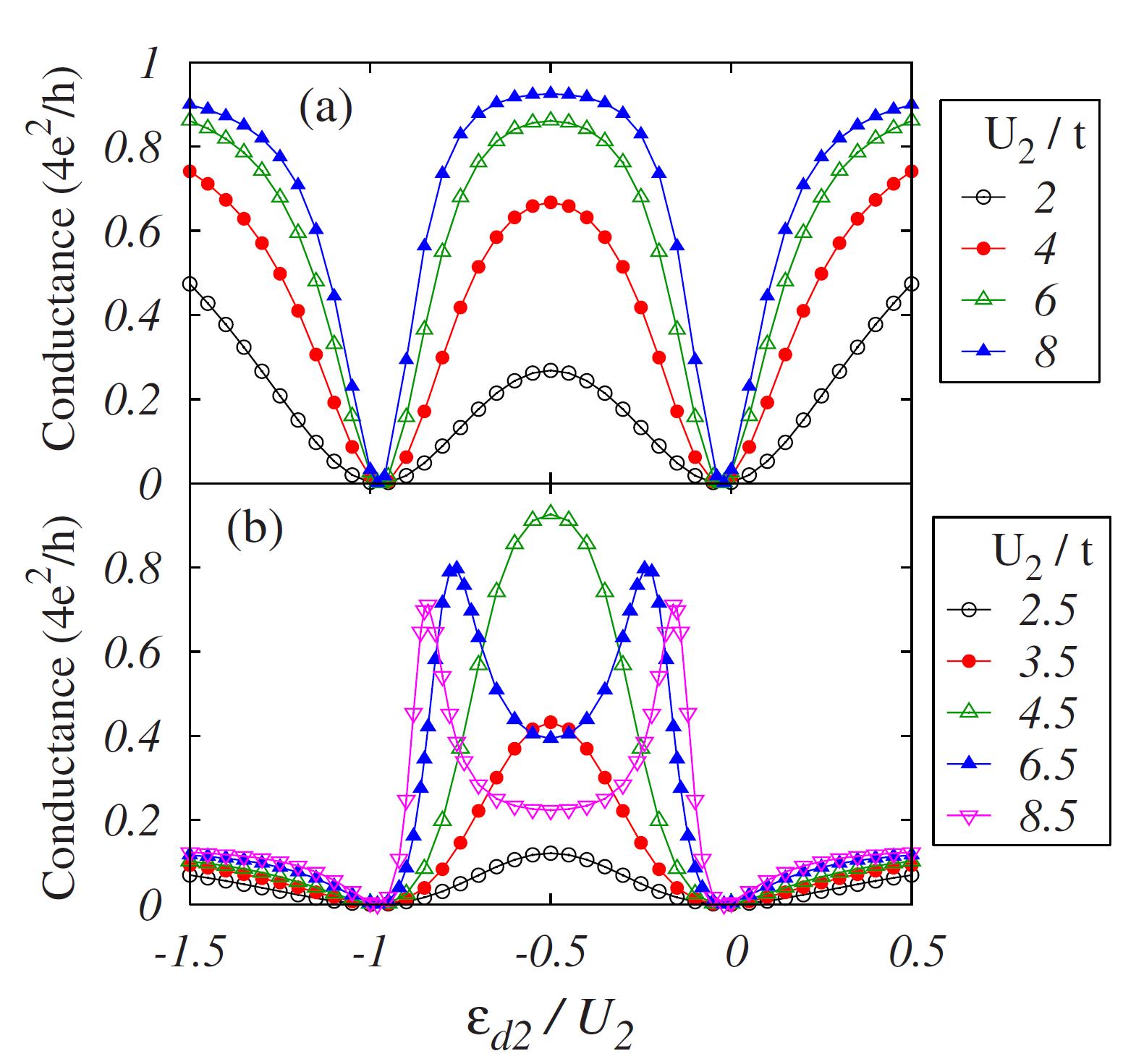}
\caption{Linear conductance for the side coupled dot between a normal and a superconducting lead
considered in Ref. \cite{tana08} as a function of the lateral dot level $\epsilon_2$ and for
different values of the corresponding Coulomb interaction parameter $U_2$. The upper panel shows a
schematic representation of the setup considered in this work. 
Reprinted figure with permission
from Y. Tanaka {\it et al.}, Physical Review B {\bf 78}, 035444, 2008 \cite{tana08}.
Copyright (2008) by the American Physical Society.}
\label{tanaka-08-fig2}
\end{center}
\end{figure}

Andreev transport in double quantum dot systems connected to normal and superconducting leads
has been studied so far in a few works. Tanaka et al. \cite{tana08} considered the case of
a T-shape geometry where a central dot is coupled to both electrodes and a second dot is only 
side-coupled to the central one, as shown in the upper panel of Fig. \ref{tanaka-08-fig2}. 
The authors used the NRG method in the $\Delta \rightarrow \infty$
limit where, as discussed in Subsect. \ref{diagonalization}, the superconducting lead acts as a simple boundary condition
for Andreev reflection. They focused in the case where interactions are neglected in the central
QD analyzing the effect of increasing $U$ in the lateral dot on the conductance through the system.
The results of Fig. \ref{tanaka-08-fig2} show that the Andreev conductance gradually approaches the unitary
limit as the side dot $U$ increases for the case $\Gamma_N = t$, where $t$ is the hopping between the
two dots. On the other hand, for smaller values of $\Gamma_N$ the evolution of the conductance with
$U$ is non-monotonous exhibiting for the symmetric case first an increase followed by a decrease. 
In this work it was also shown that the conductance in the electron-hole symmetric case can be tuned 
to the unitary limit for fixed $U$ by varying the coupling to the superconducting lead $\Gamma_S$.
This possibility is analogous to the one already discussed for the single SQDN system in Sect. \ref{NQDS}.

\begin{figure}
\begin{center}
\includegraphics[scale=0.25]{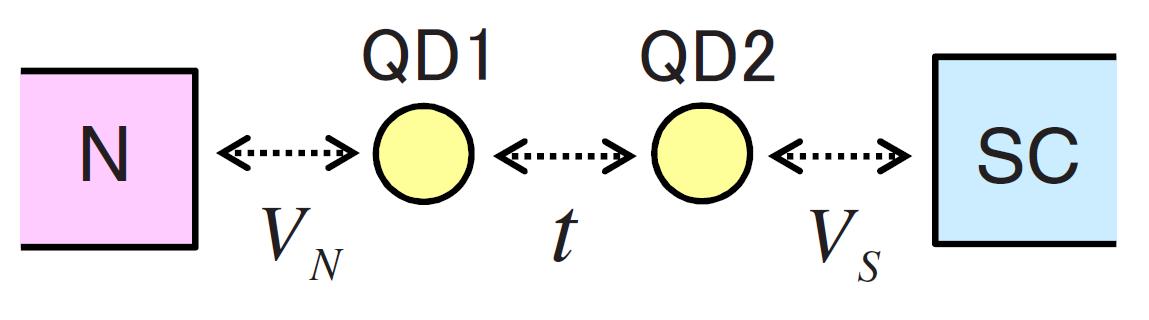}
\includegraphics[scale=0.2]{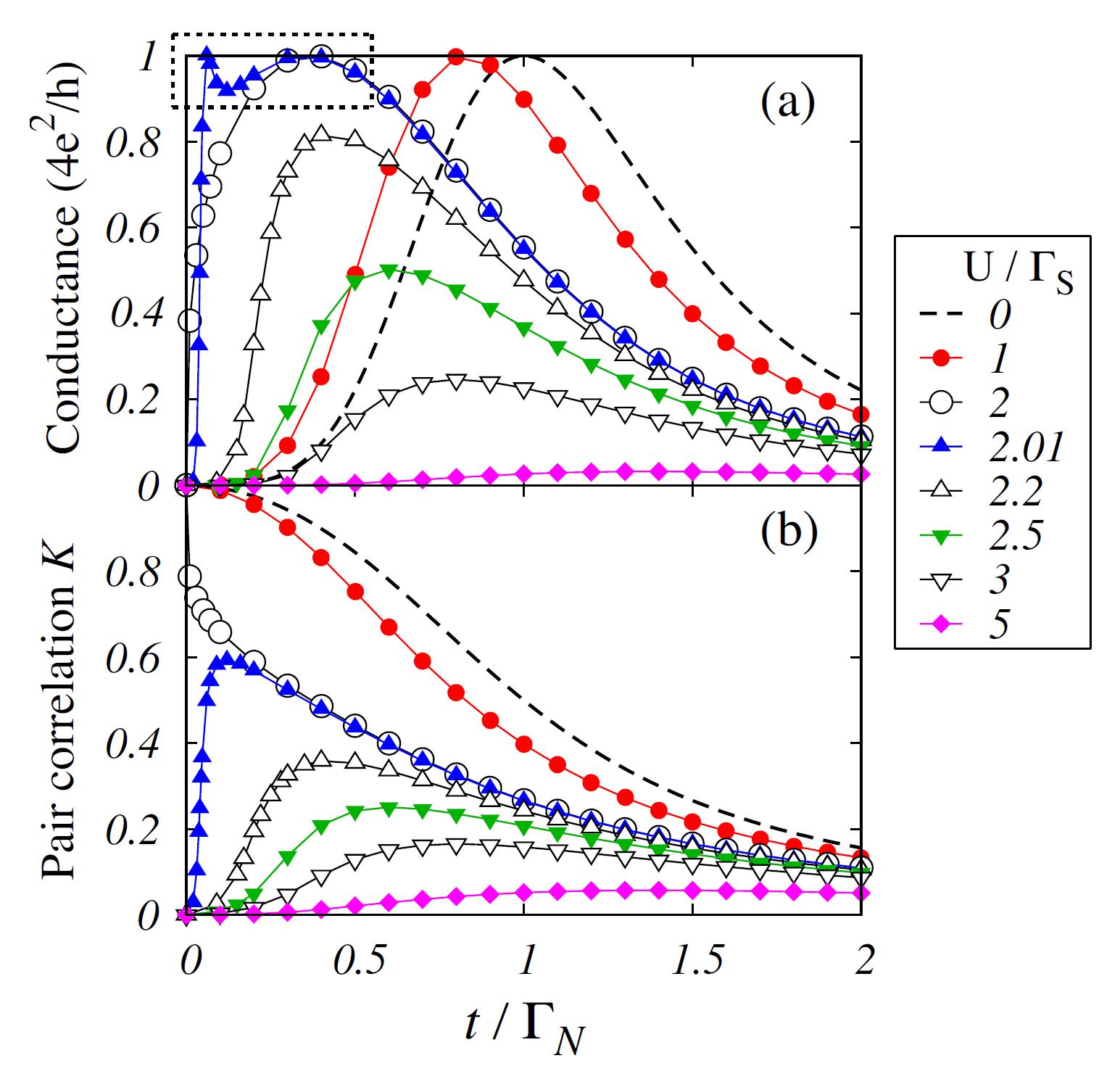}
\caption{(a) Linear conductance for the series double quantum dot between a normal and a superconducting lead
discussed in Ref. \cite{tana10} as a function of the interdot tunneling $t$ and for different values of $U/\Gamma_s$. 
Panel (b) shows the corresponding induced pairing correlation in the dots region. 
The results were obtained using the NRG method in the $\Delta \rightarrow \infty$ limit. Upper panel: setup 
considered in this work. Reprinted figure with permission
from Y. Tanaka {\it et al.}, Physical Review B {\bf 81}, 075404, 2010 \cite{tana10}.
Copyright (2010) by the American Physical Society.}
\label{tanaka-10-fig3}
\end{center}
\end{figure}

In a subsequent paper \cite{tana10}, the same authors considered the case of a double QD in series connected
to a normal and a superconducting lead (see upper panel of Fig. \ref{tanaka-10-fig3}) using the same theoretical approach. 
They first showed that for the case where the dot coupled to the normal lead is in the
electron-hole symmetric condition the problem can be mapped into an effective normal 
two impurity Anderson model in terms of the Bogoliubov operators. Furthermore, this transformation
allows to calculate the conductance using the Friedel sum rule in terms of a phase-shift given
by 

\begin{equation}
G = \frac{4e^2}{h} \left(\frac{\Gamma_S}{E}\right) \sin(\pi {\cal Q}),
\end{equation}
where $E = \sqrt{\epsilon_2^2 + \Gamma_S^2}$, $\epsilon_2$ being the level for the dot
coupled to the superconductor and ${\cal Q} = \sum_{\sigma} \left(<\gamma^{\dagger}_{1\sigma}\gamma_{1\sigma}> + <\gamma^{\dagger}_{2\sigma}
\gamma_{2\sigma}>\right)$, $\gamma_{i\sigma}$ indicating the Bogoliubov operators.
The authors identified three different regimes. For small $t$ and $U <2\Gamma_s$ a regime corresponding to
the a local superconducting singlet is found while for $U>2\Gamma_s$ the Kondo singlet state is formed. On the other hand,
for large $t$ the antiferromagnetic coupling between the dots dominates. The behavior of the 
conductance in these three different regimes is summarized in Fig. \ref{tanaka-10-fig3}. As can be observed,
for sufficiently small $U/\Gamma_s$  the conductance can reach the unitary limit, the maximum displacing
towards smaller $t$ values as $U$ increases and eventually for $U/\Gamma_s \sim> 2$ the unitary limit
cannot be reached. A surprising feature appears for $U/\Gamma_s \sim 2$ where the conductance reaches
the unitary limit for two different $t$ values, indicated by the dashed rectangle in Fig. \ref{tanaka-10-fig3}.

\begin{figure}[htb!]
\begin{center}
\includegraphics[scale=0.4]{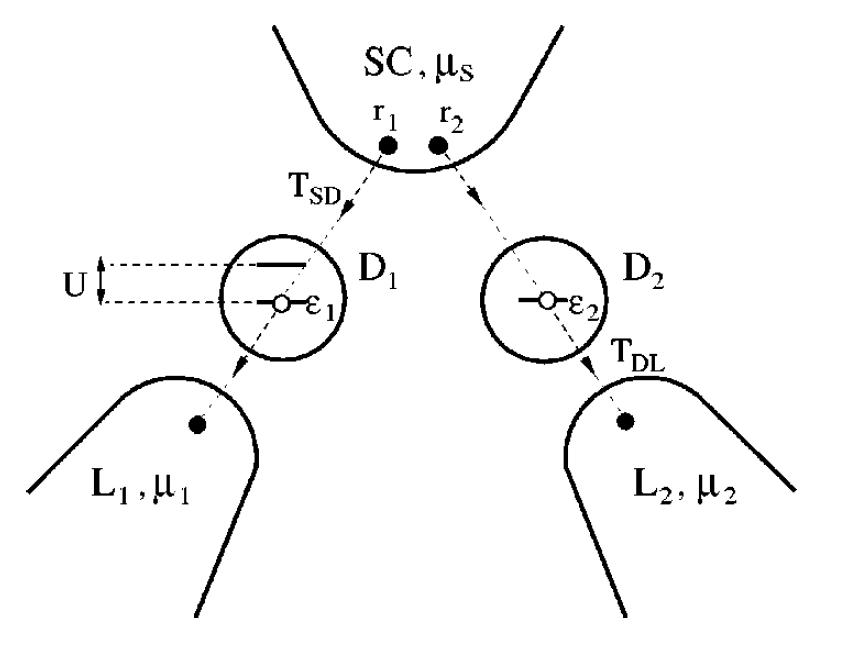}
\caption{Schematic representation of a double quantum dot coupled to a central superconducting and two
lateral normal leads in the proposal of Ref. \cite{recher01} for producing entangled electron
pairs by means of non-local Andreev processes. Reprinted figure with permission
from P. Recher {\it et al.}, Physical Review B {\bf 63}, 165314, 2001 \cite{recher01}.
Copyright (2001) by the American Physical Society.}
\label{recher-01-fig1}
\end{center}
\end{figure}

\subsubsection{Non-local Andreev transport through single or double quantum dots}

Double dots coupled to several normal and superconducting leads are receiving recently considerable
attention in connection to the possibility of producing non-local entangled electrons from the splitting
of Cooper pairs \cite{recher01,basel,takis}. The basic ideas were first put forward in Ref. \cite{recher01} 
where the multiterminal geometry of Fig. \ref{recher-01-fig1} was considered. In this configuration, when
a Cooper pair is injected from the SC lead it can either be transmitted as a whole to one of the normal
leads by a {\it local} Andreev process or split with each of the electrons in the pair transmitted 
to a different lead (corresponding to a {\it non-local} or crossed Andreev reflection process). 
The advantage of the DQD set up is twofold: on the one side it allows to tune independently the
two dot levels and on the other hand Coulomb interactions could be used to favor the splitting
processes compared to the local ones. 
While this issue is of a great current interest it goes beyond
the scope of the present review. We would just mention several works addressing the non-local
Andreev transport involving quantum dots and multiterminal configurations in Refs. \cite{konig09,konig10}.

\section{Concluding remarks}
\label{conclusions}

In this review article we have summarized the most relevant 
published work related to superconducting transport in quantum dots systems. 
The large variety of topics that we have covered give an indication of the great
activity which this field has shown in recent years. Due to the limited space
it has become necessary to restrict somehow its content. 
For this purpose we have chosen to give priority to the more basic topics
like Josephson effect and Andreev transport through single level quantum dots,
and had left aside some interesting but more specialized situations
like those involving ferromagnetic materials or unconventional superconductors.
In the same way, we have not covered in this review the response of these systems to 
external ac fields, like photon assisted 
transport in S-QD-S \cite{cho99,zhu02} or N-QD-S systems \cite{zhao98,zhao01} and
adiabatic pumping in NDQS systems \cite{sple07}. 

Even within the basic topics discussed in this review there remain
several issues which are not completely understood and deserve further analysis. 
Among them we may point out: 
1) the conflicting description of ABs within the different approximation schemes
for S-QD-S systems, as discussed in Sect. \ref{SQDS-eq};
2) a more complete analysis of the phase diagrams of double QDs which
so far has been restricted to certain parameter ranges, as commented in
Sect. \ref{multi};
3) clarifying the interplay of Kondo and Andreev transport in N-QD-S beyond the
linear regime (discussed in Sect. \ref{NQDS}, and
4) extending the analysis of the MAR regime in S-QD-S beyond the limit of
weak interactions analyzed in Sect. \ref{SQDS-neq}.

It could be expected that the intense experimental and theoretical activity
within this field will continue to grow in the next years. 
In addition to the already commented open issues there are several directions
in which the research can be oriented. 
There are, on the one hand, other transport properties to be explored 
in the systems considered in this review, specially those related
to current fluctuations. Some recent work address in fact the
full-counting statistics in a non-interacting N-QD-S system \cite{soller11},
but certainly there is a lot of open issues regarding the effect of
interactions and the non-local current correlations in multiterminal
configurations. In fact the analysis of these correlations can 
provide insight on the issue of non-local entanglement produced
by the splitting of Cooper pairs, as has been shown in the case of
diffusive samples \cite{bignon} and ballistic conductors \cite{samuelsson}.

On the other hand one could expect a renewed interest in these
systems arising from the inclusion of recently discovered 
materials, like graphene and topological insulators.
While graphene quantum dots have been already successfully produced
experimentally and combined with superconductors like in Ref. \cite{dirks11},
proposals of combining these systems with topological insulators are
still on an speculative level \cite{golub11}. We expect nevertheless
that these issues would exhibit a great development in the near future.
 
\section*{Acknowledgements}

The authors would like to acknowledge the contribution of many people to their
work within the field of superconducting transport in quantum dot systems. 
During many years they have had the pleasure to collaborate in related topics with
F. Flores, F.J. Garc\'{\i}a-Vidal, J.C. Cuevas, A. L\'opez-D\'avalos,
E. Vecino, S. Bergeret, C. Urbina, E. Scheer, N. Agra\"it, G. Rubio, J.M. van Ruitenbeek, D. Esteve, 
M.H. Devoret, P. Joyez, C. Sch\"onenberger, T. Klapwijk, R. Egger, A. Zazunov, J.D. Pillet, C. Bena, 
T. Kontos, P. Roche, F. Portier and C. Strunk. 
Financial support from Spanish MICINN through project FIS2008-04209 and by EU FP7 SE2ND project
is acknowledged.

\bibliography{review-ABS}

\end{document}